\documentclass[12pt,sort&compress]{elsarticle}
\usepackage[utf8]{inputenc}

\usepackage{bm}
\usepackage{placeins}
\usepackage{booktabs} 
\usepackage{mathrsfs}
\usepackage{amsmath}
\usepackage{amsfonts}
\usepackage{amsthm}
\usepackage{upgreek}
\usepackage{color}
\usepackage{sansmath}

\usepackage{array}

\usepackage{floatrow} 
\usepackage{graphicx}
\usepackage[tableposition=top]{caption}

\usepackage{caption}
\usepackage{subcaption}
\usepackage{float}
\usepackage{fancyhdr}
\usepackage{url}
\usepackage{multicol}
\usepackage[top=1in,bottom=1in,left=1in,right=1in]{geometry}
\usepackage[pdfborder={0 0 0},colorlinks,allcolors=blue]{hyperref}
\usepackage{txfonts}
\usepackage{setspace}
\usepackage{arydshln}
\usepackage{enumitem}
\usepackage{lipsum}
\usepackage{siunitx,booktabs}
\usepackage{nth}
\usepackage{accents} 
\usepackage{multirow} 
\usepackage{float}  
\usepackage{floatrow} 
\theoremstyle{definition}

\allowdisplaybreaks

\usepackage{prettyref}
\newrefformat{fig}{Figure~\ref{#1}}
\newrefformat{Fig}{Fig.~\ref{#1}}
\newrefformat{Eq}{Eq.(\ref{#1})}
\newrefformat{eq}{Eq.(\ref{#1})}
\newrefformat{sec}{Section~\ref{#1}}
\newrefformat{tab}{Table~\ref{#1}}
\newrefformat{Tab}{Table~\ref{#1}}
\newrefformat{alg}{Algorithm~\ref{#1}}
\newrefformat{Alg}{Algorithm~\ref{#1}}

\usepackage[dvipsnames]{xcolor}
\newcommand\hl[1]{%
	\bgroup
	\hskip0pt\color{red!80!black}%
	#1%
	\egroup
}

\usepackage{algorithm}
\usepackage{algpseudocode}
\usepackage{algorithmicx}
\definecolor{shadecolor}{cmyk}{0,0,0,0.03}

\usepackage{amssymb}

 \usepackage{lineno}
\journal{Physics of Fluids}

\begin{document}
\begin{frontmatter}



\title{{\normalsize\textit{This article has been submitted to Physics of Fluids. Final version will appear at} \url{https://aip.scitation.org/journal/PoF}}\\[0.5em]
Large eddy simulation of a utility-scale vertical-axis marine hydrokinetic turbine under live-bed conditions}


\author[sbu]{Mehrshad Gholami Anjiraki}
\author[sbu]{Mustafa Meriç Aksen}
\author[sbu]{Jonathan Craig}
\author[sbu]{Hossein Seyedzadeh}

\author[sbu]{Ali Khosronejad\corref{cor1}}
\ead{}
\cortext[cor1]{Corresponding author}

\address[sbu]{Department of Civil Engineering, Stony Brook University, Stony Brook, NY 11794, USA}

\begin{abstract}
\noindent We present a coupled large-eddy simulation (LES) and bed morphodynamics study to investigate the impact of sediment dynamics on the wake flow, wake recovery and power production of a utility-scale marine hydrokinetic vertical-axis turbine (VAT). A geometry-resolving immersed boundary method is employed to capture the turbine components, the waterway, and the sediment layer. Our numerical findings reveal that increasing the turbine tip speed ratio (TSR) would intensify turbulence, accelerate wake recovery, and increase erosion at the base of the device.
Furthermore, it is found that the deformation of the bed around the turbine induces a jet-like flow near the bed beneath the turbine, which enhances wake recovery. Analyzing the interactions between turbulent flow and bed morphodynamics, this study seeks to provide physical information on the environmental and operational implications of VAT deployment in natural riverine and marine environments.
\end{abstract}


\end{frontmatter}

\section{Introduction}

\label{sec:intro}
\noindent In recent years, even amid a troubled economic wake after the COVID-19 pandemic, renewable energy projects to increase their capacity have progressed tremendously. By the end of 2023, additional renewable energy capacity reached 507 gigawatts (GW), nearly half as much as the increase in renewable energy capacity from 2022 \cite{[1]}. Yet, this rate of improvement is not great enough to reach the goal of tripling global renewable power capacity by 2030, which the International Energy Agency (IEA) established \cite{[1]}. Global investment in renewable energy has grown from USD 576 billion in 2022 to USD 622.5 billion \cite{[2]}. Most renewable power capacity additions come from solar and wind due to their lower generation costs, which the IEA collectively projected to surpass hydropower in renewable energy generation by 2024 \cite{[3]}. In an evolving renewable energy economy, another energy resource that would complement declining hydropower generation is marine energy. Whereas traditional hydropower plants employ dams or penstocks to make an artificial water-head, hydrokinetic converters do not significantly alter the natural waterway to extract marine energy \cite{[4],[5],[6],[7]}.

For the USA alone, marine energy from waves, tides, currents, rivers, and thermal energy can offer 2,300 TWh/yr as an estimated technical resource \cite{[8]}. Marine energy from tidal streams offers the advantages of high predictability and energy yield. Since tidal energy does not depend on specific weather conditions, it is theoretically predictable until the end of its commercial lifetime \cite{[9],[10]}. On the other hand, the lack of an established supply chain raises the costs of construction, installation, and maintenance, all of which comprise the greatest obstacle to investment \cite{[11],[12]}. Other disadvantages include bed-induced turbulence and potential environmental impacts \cite{[11],[13],[14],[15]}. Notwithstanding the localized, varying nature of tidal and riverine hotspots, demonstration projects for tidal power in the USA \cite{[16],[17],[18],[19],[20]} and in Europe \cite{[21]} have evidenced the global applicability of tidal stream turbines as marine energy technology advances \cite{[10],[13]}.  It is possible to adapt the marine technologies from oceanic to riverine environments, of which the latter represents a small but significant aspect of the marine energy economy. Of the estimated 2,300 TWh/yr of technical marine energy in the USA, riverine energy comprises 99 TWh/yr, or $2.3\%$ of the United States’ electricity generation, which could power 9.3 million homes in many states without access to oceanic resources \cite{[8]}. Albeit a mere fraction of the estimated technical energy, riverine flows represent a viable and yet underexplored marine energy resource. Some sites, such as the East River in New York, have already demonstrated Verdant Power marine turbines for utility-scale testing \cite{[20]}, and many turbines, including Atlantis AS400 and SeaUrchin, have been designed in consideration of riverine environments \cite{[80]}. Overall, horizontal axis turbines (HATs) have been the predominant turbine model in tidal energy research \cite{[21]}, with about $76\%$ of tidal energy research dedicated to HATs according to the 2014 JRC Ocean Energy Status Report \cite{[22],[23]}. HATs’ predominance partly owes itself to the adaption of already successfully established HAT designs in wind farms; thus, HATs have undergone greater convergence and optimization \cite{[23],[24]}. 

In addition to horizontal-axis turbines, there are also vertical-axis turbines. Also known as cross-flow turbines, VATs aim to optimize unsteady fluid forces by confining blade kinematics to a single rotation \cite{[25]}. In spite of their lower efficiency and self-starting than HATs \cite{[26]}, VATs can more efficiently extract energy in farm layouts \cite{[27]}. Notwithstanding HATs’ prevalence in tidal energy experiments, VATs have occupied a significant portion of research and development in marine energy technologies \cite{[4]}. \citet{[28]} argued that VATs are suitable for shallow near-shore waters due to their high power density and design flexibility. Specifically, the rectangular cross-section covers a greater swept area than HATs, potentially offering higher power generation \cite{[13],[29]}. VATs provide other economic and environmental benefits: omnidirectional flow input \cite{[23],[24],[29]}, relatively easy installation \cite{[30]}, reduced noise generation \cite{[4b],[27]}, lower fish mortality \cite{[5b]}, quicker wake recovery \cite{[32]}, and greater access to remote and isolated communities \cite{[8],[13],[34]}. Even though VATs have not historically occupied as much attention and application as HATs, VATs are a promising avenue for energy extraction from riverine environments without many negative ecological and hydrological impacts, for which they seem most suitable \cite{[23],[34]}. 

As with HATs, the similarity to wind turbines facilitates the adaption of VATs’ design and function through experimentation \cite{[34]}. By as early as 1986, \citet{[35]} explored the Darrieus VAT’s dynamic stall in a water channel. In industrial usage, the Ocean Renewable Power Company deployed the first grid-connected helical VAT in the state of Maine in 2012 \cite{[36]}. Even then, \citet{[37]} underscored the dearth of experimental studies about helical VATs, which inspired their performance measurements of comparable VAT designs in a towing tank. Since then, many experimental studies about VATs have followed, primarily focused on more profound understanding of VATs’ wake flows and more recently on design optimization \cite{[25],[29],[38],[39],[40],[41],[42]}. 

Since resources such as time and cost restrict the extent of application, experimental works alone may not provide a complete visualization of VATs’ complex fluid-structure interactions \cite{[27]}. Hence, computational fluid dynamics (CFD) complements experimental studies by expounding a wider variety of potential scenarios for VATs \cite{[13]}. The strength of computational studies is that one can apply many of the same computational techniques for the general modeling of riverine environments \cite{[13]}. It is easier to work with unidirectional riverine flow than bidirectional tidal flows; however, tidal stream turbines’ deployment may significantly impact rivers’ morphology by altering sediment dynamics. More and more, computational studies have significantly advanced our understanding of vertical axis wind turbine (VAWT) performance and wake dynamics through a variety of computational and experimental approaches, which could inform growing CFD research on marine VATs. By exploring unsteady flow around a VAWT through the large eddy simulation (LES) turbulence model and the sliding mesh method, \citet{[44]} discovered a strong correlation between the LES results and momentum theory predictions at high TSRs. In another study, \citet{[45]} measured better agreement between LES and experimental results of a VAWT, which they found as superior to the limited vortex modeling in the unresolved Reynolds-averaged Navier-Stokes (URANS) method. \citet{[46]} utilized both experimental data and LES for a study on VAWT wake structure at high Reynolds numbers, which illustrated that lower TSRs resulted in more asymmetric wakes and larger vortices due to stronger dynamic stall effects. 

\citet{[47]} coupled an actuator line (AL) method with the LES to determine the optimal turbine solidity and TSR for maximizing the power coefficient, followed by a wake analysis to identify downstream regions of maximum velocity deficit and turbulence intensity.  \citet{[48]} likewise employed a coupled LES-AL approach to study wake deficits behind VAWTs in the atmospheric boundary layer. Saliently, they found that coarse spatial resolution in simulations of large wind farms could not resolve the flow around individual HAWTs or VAWTs. With the LES modeling for a VAWT, \citet{[49]} examined the effects of wind velocity, blade airfoil type, and turbulence intensity on VAWT output under fixed- and variable-pitch conditions, also finding a fair agreement with experimental data. \citet{[27]} used LES coupled with the Immersed Boundary Method (IBM) to model the performance of Darrieus VATs in both laminar and turbulent flows, comparing these results with body-fitted methods, RANS-based models, and experimental data across different TSRs. \citet{[50]} analyzed VAWTs with varying array configurations and TSRs through LES and IBM, by which they highlighted the significant role of the blockage ratio in improving downstream turbine performance. In another study with LES-IBM, \citet{[51]} investigated wake recovery as influenced by dynamic solidity, where they concluded that higher dynamic solidity corresponded to shorter downstream distances of momentum recovery.

Of course, design considerations are essential in advancing VATs’ optimization efforts. Since there is growing interest in optimizing VAT design \cite{[29]}, many models have become promising candidates for commercial adoption, namely, the Darrieus and Savonius models, which can feature variations between straight, skewed, and helical blades \cite{[34]}. Each model exhibits advantages and disadvantages: with a drag-based configuration, the Savionus turbine has a simpler design and a higher self-start but lower efficiency. With a lift-based configuration, the Darrieus turbine has a higher power coefficient and TSR but low self-starting and torque pulsations \cite{[34],[52]}. Moreover, one must consider the effects of other parameters, including solidity and blade profile. \citet{[25]} studied an angular rotation rate controller that optimizes the blade’s angle of attack to maximize power extraction without adding extra degrees of freedom. \citet{[34]} reviewed various rotor types in hydrokinetic VATs reporting that lift-based models exhibit superior performance parameters such as power coefficient. \citet{[53]} designed both a novel drag-driven turbine and a conventional Savonius vertical axis tidal stream turbine (VATT) to compare their power characteristics, hypothesizing that enhanced lift force improves the performance of the Savonius VATT. \citet{[54]} developed a self-starting VAWT rotor with a high power coefficient. \citet{[55]} conducted a numerical comparison of various vertical-axis 
turbines across different TSRs and incoming velocities using the SST $k-\omega$ turbulence model. Of the tested models, they found that the Darrieus rotor had the highest power efficiency. \citet{[24]} compared straight and helical blade designs by simulating flow-driven rotors to assess the self-starting capability, torque fluctuations, and RPM performance of Darrieus VATs.

A critical intersection of design consideration and environmental impact is interactions between VATs and bed morphodynamics, which is a pressing challenge in general research on MHK turbines. Within the last two decades, the uncertainty of MHK turbines’ impact on morphodynamics in riverine and tidal environments has distinguished itself as a crucial area of investigation \citet{[56]}. Ecologically, MHK turbines may modify sediment composition and deposition patterns to displace benthic organisms that stabilize long-term sedimentary habitats \cite{[57]}. Since sediment transport entrains nutrients, contaminants, and organic matter, changes in sediment transport entail changes in water quality from water treatment to recreational usage \cite{[58],[59],[60],[61]}. Although early modeling studies offered informative insights on changes in ecology as well as morphodynamics and hydrodynamics, they lacked sufficient field data to validate their impact as well as emphasis on sediment transport \cite{[62]}. Too much focus on design considerations, such as optimization of power production, has also diverted attention from sediment dynamics as a crucial part of environmental studies.  \cite{[63]}. As a general trend, the analogy to other aquatic structures, including bridge piers and offshore wind turbines, has helped to inform modeling expectations of MHK turbines’ potential alterations to sediment transport, particularly in the way of local scour around any supporting structure and more recently in the influence from debris accumulation \cite{[64],[65],[66],[67],[68],[69]}. 

Many studies have reported a relationship between laboratory-scale turbines’ bed-induced turbulence and altered sediment transport. For example, according to \citet{[71]}, the installation of turbines by anchors and mooring may affect sediment transport in the short term by raising turbidity and releasing buried contaminants and, in the long term, by wake turbulence from rotor operation. \citet{[70]} reported that the combination of sediment transport and cavitation would increase fatigue on marine machinery, including turbines, which would entail higher operation and maintenance costs. \citet{[72]} correlated HATs’ presence with greater flow intensity and sediment transport, and they found that lower tip clearance induces more erosion. \citet{[73]} reported that the sediment transport reduces turbine performance and that the larger rotor and steeper bedform lead to an increase in interactions between turbines and sediment. Additionally, \citet{[74]} focused on the tip clearance’s role in HATs through experiments correlating reduced clearance with more significant scour processes. \citet{[75]} experimentally examined the impact of HATs on suspended and rigid beds, in which they reported no wake flow recovery in the far field. \citet{[68]} investigated the influence of modified morphodynamics and debris accumulation on the power production of a utility-scale HAT numerically. \citet{[62], [76]} conducted a series of laboratory experiments on MHK turbine-sediment interactions using a small-scale model of HAT.  They formulated a relationship between scour depth and drag force exerted by MHK turbine structures, in which they detected both local and non-local impacts on bed morphodynamics \cite{[62]}. They also discussed the effect of asymmetrically sited HAT on morphodynamic processes \cite{[76]}. Overall, axial-flow MHK turbines not only locally altered sediment transport but also could more broadly influence bed forms’ shape and migration \cite{[62],[72],[73]}. As \citet{[62]} noted, even though the migrating bedforms did not damage the operating turbines, the modified sediment distribution from scour and deposition might threaten riverine morphology. Nonetheless, to our knowledge, no prior study has investigated sediment dynamics and its related environmental processes around a utility-scale vertical axis hydrokinetic turbine.

The past research efforts on the VATs interactions with sediment transport have only recently emerged. In an early computational analysis, \citet{[77]} analyzed differences in bed shear stress induced by various VATs, and found that such turbines exert the greatest influence on surrounding morphodynamics as well as downstream susceptibility to shear stresses. \citet{[78]} introduced a new installation concept for a drag-driven hydrokinetic VAT, designed for operation under live bed conditions with low bedload transport. \citet{[66]} analyzed a VAT exposed to suspended sediment particles to identify areas most prone to erosion and explored methods for their protection. For the complex nature of VATs’ interactions with morphodynamics, some studies proposed strategies to deter such less investigated interactions.  For example, \citet{[79]} suggested deploying the turbines at mid-depth, where the turbine seemingly interacts less with the erodible bed. With too few studies about interactions between VATs and sediment transport, the bilateral nature of VAT-sediment effects has remained an inexorable obstacle in accurate modeling within experimental and computational studies. Further computational modeling deems essential to address the knowledge gap related to sediment dynamics and turbine interaction and advance research on VATs to supplement the growing body of investigation on the matter and alleviate the practical challenges of tidal farm projects. 

The objective of this numerical study is two-fold. First, we seek to examine the impact of the utility-scale  VAT at various tip-speed ratios on the evolution of the live bed of the channel under live-bed conditions. Through a series of numerical simulations, we investigated the sediment dynamics of the live bed by considering a range of sand particle sizes. Examining the simulation results of the bed evolution of the channel,  we attempt to analyze the formation, growth, and propagation of sand waves that are numerically captured around the turbine. Secondly, this study intends to investigate the impact of evolving bed topography on the turbine's wake flow field and performance, i.e., power generation. In addition to the simulations under live-bed conditions, we also carried out numerical simulations of the turbine by considering a rigid bed for the channel.  In a series of quantitative comparisons, in which the latter simulations serve as a benchmark, we attempt to gain insight into the two-way interaction of live bed and the utility-scale VAT. By elucidating the two-way interactions between turbine-induced flow and channel morphodynamics, we seek a deeper understanding of VATs’ environmental impacts.

The numerical simulations are performed using our in-house code, the Virtual Flow Simulator (VFS-Geophysics) code \cite{[107], [147]}. The flow and bed deformation is resolved using the coupled hydrodynamics and bed morphodynamics modules of the code. The turbulence is captured via LES method. Given the high computational cost of LES at high Reynolds numbers, a wall model approach is adopted to resolve the flow field adjacent to the solid-water and sediment-water interfaces. The turbine and its moving and stationary structural components are resolved using the immersed boundary method \cite{[68], [143]}. Also, the complex dynamic topology of the bed deformation is handled by the immersed boundary method \cite{[101], [113]}. The instantaneous bed evolution is captured by solving the sediment mass balance equation within the bed load layer. To march the computations of sediment and flow field in time, an efficient dual time-stepping technique is employed \cite{[69], [128]}. Lastly, an effective avalanche model, which locally satisfies the sediment mass balance, is adopted to ensure that the local slope of the computed bed topography at each time step is physically limited \cite{[101]}. 

The structure of the paper is outlined as follows. \prettyref{sec:2} introduces the governing equations for the hydrodynamic and morphodynamic. \prettyref{sec:3} describes the setup of the test cases, covering turbine, flow, and channel properties, along with details of the sediment transport model. \prettyref{sec:4} presents the results of the test cases and provides a discussion. Finally, \prettyref{sec:5} highlights the key findings of this study and their implications for future research.

\section{Governing equations}
\label{sec:2}
\subsection{The hydrodynamic model}
\label{sec:2.1}
\noindent The hydrodynamics model solves the spatially filtered Navier-Stokes equations for incompressible flow in non-orthogonal generalized curvilinear coordinates. In compact Newton notation, with repeated indices indicating summation, the equations are expressed as follows \cite{[101], [135]}:

\begin{equation}
    J\frac{\partial U^j}{\partial\xi^j}=0
    \label{eq:1}
\end{equation}

\begin{equation}
    \frac{\partial U^i}{\partial t}=\frac{\xi_l^i}{J}\left(\frac{\partial}{\partial\xi^j}\left(U^ju_i\right)+\frac{1}{\rho}\frac{\partial}{\partial\xi^j}\left(\mu\frac{G^{jk}}{J}\frac{\partial u_i}{\partial\xi^k}\right)-\frac{1}{\rho}\frac{\partial}{\partial\xi^j}\left(\frac{\xi_i^jp}{J}\right)-\frac{1}{\rho}\frac{\partial\tau_{ij}}{\partial\xi^j}
\right)
    \label{eq:2}
\end{equation}

\noindent where, the Jacobian of geometric transformation, $J=\left|\partial\left(\xi^1,\xi^2,\xi^3\right)/\partial\left(x_1,x_2,x_3\right)\right|$, is used for transforming the Cartesian coordinate system to curvilinear. $U^i=(\xi_m^i/J)\ u_m,$ shows the contravariant volume flux, where $\xi_l^i=\partial\xi^i/\partial x_l $. The $i$-th filtered velocity component in cartesian coordinate is shown as $u_i$, $\mu$ represents the dynamic viscosity of the fluid (i.e., water), and also $G^{jk}=\xi_l^j\ \xi_l^k$, shows the contravariant metric tensor, the background density (i.e., water density) is shown as $\rho$ ($=1000 \ kg/m^3$), and the pressure term is defined as $p$. The subgrid-scale stresses are modeled with dynamic Smagorinsky  in the LES turbulence model and defined as \cite{[100], [103], [104]}:

\begin{equation}
     \tau_{ij}=-2\mu_{\mathrm{t}} \overline{S}_{{ij}}+\frac{1}{3}\tau_{kk}\delta_{ij} 
     \label{eq:3}
\end{equation}
\begin{equation}
    \mu_{\mathrm{t}} = C_{\mathrm{s}}\Delta^2 \left |\overline{S} \right|
    \label{eq:4}
\end{equation}

\noindent where  $\mu_{\mathrm{t}}$ is the viscosity of eddies, the tensor of filtered strain-rate is represented as $\overline{S}_{{ij}}$, and \( \delta_{ij} \) is Kronecker delta. The Smagorinsky constant is shown as $C_\mathrm{s}$, and $\left |\overline{S} \right| = \sqrt{2 (\overline{S_{ij}} \overline{S_{ij}} )}$. The filter size ($\Delta$) is defined as the cubic root of the volume of the cell as follows \cite{[100]}:

\begin{equation}
    \Delta = J^{-1/3}
    \label{eq:5}
\end{equation}

\noindent where $J$ represents the volume of the cell.

\subsection{Turbine modeling}
\label{sec:2.2}
\noindent To model the interaction between the flow and the turbine, two primary approaches are commonly employed in the literature \cite{[143]}. The first is turbine parameterization, which utilizes actuator-based models to represent the turbine's influence by introducing turbine-induced lift and drag forces as momentum sinks in the governing equations \cite{[136], [137], [138]}. The second is the turbine-resolving approach \cite{[139], [140], [141], [142], [143]}, which explicitly resolves the flow–blade interactions by employing a sufficiently refined computational grid. While the parameterization method is computationally efficient for modeling turbine arrays, we employ the turbine-resolving approach in this study, which offers greater accuracy in capturing detailed flow physics around individual turbines due to its less modeling and parametrization \cite{[143]}.

To achieve this, the sharp-interface curvilinear immersed boundary (CURVIB) method is implemented \cite{[100], [68], [135], [147]} to handle fluid–structure interactions (FSI) for both solid transient (e.g., evolving bed, and turbine blades and struts) and stationary objects (e.g., the channel, rigid bed, and turbine shaft). This method discretizes the solid objects into unstructured mesh and immerses them within the structured background mesh of the fluid domain. Instead of conforming to the turbine geometry, CURVIB treats the boundaries as sharp interfaces, reconstructing boundary conditions at the grid points adjacent to the solid surfaces using interpolation along the local normal direction \cite{[143], [144], [145]}.

The computational grid nodes are categorized based on their location relative to solid objects. Nodes located inside solid objects and outside the fluid domain are classified as exterior nodes and are excluded from the computations. Nodes within the fluid domain where the governing equations are solved are referred to as interior (fluid) nodes. Finally, immersed boundary (IB) nodes are those positioned within the fluid but in the vicinity of the solid boundaries \cite{[143]}.

A critical aspect of the CURVIB method is the efficient classification of grid nodes, particularly for transient objects. This requires an adaptive search algorithm to reclassify nodes as the location or shape of the immersed bodies changes during the simulation \cite{[143]}. The ray-tracing method is employed for this purpose \cite{[146]}. For stationary objects, the ray-tracing algorithm is applied only at the beginning of the simulation (i.e., the first time step). However, for transient objects, the algorithm is executed at every time step to ensure proper node reclassification for the subsequent time step \cite{[143]}. 
Finally, a wall model within the CURVIB framework is employed to effectively reconstruct velocity at IB nodes while maintaining computational efficiency (for further details, see \cite{[102], [112], [143]}).

\subsection{Bed morphodynamics}
\begin{figure} 
  \includegraphics[width=1\textwidth]{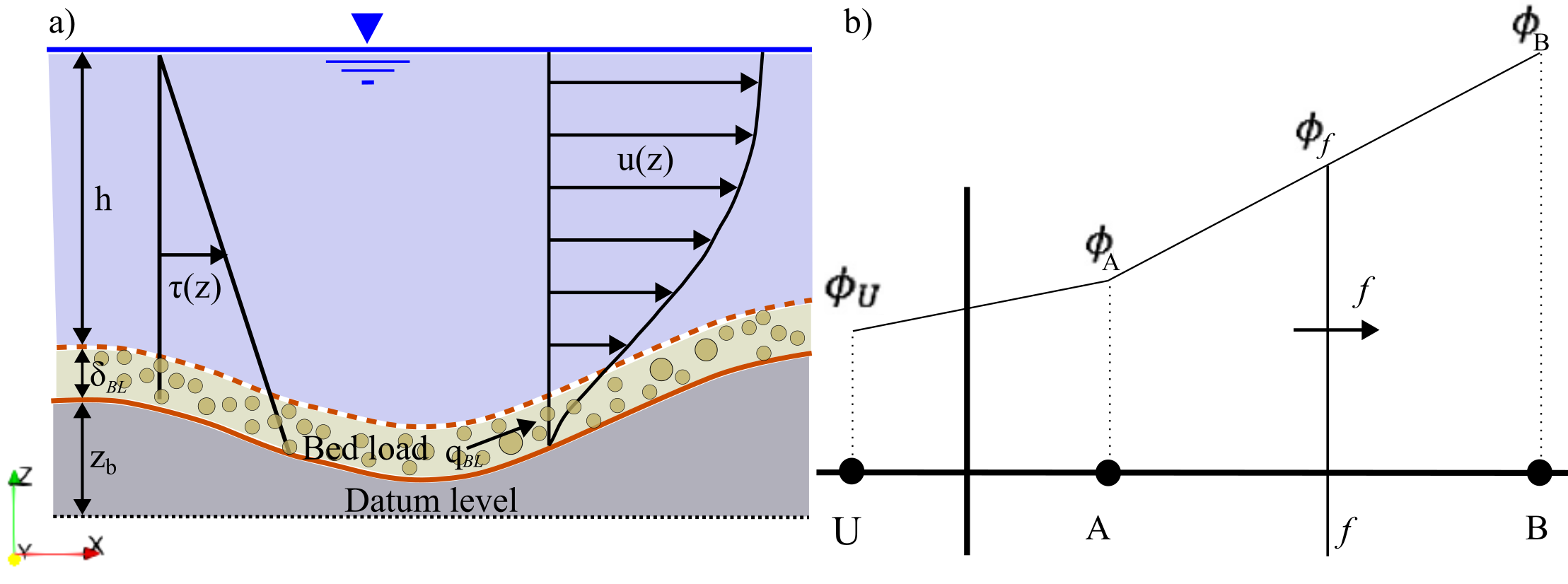}  
  \caption{(a) Illustration of the flow domain and the bed-load layer in an open channel over a live bed. The bed-load layer has a thickness of $\delta _b$, while the flow depth is $h$ (assuming $\delta _b \ll h$). The velocity distribution is denoted by $u(z)$, and the shear stress distribution is represented by $\tau(z)$. Small brown circles represent the sediment particles inside the bed load layer. (b) schematic of a non-uniform unstructured grid illustrating the location of a cell face ($f$) and its relation to the central ($A$), upstream ($U$), and downstream ($B$) cells \cite{[112]}.}
  \label{Fig:1}
\end{figure}

\noindent The non-cohesive bed material include a range of sizes of sand. Sediment particles are transported in three primary modes: rolling or sliding in continuous contact with the bed, saltating or hopping along the bed, or being suspended in the flow. The modes of transport depend on the bed-shear velocity relative to the particle's critical motion threshold. According to \citet{[105]}, sediment transport is classified into: bed load (particles move by rolling and saltating), suspended load (particles are lifted into suspension by turbulence when the bed-shear velocity exceeds their fall velocity, remaining suspended without contacting the bed for a considerable period of time), and wash load (fine particles transported without deposition).
Assuming the elevation of the top part of the bed load layer is defined as $z_{\mathrm{b}}$ (see, \prettyref{Fig:1}(a)), the temporal variation of this elevation is governed by the non-equilibrium mass balance equation, so-called Exner-Polya equation, as follows \cite{[106]}:

\begin{equation}
(1-\gamma)\frac{\partial Z_{\mathrm{b}}}{\partial t} + \nabla \cdot \mathbf{\bm{q}}_{\mathrm{BL}} = {D_b}-{E_b}
\label{eq:6}
\end{equation}

\noindent where $\gamma$ is the sediment porosity, which is taken 0.4 in this work. $D_b$ is the rate of sediment deposition (i.e. the net rate of sediment deposits from the suspended sediment onto the bed) and $E_b$ is the rate of sediment deposition (i.e., the net rate of sediment entrains from the bed into the flow) \cite{[107]}. Once again, since the suspended load is not captured in this work, $D_b$ and $E_b$ reduces to zero. The divergence operator is shown with $\nabla$, and $\bm{q}_{\mathrm{BL}}$ is the vector of bed load flux. Thus, the term $\nabla \bm{q}_{\mathrm{BL}}$ shows the net flux of sediment, transported inside the bed load layer. \prettyref{eq:7} defines the vector of bed load flux as follows \cite{[108]}:
\begin{equation}
\mathbf{\bm{q}}_{\mathrm{BL}} = \psi \, {\|d_{\mathrm{s}}\|} \, {\|\delta_{\mathrm{BL}}\|} \, \mathbf{\bm{u}}_{\mathrm{BL}}
\label{eq:7}
\end{equation}

\noindent where $d_s$ represents the edge length of each triangular bed mesh, $\delta_{\mathrm{BL}}$ is the thickness of bed load layer (= $0.005$ in this work), and the flow velocity parallel to the bed surface is shown with $\mathbf{\bm{u}}_{\mathrm{BL}}$. Sediment concentration, $\psi$, is defined as follows:

\begin{equation}
\psi = 0.015 \frac{d_{50}}{\delta_{\mathrm{BL}}} \frac{T^{3/2}}{D_*^{3/10}}
\label{eq:7_1}
\end{equation}

\begin{equation}
D_* = d_{50} \left[ \left( \frac{\rho_{\mathrm{s}} - \rho}{\rho v^2}g \right)^{1/3} \right]
\label{eq:9}
\end{equation}
where $d_{50}$ represents the mean grain size, $\rho_{\mathrm{s}}$ is the density of the sediment particle ($=2650 kg/m^3$), and $\nu$ is the kinematic viscosity of the water. The non-dimensional excess shear stress, $T$, is defined as follows:

\begin{equation}
T = \frac{\tau_* - \tau_{*\mathrm{cr}}}{\tau_{*\mathrm{cr}}}
\label{eq:10}
\end{equation}

\noindent Where $\tau_*$ is the bed shear stress and defined as follows \cite{[105]}:
\begin{equation}
    \tau_{\ast}= \rho {u_{\ast}}^2
    \label{eq:11}
\end{equation}
where $u_{\ast}$ is shear velocity. To find the velocity and shear stress at immersed body (IB) nodes (i.e., the nearest cell of the background grid nodes from the wall or sediment-water interface), we use a wall model approach as follows \cite{[101], [108]}:

\begin{equation}
\frac{u}{u_{\ast}} = \left\{
\begin{matrix}
y^{+} & y^{+} \leq 11.53 \\
\frac{1}{\kappa} \ln\left(Ey^{+}\right) & y^{+} > 11.53 \\
\end{matrix}
\right.
\label{eq:12}
\end{equation}

\noindent and,

\begin{equation}
y^{+} = yu_{\ast}/\nu
\label{eq:13}
\end{equation}

\noindent where $u$ represents the velocity magnitude at a distance of $y$ away from the wall (or sediment-water interface), and the von Kármán constant is shown with $\kappa$, which is set equal to $0.41$. $E$ represents the roughness parameter defined as follows:

\begin{equation}
E = \exp(\kappa(B - \Delta B))
\label{eq:14}
\end{equation}

\noindent where,

\begin{equation}
\Delta B = \begin{cases}
0 & \text{if } k_{\mathrm{s}}^{+} < 2.25 \\
\left[B - 8.5 + \left(1/\kappa\right)\ln(k_{\mathrm{s}}^{+})\right]\sin\left[0.4258\left(\ln(k_{\mathrm{s}}^{+}) - 0.811\right)\right] & \text{if } 2.25 < k_{\mathrm{s}}^{+} < 90 \\
B - 8.5 + \left(1/\kappa\right)\ln(k_{\mathrm{s}}^{+}) & \text{if } k_{\textbf{s}}^{+} \geq 90 \\
\end{cases}
\label{eq:15}
\end{equation}

\noindent and,

\begin{equation}
k_{\mathrm{s}}^+=k_{\mathrm{s}}u_\ast/\nu
\label{eq:16}
\end{equation}

\noindent where $B=5.2$. The boundary effective roughness height ($k_s$), is usually assumed to be greater than the mean grain size. In this work, $k_s$ is taken as three times the mean grain size (i.e. $k_s = 3d_{50}$) \cite{[107]}. For the smooth wall boundaries, $k_{\mathrm{s}}^+$ is less than $2.25$. The solid surfaces (including the channel, turbine components, and sediment layer) are modeled as hydraulically smooth wall boundaries. Thus, $\Delta B$ in \prettyref{eq:14} reduces to zero \cite{[101]}. Getting back to \prettyref{eq:10}, $\tau_{*\mathrm{cr}}$ is the critical bed shear stress, which first obtained for the flat-bed as follows \cite{[109]}:

\begin{equation}
\tau_{\ast c0} = \frac{0.3}{1 + 1.2D_*} + 0.055 \left[ 1 - \exp\left(-0.02D_*\right) \right]
\label{eq:18}
\end{equation}

\noindent The critical bed shear stress for the flat-bed ($\tau_{\ast c0}$) is then corrected for both longitudinal and spanwise directions, as follows \cite{[101], [110], [111]}:

\begin{equation}
\tau_{\ast c r} = \tau_{\ast c0} \left[ \frac{\sin\left(\phi + \alpha\right)}{\sin\left(\phi\right)} \right]
\label{eq:17}
\end{equation}

\noindent where $\alpha$ is the local bed slope at the center of each triangular bed cell, and $\phi$ is the sand's angle of repose.

Once all parameters are obtained in cell centers, they are transferred to cell faces to calculate the bed load sediment flux ($\mathbf{\bm{q}}_{\mathrm{BL}}$) using a convective scheme, so-called second-order GAMMA differencing scheme. This scheme is a combination of two schemes, the first-order upwind scheme and second-order central differencing scheme. Let’s assume that $\phi_A$ is the desired variable at the cell center, which in this case is the components of velocity and concentration of the sediment layer, $\phi_B$ is the neighbor cell of $A$, and $f$ is the face between point $A$ and $B$ (see, \prettyref{Fig:1}(b)). The GAMMA scheme is used to transfer $\phi_A$ to face $f$ as follows \cite{[112]}:

\noindent ($i$)	$\overline{\phi_A}$, the normalized value of $\phi_A$, is obtained as follows:

\begin{equation}
\overline{\phi_A} = 1 - \frac{\phi_A - \phi_B}{2 (\nabla \phi)_A \cdot \mathbf{d}}
\label{eq:19}
\end{equation}

\noindent where $\mathbf{d}$ is the connector vector between points $A$ and $B$.

\noindent ($ii$) The variable at face $f$, $\phi_f$, is then obtained with three separate conditions as follows:

\begin{equation}
\overline{\phi_f} = \begin{cases}
\overline{\phi_A} & \; \text{if }\; \overline{\phi_A} \leq 0 \; \text{(first order upwind scheme)}\\ 
\overline{\phi_A} & \; \text{if }\; \overline{\phi_A} \geq 1 \; \text{(first order upwind scheme)}\\
f_x \phi_A + \left(1 - f_x\right) \phi_B & \; \text{if }\; \beta_m \leq \overline{\phi_A} \leq 1 \; \text{(central differencing scheme)}\\
\phi_f = \left(1 - \lambda \left(1 - f_x\right)\right) \phi_A + \lambda \left(1 - f_x\right) \phi_B & \; \text{if }\; 0 \leq \overline{\phi_A} \leq \beta_m \; \text{(central differencing scheme)}\\
\end{cases}
\label{eq:20}
\end{equation}

\noindent where the blending region size between the two schemes -- i.e., the second-order central differencing and the first-order upwind -- is controlled with $\beta_m$ ($=0.33$). The factor of linear interpolation ($f_x$) and non-linear blending ($\lambda$), are defined as \prettyref{eq:21}, and \prettyref{eq:22}, respectively:

\begin{equation}
f_x = \frac{\overline{fB}}{\overline{AB}}
\label{eq:21}
\end{equation}

\begin{equation}
\lambda = \frac{\overline{\phi_A}}{\beta_m}
\label{eq:22}
\end{equation}

Once all parameters are calculated, one could then compute the new bed elevation. However, the obtained bed topography might contain local slopes that are greater than the angle of repose of the bed material, rendering the obtained bed topology non-physical. In other words, a last check is required to ensure that the new bed elevations, calculated based on the mass-balanced equation (\prettyref{eq:6}), have slopes less than or equal to the angle of repose. To achieve this, we employ a mass-balanced sand slide model, briefly explained here (see \cite{[107], [101], [112], [113], [135]} for more details). Each triangular cell has three neighbor cells. The slopes between each cell center and its neighboring cell centers are calculated at each time step. The slopes exceeding the angle of repose are flagged. We then redistribute the excess mass of each flagged cell among its neighboring cells to ensure that the slope of the flagged cell does not exceed the angle of repose. One issue is that, after redistributing the excess mass, the neighboring cells may develop slopes with their adjacent cells that exceed the angle of repose. To address this, the sand slide model is iteratively applied to all bed cells until the steepest local slope is less than $99$ percent of the angle of repose.

\subsection{The coupled hydro- and morpho-dynamics}
\noindent There are two main FSI coupling approaches to capture hydro- and morpho-dynamic interactions: loose coupling and strong coupling \cite{[108]}. In strong coupling, boundary conditions at the sediment-water interface are updated iteratively within each time step, making this method implicit in time. Although strong coupling offers greater stability, it requires significantly more computational power. In contrast, loose coupling updates boundary conditions at the same interface based on the solutions from the previous time step, making this method explicit in time and eliminating the need for additional iterations. The choice between these approaches depends on the problem's complexity and engineering considerations. For this study, we opted for the loose coupling method due to its favorable computational efficiency and its demonstrated robustness for applications similar to the present work \cite{[101], [107], [111], [112], [113], [135]}.

To couple hydro- and morpho-dynamics, we use CURVIB method as discussed in \prettyref{sec:2.2}. This involves dividing the problem into two distinct domains and solving their governing equations separately while incorporating boundary conditions from the other domain. More specifically, when solving the hydrodynamic equations (see \prettyref{eq:1} to \prettyref{eq:5}), we use updated bed elevations and bed vertical velocities as boundary conditions at the sediment-water interface. Conversely, when solving the morphodynamic equations (see \prettyref{eq:6} to \prettyref{eq:23}), we use the flow velocities and shear stresses obtained from the hydrodynamic solver as boundary conditions at the sediment-water interface.

Another important consideration is the great disparity between the time scale of hydrodynamic and morphodynamic events. For instance, while the flow field's time scale is within seconds to minutes, the time scale of bed evolution can span hours or even days. To address this issue, we adopted a dual time-stepping method \cite{[128]} that allows to solve the hydrodynamic and morphodynamic equations using different computational time steps. As a result, the time step for the morphodynamic solver is two orders of magnitude larger than that of the flow solver.

A critical component of coupling flow and sediment dynamics models is to ensure global mass conservation in response to changes in bed elevation and, thus, volume. To accomplish this, we can first calculate the rate of volume change ($\partial \psi/ \partial t$) at each time step precisely since we calculated the exact elevations of the bed for the current and previous time steps. With this rate of volume change determined, we then adjust the outlet boundary flux ($Q_{\text{out}}$) so that it equals the sum of the inlet boundary flux ($Q_{\text{in}}$), and the rate of volume change, as outlined below \cite{[110], [112]}:

\begin{equation}
Q_{\text{out}} = Q_{\text{in}} + \frac{\partial \psi}{\partial t}
\label{eq:23}
\end{equation}

The velocity field at the outlet plane, calculated using Newman boundary conditions, is then adjusted to ensure the outlet boundary flux is equal to that obtained from \prettyref{eq:23} (see \cite{[101], [112]} for more details). In summary, the procedure for coupling hydro- and morpho-dynamics using the loose coupling approach can be described as follows. To determine the unknowns at time step $n+1$, the hydro-dynamic equations (\prettyref{eq:1} to \prettyref{eq:5}) are solved using the boundary conditions of the bed (i.e., the vertical velocity of the bed and bed elevations) from the previous time step $n$. The resulting flow velocity and shear stress at time step $n+1$ then serve as the new boundary conditions for the morphodynamic solver (\prettyref{eq:6} to \prettyref{eq:22}) at the same time step ($n+1$). Subsequently, the sand slide model is applied to adjust bed changes, ensuring that the slope of bed cells does not exceed the angle of repose. The updated bed elevations and vertical velocity at time step $n+1$ are then used as the new boundary conditions for the flow solver at time step $n+2$ \cite{[107]}.

\section{Test case description and computational details}
\label{sec:3}
\noindent  This section provides detailed information about test cases and outline the numerical simulation parameters, including flow characteristics, turbine characteristics, as well as the channel and sediment particle properties. 

\begin{figure}[H]
  \includegraphics[width=1\textwidth]{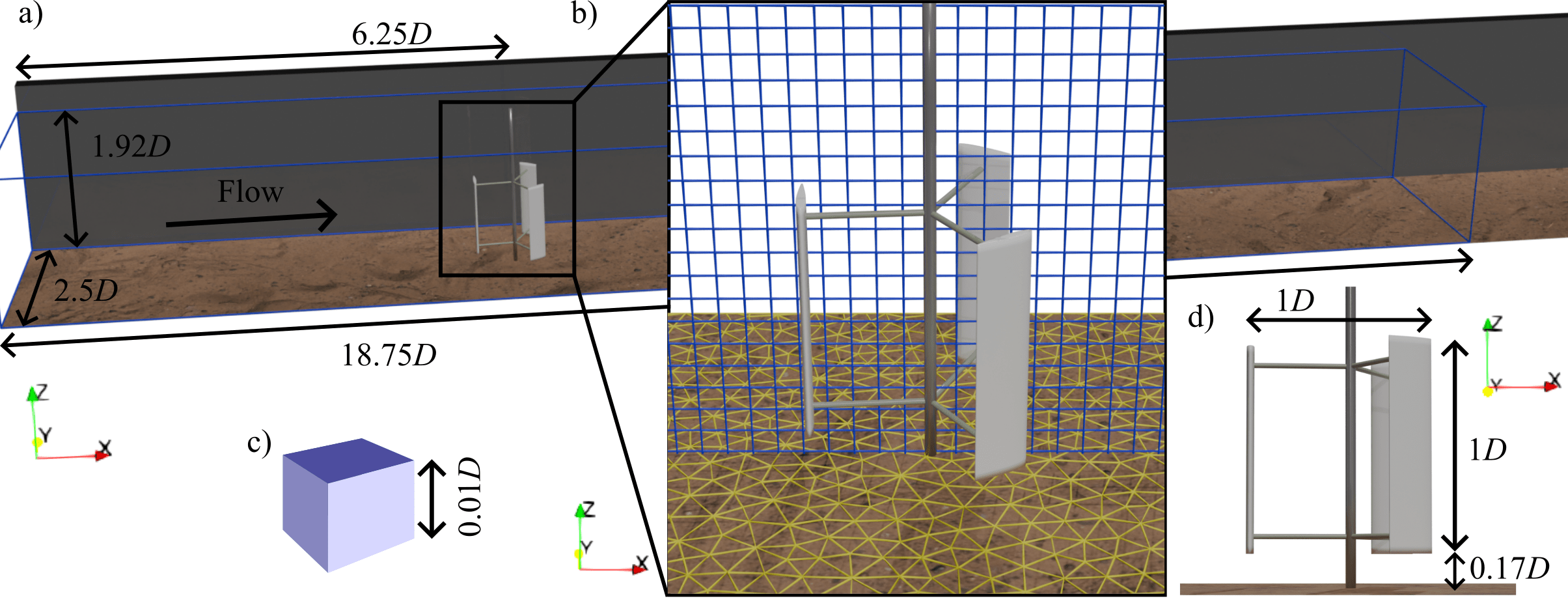}  
  \caption{Schematic of the channel and turbine in which dimensions are normalized by the rotor diameter ($D=2m$). The turbine is located at $6.25D$ downstream from the channel inlet. The flow direction aligns with the positive x-axis, while the z-axis indicates the vertical direction. The channel has a total length of $18.75D$, a width of $2.5D$, and a flow depth of $1.92D$. In (b), the blue mesh shows the structured grid system to discretize the flow field, while the unstructured triangular cells (in yellow) discretize the deforming bed. For clarity of the visual, the former and latter grid systems are coarsened by a factor of $10$ and $5$, respectively. In (c), we demonstrate a fluid cell and (d) demonstrates dimensions of the turbine.}
  \label{Fig:2}
\end{figure}

The channel's width and length are $5m$ and $37.5m$, respectively (\prettyref{Fig:2}(a)). The flow depth is $h=3.84m$. The utility-scale turbine is a three-blade H-Darrieus vertical axis turbine with a NACA0015 hydrofoil \cite{[29], [30]} and a diameter of $D=2m$, a chord length of $C=0.5m$, and a blade height of $2.0m$. This results in a geometric solidity of $\sigma=0.24$ ($= N_b C/\pi D$, where $N_b$ is the number of blades). The turbine is positioned at the centerline and $6.25D$ downstream of the channel's inlet (see, \prettyref{Fig:2} (a) and (d)). The bulk velocity is set equal to $U_{\infty} = 1.5 \ m/s$, resulting in a Reynolds number of $Re=3 \times 10^6$ ($=U_{\infty} D / \nu$).

The flow domain is uniformly discretized into $1801$ nodes in the streamwise direction, $261$ nodes in the spanwise direction, and $229$ nodes in the vertical direction (\prettyref{tab:1}), providing a spatial resolution of $0.01D$ (\prettyref{Fig:2}(c)) with a total number of $107$ million computational grid nodes. The minimum grid spacing in the vertical direction, scaled by inner wall units, is $600$. The flow solver’s non-dimensional time step is set to $\Delta t_{*} = 0.0005$ ($= \Delta t U_{\infty}/{D}$), where $\Delta t = 0.00067s$ is the physical time step (\prettyref{tab:1}). This ensures that the Courant-Friedrichs-Lewy (CFL) number is less than one at all times.  

The mean grain sizes of $d_{50} = 0.35,\ 0.7,\ 1.1,\ 1.4 mm$ were considered in this work. This range spans from medium sand with $d_{50}=0.35mm$ to very coarse sand with $d_{50}=1.4mm$ \cite{[134]}. The porosity of sediment material is set to $\gamma=0.4$, and the angle of repose of the bed material is set to $\phi=40^{\circ}$. The non-dimensional time step of the sediment transport computations is $\Delta t_s = 0.05$, and the spatial step of the morphodynamics solver, normalized with rotor diameter, is $\Delta s=0.023$. Additionally, three different angular velocities ($\Omega$) of $2.4,\ 3.0,\ 3.6 \ rad/s$ are considered for the turbine, leading to tip speed ratios of $TSR = 1.6,\ 2.0,\ 2.4$ ($ \ = \Omega D/2U_{\infty}$). Combining these three TSRs with four different mean grain sizes, under rigid and live bed conditions, a total of 15 numerical experiments are conducted (\prettyref{tab:2}).

\begin{table}[H]
\centering
\begin{tabular}{p{8cm} p{3cm}}
\hline
\multicolumn{2}{c}{\textit{Hydrodynamics solver}} \\
\hline
$N_x, N_y, N_z$ & $1801 \times 261 \times 229$ \\
$\Delta x, \Delta y, \Delta z$ & $0.01D$ \\
$\Delta t_{*}$ & $0.0005$ \\
$z^+$ & $600$ \\
\hline
\multicolumn{2}{c}{\textit{Morphodynamics solver}} \\
\hline
$\Delta t_s$ & $0.05$ \\
$\Delta s$ & $0.023D$ \\
$\gamma $ & $0.41$ \\
$\rho_s (kg/m^3)$ & $2650$ \\
$\phi$ & $40^{\circ}$ \\
$d_{50}(mm)$ & $0.35 - 1.4$ \\
\hline
\end{tabular}
\caption{Computational grid systems for the flow and morphodynamics solvers. The grid consists of $N_x, N_y$, and $N_z$ computational nodes in the streamwise, spanwise, and vertical directions, respectively. The spatial resolutions for the flow solver, normalized by the rotor diameter $D$, are denoted as $\Delta x$, $\Delta y$, and $\Delta z$. $\Delta s$ represents the normalized spatial step for the morphodynamics solver. The minimum vertical grid spacing in wall units is expressed as $\Delta z^{+}$. The non-dimensional time step for the flow solver is $\Delta t_{*} = \Delta t U_{\infty}/{D}$, where $\Delta t$ is the dimensional time step. For the morphodynamic solver, the non-dimensional time step is $\Delta t_{s}$. $\gamma$ is the sediment porosity, and $\phi$ is the angle of repose, $\rho_s$ is the sediment density, and $d_{50}$ is the sediment mean grain size.}
\label{tab:1}
\end{table}

In cases under the live bed conditions (i.e., cases $4$ to $15$), we initially treated the bed as frozen and ran the flow solver until the instantaneous flow fully developed. This was monitored by observing and ensuring that the total kinetic energy of the flow is leveled off. Subsequently, we activated the sediment transport module, allowing the bed to deform. Importantly, while water surface fluctuations around immersed solids may cause slight modulations in the wake flow, this study assumes a rigid-lid water surface to reduce the computational cost associated with tracking water surface fluctuations \cite{[68], [69]}.

\begin{table}[H]
\setlength{\tabcolsep}{15pt}      
\centering
\begin{tabular}{c c c c c c c}
\hline
\textit{Test case} & \textit{Mobility} & \textit{TSR} & \textit{$d_{50}$ (mm)} \\
\hline
$1$  & Rigid & $1.6$ & -  \\
$2$  & Rigid & $2.0$ & -  \\
$3$  & Rigid & $2.4$ & - \\
$4$  & Live & $1.6$ & $0.35$ \\
$5$  & Live & $2.0$ & $0.35$ \\
$6$  & Live & $2.4$ & $0.35$ \\
$7$  & Live & $1.6$ & $0.7$ \\
$8$  & Live & $2.0$ & $0.7$ \\
$9$  & Live & $2.4$ & $0.7$  \\
$10$ & Live & $1.6$ & $1.05$ \\
$11$ & Live & $2.0$ & $1.05$ \\
$12$ & Live & $2.4$ & $1.05$ \\
$13$ & Live & $1.6$ & $1.4$  \\
$14$ & Live & $2.0$ & $1.4$ \\
$15$ & Live & $2.4$ & $1.4$ \\
\hline
\end{tabular}
\caption{Test Cases Descriptions. Cases $1$ to $3$ are conducted over rigid bed, and cases $4$ to $15$ are conducted over live bed. TSR denotes tip-speed ratio, $d_{50}$ is sediment mean grain size, $U_{\infty}$ bulk velocity, $Re$ is the Reynolds number based on rotor diameter, and $h$ is water depth.}
\label{tab:2}
\end{table}

A separate precursor simulation with the periodic boundary condition in the streamwise direction was conducted to generate an instantaneously-converged turbulent flow in the rigid-bed channel without any turbine. The fully developed turbulent flow was achieved by monitoring the total kinetic energy until it plateaued. The instantaneous turbulent flow over a cross plane, at the mid-length, of this simulation was extracted and imposed as the inlet boundary condition for the test cases.  Also, the Neumann boundary condition was applied at the outlet of the channel. 

The numerical simulations for each case were performed on a Linux cluster equipped with 192 processors (AMD Epyc). On average, 108,000 CPU hours were needed for each rigid bed case to achieve statistically converged flow fields. In contrast, each coupled flow and sediment simulation under the live bed conditions required approximately 165,000 CPU hours to reach an equilibrium bed topology.

\section{Results and discussions}
\label{sec:4}
\noindent In this section, we analyze the hydrodynamics and morphodynamics simulation results. We begin by presenting and discussing the instantaneous and time-averaged hydrodynamic results over the rigid bed. Subsequently, the simulation results for the hydrodynamics and bed morphodynamics under the live-bed conditions will be discussed.  Finally, we examine the simulation results for the turbine wake recovery and performance to provide new insights into the effects of sediment dynamics on the efficiency of the utility-scale vertical-axis turbine.

\subsection{Wake flow under rigid-bed conditions}
\label{subsec:Wake analysis over Rigid Bed}
\begin{figure} [H]
  \includegraphics[width=1\textwidth]{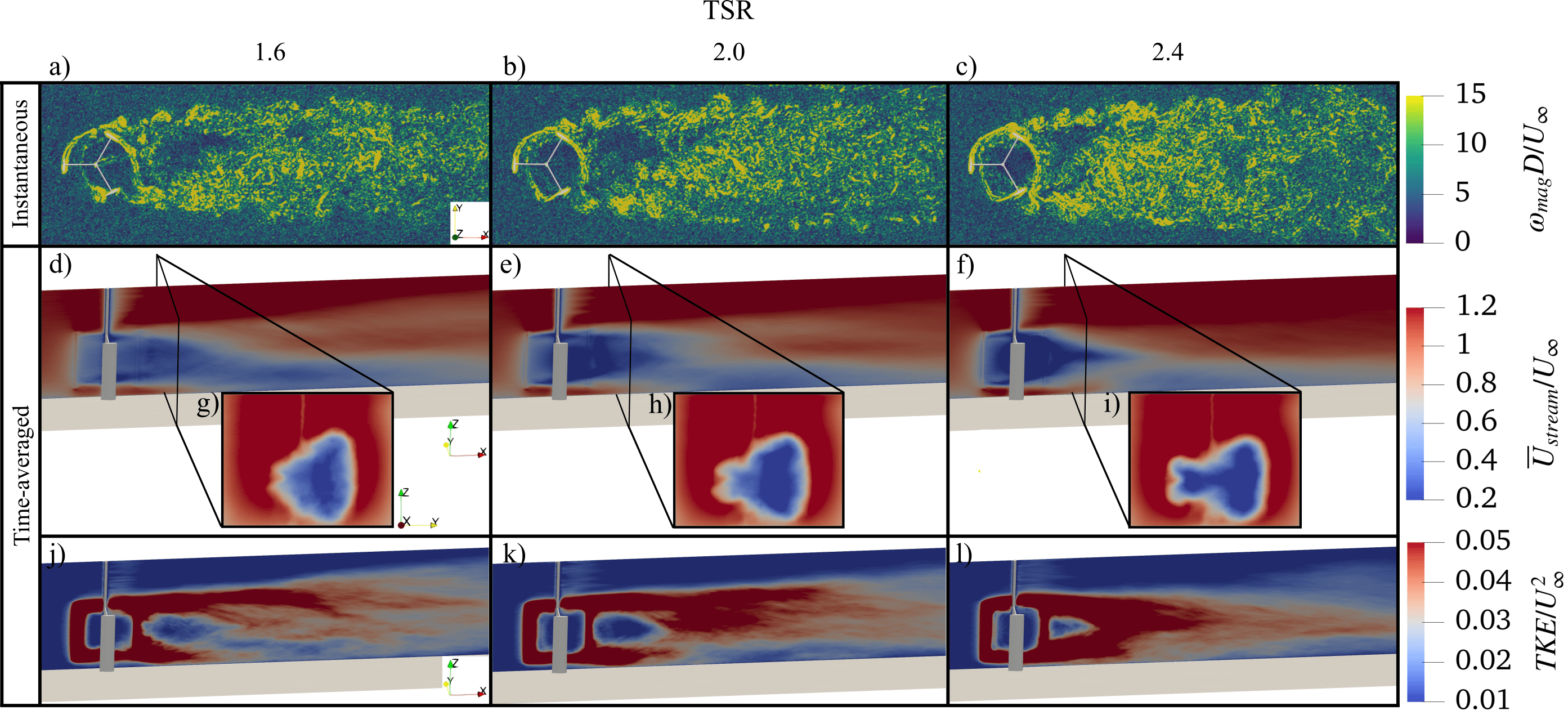}  
  \caption{Color maps of hydrodynamic results under rigid-bed conditions (i.e., cases $1$ to $3$). (a) to (c) depict the nondimensional instantaneous vorticity magnitude from the top view at the mid-depth elevation of the turbine blades. (d) to (f) show 3D views of the non-dimensional mean streamwise velocity. (g) to (i) illustrate the nondimensional mean streamwise velocity on the cross plane located $1D$ downstream of the turbine. (j) to (l) present the nondimensional TKE along longitudinal planes at the centerline of the channel. (d) to (f) and (i) to (l) are shown on the longitudinal slice at the channel's centerline. Flow is from left to right.}
  \label{Fig:3}
\end{figure}

\noindent In \prettyref{Fig:3}(a) to (c), we plot the nondimensional instantaneous vorticity magnitude from top view under the rigid-bed conditions for different TSRs. These top-view slices are taken from the mid-depth elevation of the turbine blades (i.e., $0.67D$ above the bed), showing details of the wake dynamics and turbine-induced turbulence structures in the wake of the turbine. As seen, the effect of dynamic stall is evident in all cases, as flow separation and high vorticity are observed around the blades \cite{[120]}. At the lowest TSR of $1.6$, the vorticity field is characterized by relatively less pronounced diffuse near-wake vortices (\prettyref{Fig:3} (a)). The slower blade rotation at this TSR results in relatively weak shear layers. Thus, the downstream wake is characterized by reduced vorticity intensity and a rather dispersed series of flow coherent structure extending further downstream. As TSR increases to $2.0$ (\prettyref{Fig:3}(b)) and $2.4$ (\prettyref{Fig:3}(c)), the higher rotational speed increases the frequency of blade passage which, in turn, intensifies both the flow-blade and blade-to-blade interactions \cite{[46]}. This results in stronger and slightly wider shear layer, a wider range of coherent vortices with higher intensity, and a more energetic wake structure.

Notably, as TSR increases, the energetic vortices shift to the near-field region, intensifying turbulence in this near-wake area. This transition from diffused to more structured vortices indicates enhanced turbulence production, which aligns with the findings reported in \citet{[46]} that demonstrated that TSR is proportional with the energy content of the near-wake region. Additionally, it can be observed that at lower TSRs, the suction side of the blades at the most upstream locations correspond with slightly higher vorticity. This can be attributed to the wider range and higher angles of attack, as well as the more pronounced dynamic stall phenomena that blades experience. This observation is consistent with the findings of previous studies \cite{[46], [50], [116], [121]}. Also, it can be seen that the influence of the turbine shaft on the wake flow is less pronounced at higher TSRs.  For instance, at TSR $= 2.4$ (see \prettyref{Fig:3}(c)), the vortices downstream of the shaft and within the rotor region are diminished, indicating that the shaft's effect becomes less significant at higher TSRs. Moreover, at TSR $=1.6$, each blade predominantly interacts with its own wake. However, as TSR increases, the blades interact not only with the wake generated by their own rotation but also with those shed by other blades, as seen in \prettyref{Fig:3} (c). Additionally, with increasing TSR, the asymmetrical distribution of vortices gradually diminishes, transitioning to a more symmetrical pattern. These observations agree with the findings of other researchers, e.g., see \citet{[46]}.


In \prettyref{Fig:3}(d) to (l), we plot the first- and second-order turbulence statistics of the wake flow under rigid-bed conditions. \prettyref{Fig:3}(d) to (f) depict color maps of time-averaged streamwise velocity along longitudinal vertical planes along the centerline of the channel, while \prettyref{Fig:3}(g) to (i) are taken from cross-planes located $1D$ downstream of the turbine.  As seen in all cases, the wake flow marks a prominent near-bed high-momentum region that could increase local erosion. We note that as the TSR increases, the blade passage frequency increases and, thus, a stronger blade-flow interaction is expected. As seen in \prettyref{Fig:3}(d) to (f), at higher TSRs, the wake region becomes more localized in the near-wake with a greater velocity deficit. Further downstream, it is observed that the wake structure of the test case with higher TSR dissipates faster, allowing for a faster recovery of the streamwise velocity. In other words, the results show that an increase in TSR leads to an increased momentum deficit and enhanced wake recovery, which is consistent with findings of \citet{[117]}. Further, as seen from the contours of the mean streamwise velocity over the cross planes of \prettyref{Fig:3}(g) to (i), the wake of the turbine is asymmetrical owing to the counter clock-wise rotation of the VAT, which was previously reported by \citet{[38]} and \citet{[46]}. As seen in these figures, increasing the TSR makes the wake region larger, creating a relatively wider area with elevated turbulence around the turbine.

Figure \ref{Fig:3}(j) to (l) show the turbulence kinetic energy (TKE) distribution along the longitudinal plane at the channel's centerline.  As seen,  increasing the TSR renders the high TKE region to concentrate near the mid-depth of the flow and closer to the turbine and the near-field. As reported by \citet{[117]},   such TKE pattern could be attributed to the increased rotational speed of the turbine blades, which amplifies localized turbulence generation, at higher TSR. Given the link between the regions with high TKE and sediment erosion, and as discussed below, such observation could have important consequences for the live-bed channel.


\subsection{Wake flow under live-bed conditions}
\label{subsec:Wake analysis under live bed conditions}
\noindent Herein, we focus on the hydrodynamics results of the coupled flow-morphodynamics simulation followed by concurrently occurring sediment dynamics and bed deformations in the next section. The initial flow condition for the coupled flow and morphodynamics simulations under the live-bed conditions is obtained by freezing the bed until the instantaneous wake flow is statistically converged. Then, the coupled flow and morphodynamics simulation starts over a flat live-bed channel. Soon after, the bed material starts interacting with the turbulent flow, deforming the bed topology. The couple simulation is continued until the live bed of channel reaches a dynamic equilibrium. Such a state of the bed is achieved when the bed is covered with migrating sand waves and the maximum scour depth and height of the sand bar stay nearly constant. At dynamic equilibrium state of the bed, the coupled simulation is stopped. Freezing the bed geometry at its dynamic equilibrium, the flow solver is activated to generate wake flow field data. The so-obtained instantaneous data are then time-averaged to compute turbulence statistics.

\begin{figure} [H]
  \includegraphics[width=1\textwidth]{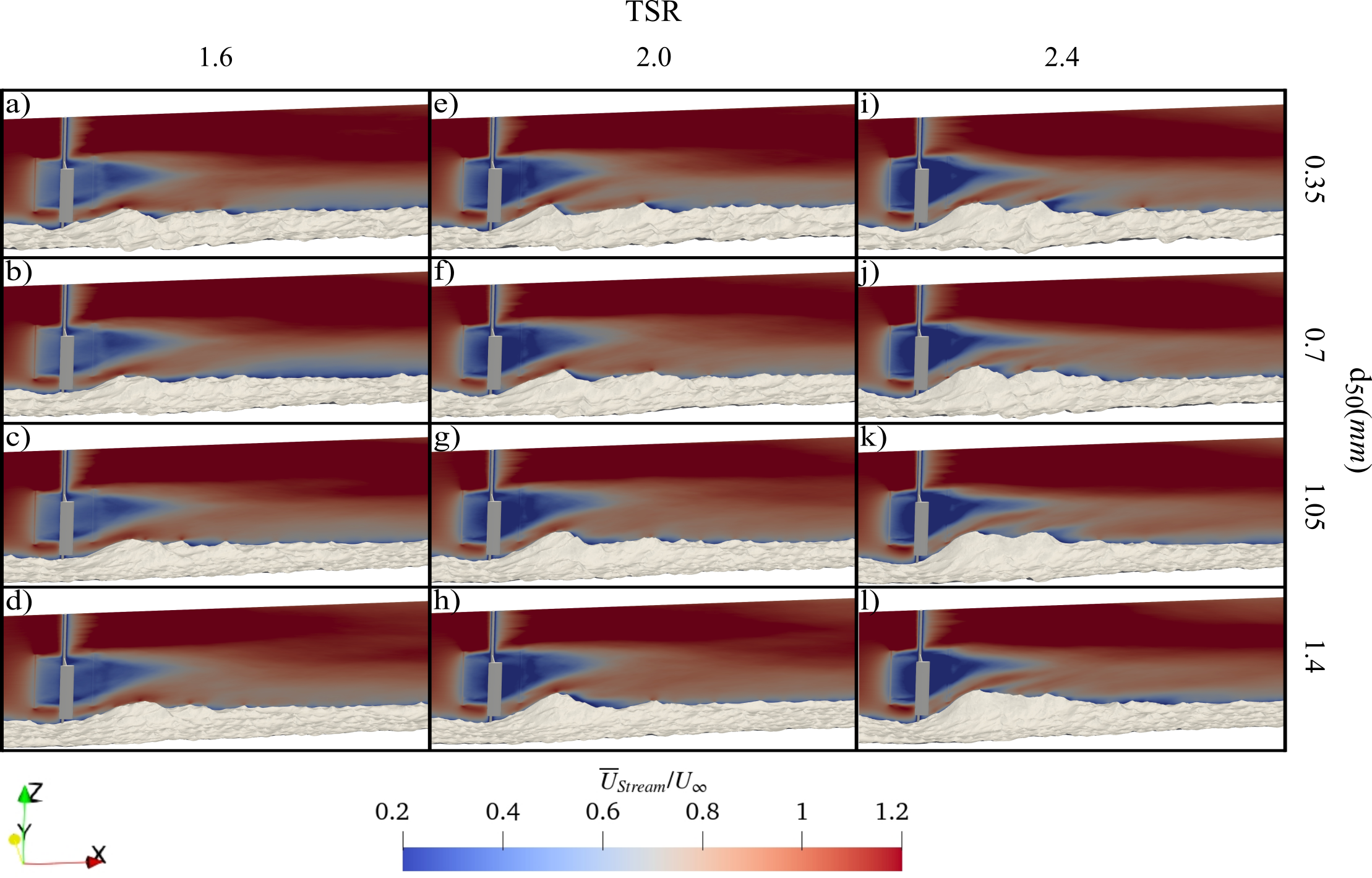}  
  \caption{Color maps of the computed mean streamwise velocity component normalized by the bulk velocity (=1.5 m/s) at the equilibrium state of the live bed (i.e., test cases $4$ to $15$). The color maps are shown over vertical planes along the centerline of the channel. The first, second, third, and fourth rows correspond to the bed material with median grain sizes of $d_{50}=0.35mm$, $0.7mm$, $1.05mm$, and $1.4mm$, respectively. While the first, second, and third columns correspond to TSR of $1.6$, $2.0$, and $2.4$, respectively. Flow is from left to right.}
  \label{Fig:7}
\end{figure}

We plot in \prettyref{Fig:7} dimensionless time-averaged streamwise velocity ($\overline{U}_{stream}/U_{\infty}$) at the dynamic equilibrium state of the live bed for various TSRs and sediment grain sizes. This figure corresponds to \prettyref{Fig:3}(d) to (f) for the rigid bed channel. Comparing the wake flow fields under the two bed conditions, the flow momentum in the near-bed region around the turbine under live-bed conditions is significantly lower than that of rigid-bed conditions. Such reduction in the near-bed flow velocity seems to be a direct result of the bed deformation. The bed topography seems to adjust itself to reach a minimal bed movement at the dynamic equilibrium. Also, as the vertical distance from the deformed bed increases, the influence of the live-bed condition on the mean streamwise velocity within the rotor region gradually diminishes. As a result, at approximately $0.3D$ above the deformed bed, the mean flow field roughly resembles the results observed under rigid-bed conditions.

As seen from the near wake results of \prettyref{Fig:7}, the wake flow exhibits a significantly shorter wake region and faster velocity recovery compared to those discussed over the rigid bed. The observed quick recovery could be attributed to the significant heterogeneity in the velocity field because of the live bed conditions.  Namely, the bed deformations, mainly scour geometry around the turbine, seem to introduce localized turbulence and enhance momentum exchange, facilitating quick flow recovery. More specifically, a jet flow emerges as a result of the bed deformations beneath of the turbine, accelerating the recovery of flow by injecting momentum into the wake core region and redistributing the flow energy.
Additionally, unlike the rigid-bed conditions in which the TSR of the turbine played a key role in the longitudinal size of the turbine wake, the live-bed simulation results indicate that the size of the wake region is relatively less influenced by the TSR. In other words, the downstream distance required for flow recovery over the deformed bed is nearly similar for different TSRs as the wake recovery under live-bed conditions seems to be mainly dominated by the strong near-bed jet flow injecting momentum into the wake core. We note that the intensity of wake deficit is still a function of TSR under both bed conditions, as discussed below. Lastly, as seen in this figure, comparing the wake flow field over the sediment layers with different grain sizes considered in the present study, it is found that the sediment grain size does not modulate the turbine wake recovery.

\begin{figure} [H]
  \includegraphics[width=1\textwidth]{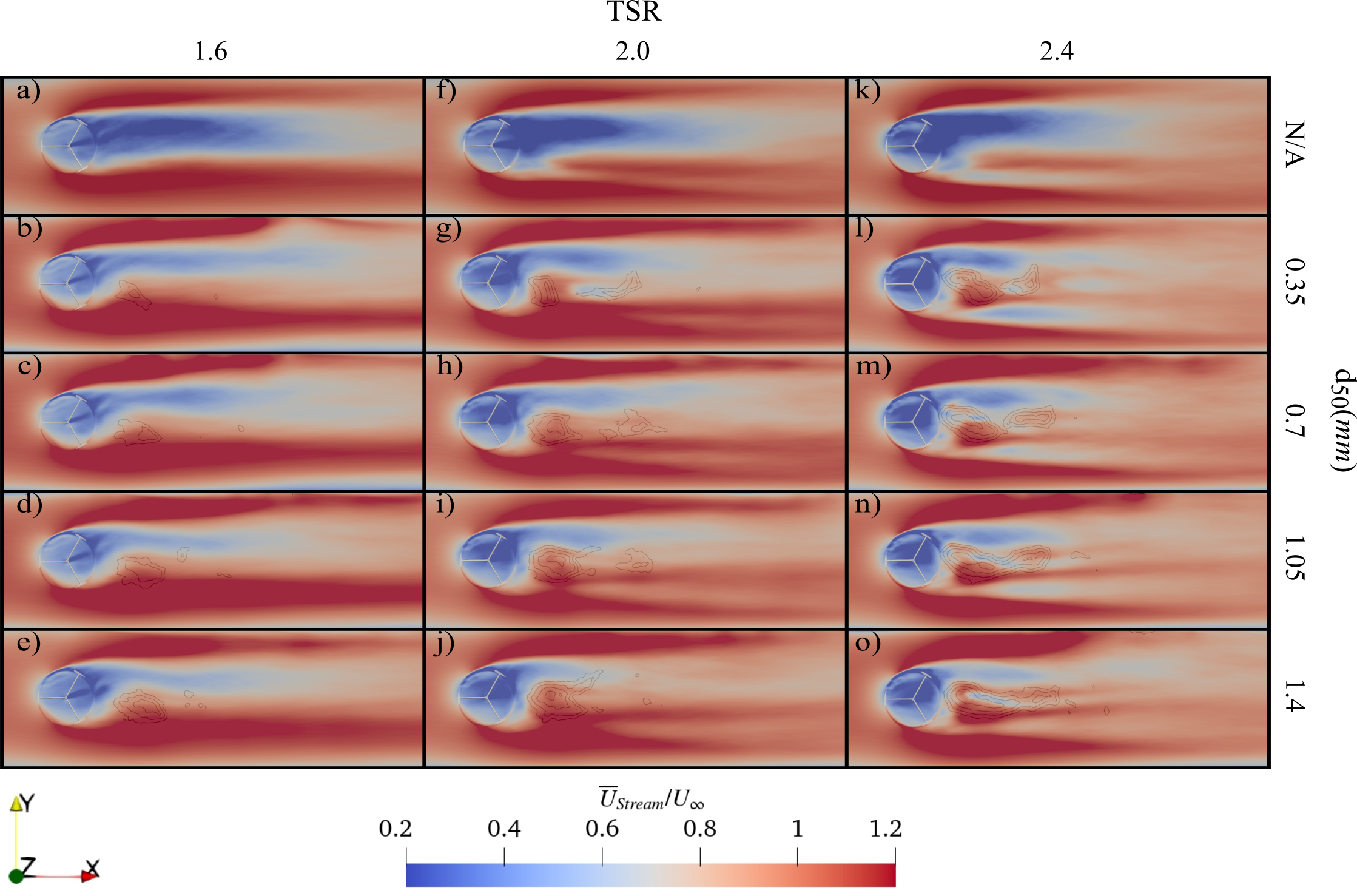}  
  \caption{
  Color maps of the computed mean streamwise velocity component normalized by the bulk velocity (=1.5 m/s) at an elevation of $z=0.4D$ above the bed and from the top view. The color maps are shown over horizontal planes and under rigid- (i.e., the first row corresponding to test cases 1 to 3) and live-bed (i.e., second to fifth rows corresponding to test cases 4 to 15) conditions. The first, second, and third columns correspond to TSR of $1.6$, $2.0$, and $2.4$, respectively. The second to fifth rows correspond to the bed material with median grain sizes of $d_{50}=0.35mm$, $0.7mm$, $1.05mm$, and $1.4mm$, respectively. The color maps of the live-bed conditions are superimposed over the contour lines of the bed elevation at the equilibrium state of the bed topography ranging from $0.12D$ to $0.3D$ in each case. Flow is from left to right.}
  \label{Fig:8}
\end{figure}

\begin{figure} [H]
  \includegraphics[width=1\textwidth]{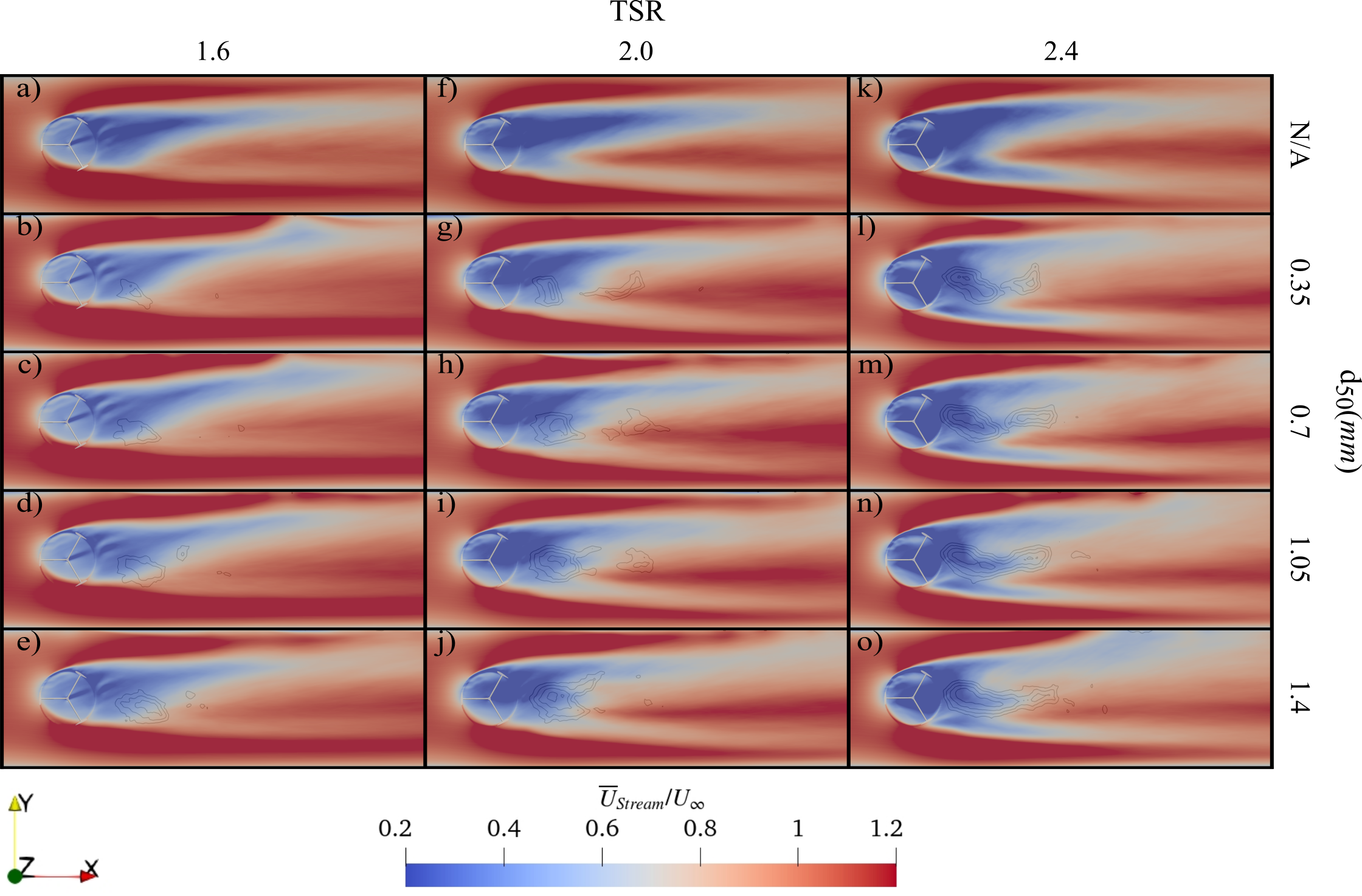}  
  \caption{Color maps of the computed mean streamwise velocity component normalized by the bulk velocity (=1.5 m/s) at an elevation of $z=0.8D$ above the bed and from top view. The color maps are shown over horizontal planes and under rigid-bed (i.e., the first row corresponding to test cases 1 to 3) and live-bed (i.e., second to fifth rows corresponding to test cases 4 to 15) conditions. The first, second, and third columns correspond to TSR of $1.6$, $2.0$, and $2.4$, respectively. The second to fifth rows correspond to the bed material with median grain sizes of $d_{50}=0.35mm$, $0.7mm$, $1.05mm$, and $1.4mm$, respectively. The color maps of the live-bed conditions are superimposed over the contour lines of the bed elevation at the equilibrium state of the bed topography ranging from $0.12D$ to $0.3D$ in each case. Flow is from left to right.}
  \label{Fig:10}
\end{figure}

To further examine the effect of the live bed conditions on the turbine wake flow dynamics, we depict in \prettyref{Fig:8} and \prettyref{Fig:10} the wake flow field of various cases over horizontal planes at vertical distances of $0.4D$ and $0.8D$ above the bed, respectively. These figures plot the color maps of normalized mean streamwise velocity components. The color maps are superimposed over the contour lines of the bed topography. We note that the first row of the figures corresponds to the flatbed of the channel under the rigid-bed conditions, thus lacking the contour lines. It should be noted that the concentric shapes of contour lines represent sand bar formations or deposition regions with relatively higher bed elevation, which often occur in the near-wake region. 

As observed in \prettyref{Fig:8}, which is taken slightly above the peak of the sand bar, i.e., $z/D=0.4$ above the bed, there exists a relatively weak correlation between turbine's TSR and wake deficit in the rotor area and near-wake region. In particular, as TSR increases, the wake deficit in the rotor region becomes slightly more pronounced, owing to the blade-flow interactions. This trend can be readily seen under both the rigid- and live-bed conditions. As seen in the first row of \prettyref{Fig:8} under the rigid-bed conditions, the near-wake flow field can be characterized by a persistent low-velocity region. Under the live-bed conditions, however, the sand bar deposition downstream of the turbine plays a key role in the wake flow field (see second to fifth rows of \prettyref{Fig:8}). The sand bar deposition in the near wake seems to generate localized jet flows that introduce extra momentum into the wake core, accelerating the wake recovery process. Further, the jet flow disrupts the downstream wake and contributes to an asymmetry in the wake structure, as previously reported in \citet{[69]}. This behavior contrasts with the symmetric wake flow pattern observed over the rigid bed. The asymmetrical effect caused by the sediment dynamics seems to intensify as the turbine's TSR increases. This behavior could be attributed to the interplay of the sand bar topology and the turbine's TSR, i.e., the higher the turbine's TSR, the more pronounced asymmetrical topology of the sand bar, which, in turn, leads to a more asymmetrical wake flow. Further, as seen in this figure, the median grain size of the bed material seems to have minimal, if any, effect on the wake flow field at this elevation above the bed.

At $0.8D$ above the bed (see \prettyref{Fig:10}), the effect of bed deformation significantly diminished and, therefore, the wake deficit patterns resemble those of the flow field over the rigid-bed conditions. The simulation results at this elevation above the bed demonstrate the increase of the wake deficit with the turbine's TSR. Also, the sediment grain size seems to have a minimal impact on the flow and wake deficit of the turbine. 

\begin{figure} [H]
  \includegraphics[width=1\textwidth]{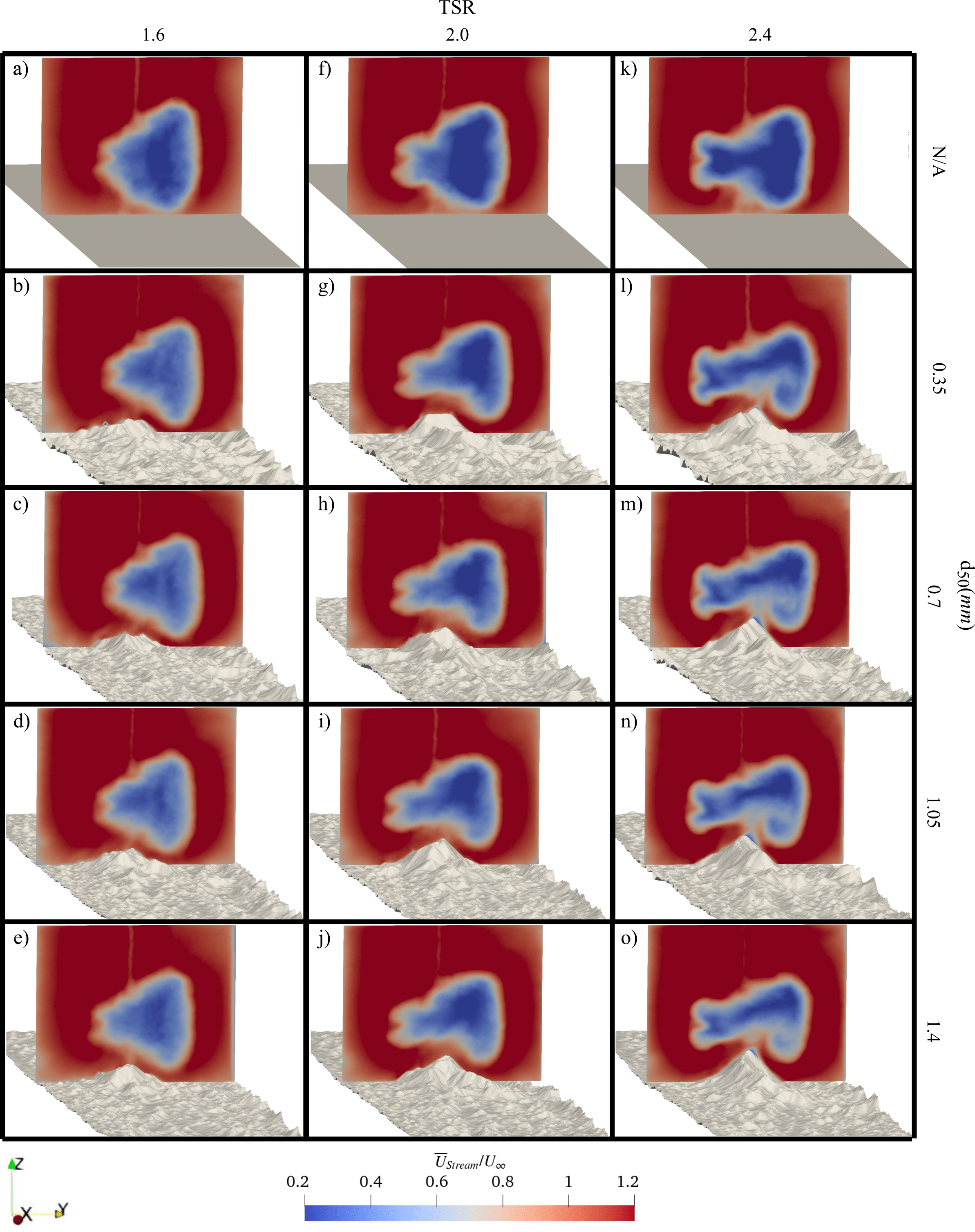}  
  \caption{Color maps of mean streamwise velocity (normalized with the bulk velocity) over cross planes located $1D$ downstream of the turbine at the equilibrium state. The first row corresponds to the rigid bed cases, while the second to fifth rows correspond to the live bed cases with $d_{50}=0.35mm$, $0.7mm$, $1.05mm$, and $1.4mm$, respectively. Moreover, the first column corresponds to TSR$=1.6$, the second column corresponds to TSR $=2.0$, and the third column corresponds to TSR$=2.4$. The bottom surfaces show the details of bed geometry in various cases at equilibrium.}
  \label{Fig:11}
\end{figure}

In \prettyref{Fig:11}, we plot color maps of mean streamwise velocity on cross planes across the channel and located $1D$ downstream of the turbine. The plots in the first row correspond to the rigid bed conditions, while the rest of the rows show the wake flow under live-bed conditions. This figure provides a qualitative comparison between the two bed conditions. As seen, the velocity deficit of the cases over the rigid-bed (see the dark blue area) is greater than that of the live-bed cases. Moreover, relatively speaking, the area with high momentum deficit is farther away from the live bed, owing to the effect of jet-like flow beneath the turbine. While the high momentum deficit region is quite close to the rigid bed. Although the equilibrium bed topology of the live-bed cases with various grain sizes is different, the variation in the sediment particle size seems to have a negligible impact on the wake flow.


\begin{figure} [H]
  \includegraphics[width=1\textwidth]{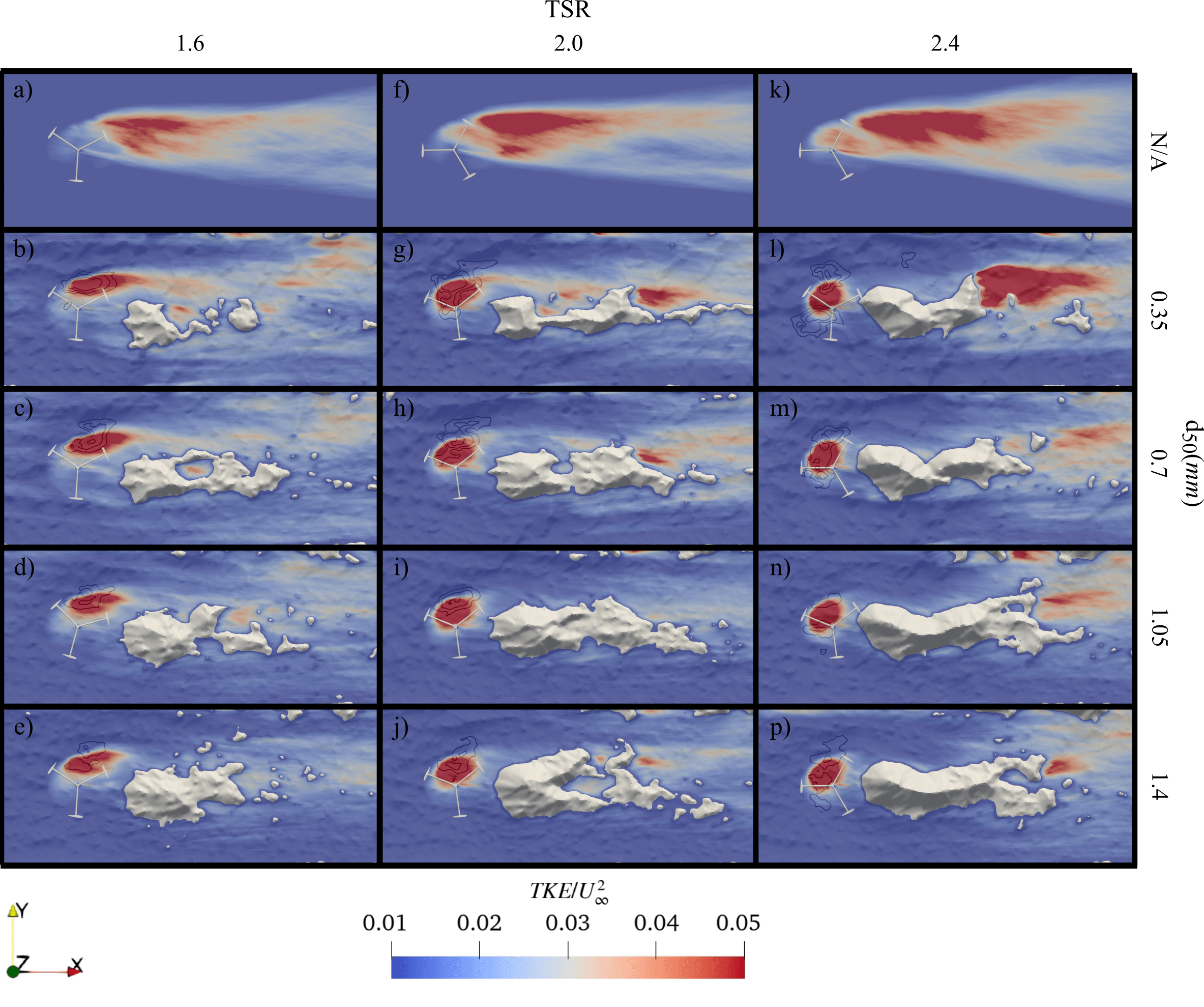}  
  \caption{Color maps of TKE, normalized by $(U_{\infty})^2$, from the top-view and over near-bed planes, $z=0.05D$. The top row presents the rigid bed conditions and the rest relate to the live bed conditions at equilibrium state. The contour lines, which are shown over the bed surface near the high TKE region, depict regions of deep scour with bed elevations ranging from $-0.2D$ to $-0.25D$. Flow is from left to right.} 
  \label{Fig:12}
\end{figure}

To explore the link between the scour and turbulence past the turbine, we plot in \prettyref{Fig:12} color maps of TKE, normalized by $(U_{\infty})^2$, at slightly above the bed, i.e., $Z=0.05D$. The horizontal planes are superimposed on the deformed bed of each test case. The gray areas correspond to the sand bars whose height is greater than $0.05D$. The dark red and blue regions of the color maps show the high and low TKE regions, respectively. The contour lines of bed elevation near the deep scour areas are displayed on the deformed bed surface. These contour lines can be seen near the high TKE region corresponding to the bed elevation ranging from $-0.2D$ to $-0.25D$. 

As seen in \prettyref{Fig:12}, the deepest scour areas closely overlap the region with high TKE, illustrating the critical contribution of turbulent fluctuations on the bed material erosion around the turbine. This finding is consistent with the findings of \citet{[122]}, who studied the flow and bed deformation around the base of stationary hydraulic structures in large-scale waterways. The bed material transported from the scour region is mainly deposited a short distance downstream, where the TKE is relatively lower, forming the sand bars (shown in gray).

\subsection{Sediment dynamics}
\label{subsec:Bed deformation analysis}
\noindent Herein, we discuss the bed morphodynamics results of the coupled flow and bed simulations. In all cases under the live-bed conditions, soon after the morphodynamics module is activated, the live bed of the channel starts evolving. Overt time, the entire channel bed experiences the formation of small-scale sand waves, i.e., ripples. The numerically captured sand waves grow in size and migrate downstream. As sand waves of different size develop and migrate downstream, the near wake region of the channel bed also undergoes a relatively large deformation. Namely, the near wake zone of the bed can be characterized by a dominant scour hole with a sand bar immediately downstream of it. The scour depth and sand bar's height continuously increase until they reach a dynamic equilibrium. By monitoring the successive fluctuation of the bed elevation in the scour and deposition zones, the dynamic equilibrium state is numerically identified when the changes in the bed elevations in two consecutive time-steps are less than $1 \%$ \cite{[69]}. 

\begin{figure} [H]
  \includegraphics[width=1\textwidth]{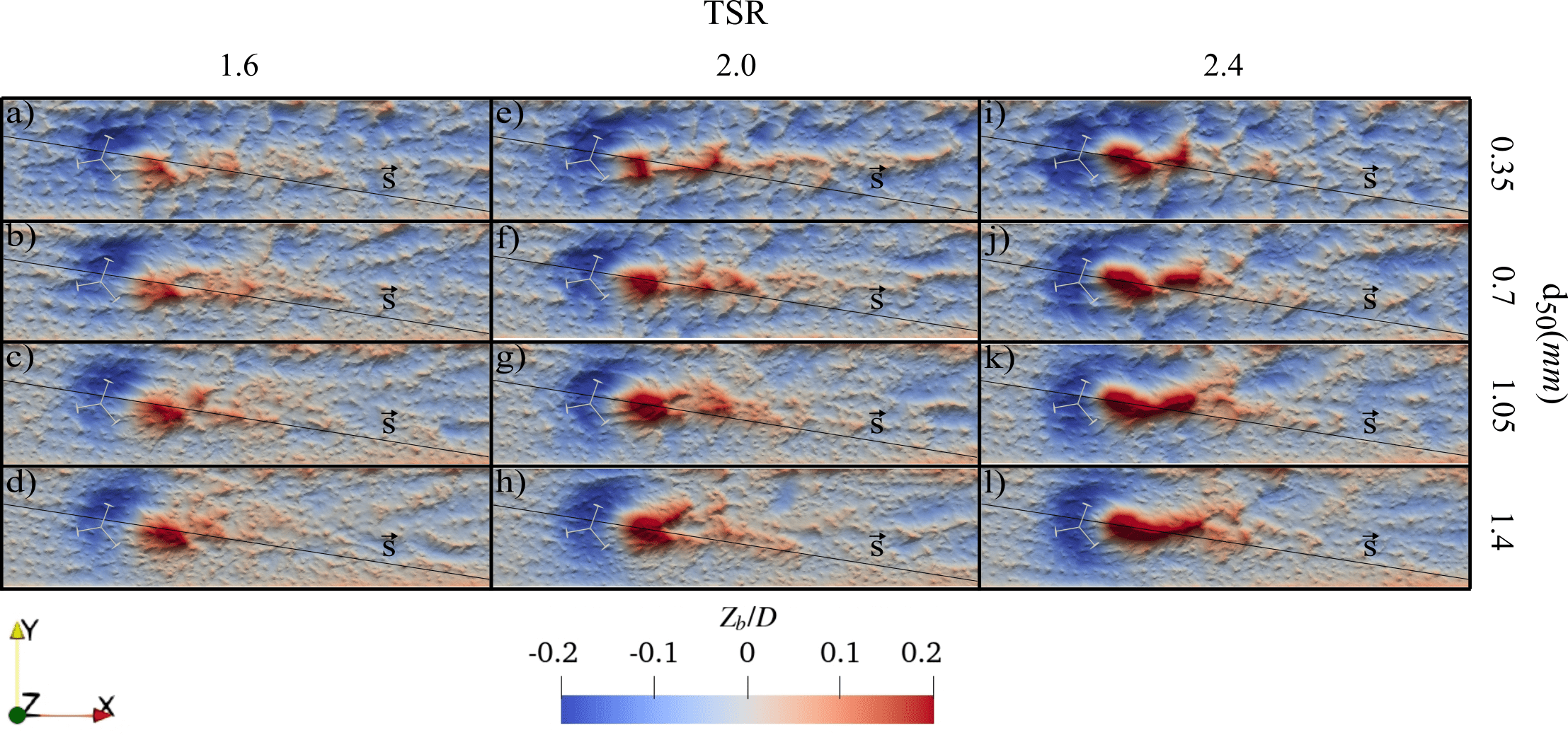}  
  \caption{Color maps of bed elevation, normalized with rotor diameter ($D=2m$), from the top view at dynamic equilibrium state. Each column depicts the results of a specific TSR, while each row corresponds to the results of a specific median grain size. The main scour zones can be observed around the turbines, followed by the sand bars that are captured immediately downstream from the turbine. The black lines mark the coordinates of the 's' vector along which the bed elevation profiles of Fig. 10 are extracted. Flow is from left to right. }
  \label{Fig:4}
\end{figure}

In \prettyref{Fig:4}, we plot the equilibrium bed geometry of the channel for various cases under live-bed conditions. At the first glance, the numerically captured contours of the bed elevation of the bed, normalized by the diameter of the rotor, reveal a number of salient features. First, the entire live bed of the flume is covered with sand waves with a range of amplitude. The smaller sand waves are at their early stages of development while the larger sand waves are at or near their full-grown size as they migrate downstream. The numerically observed sand waves are $7.3cm$ to  $17.3cm$ high and $1.1m$ to $1.8m$ long. We should note that it is critical to investigate the dynamics of these sand waves because their interaction with the scour hole and the sand bars could have important consequences for the performance and operation of the turbine. Namely, as sand waves migrate downstream, they typically cover the surface of the sand bars and scour holes. Therefore, the passage of sand waves can temporarily increase the local bed elevation over the peak of a sand bar, positioning it within reach of the rotating blades of the turbine.

Secondly, a scour region is observed around the base of the turbine. The maximum depth and horizontal expansion of the scour holes range from $0.4m$ to $0.5m$ and $3.8 m$ to $5.6m$, respectively. Depending on the TSR of the turbine, the pattern of the scour hole can be symmetrical with respect to the turbine. Namely, at lower TSR, the scour hole is more asymmetrical. However, in cases with higher TSR, the scour pattern becomes increasingly more symmetrical owing to the turbine-induced highly turbulent flow beneath the turbine and near the bed. This finding is consistent with the observations of flow and sediment interactions reported by \citet{[69]}. Given the proximity of the main scour zone to the base of the turbine, we argue that the depth and pattern of the scour hole should be considered as a factor in the design of the turbine foundation in natural environments.  

The third salient feature of the bed topology pertains to the sand bar formation. As seen, in all cases, a dominant sand bar forms downstream of the scour hole. The maximum height of the sand bars ranges from $0.5 m$ to $ 0.8 m$. The pattern and peak height of the sand bar are mainly dependent upon the TSR of the turbine. As seen, as TSR increases, the sand bar's height and length increase. As the sand bar's height increases, it could come into contact with the rotating turbine blades. In other words, as the peak of the sand bar rises it could end up within the reach of the rotating blades and eventually lead to collision of bed material and blades. Such collisions were somewhat observed in our simulation for the cases with TSR of 2.4. These blade-sand collisions, which were captured successfully with the IB method, can have important consequences in terms of damaging turbine components and operational malfunction. Overall, our simulation results emphasize the critical role of TSR in shaping the equilibrium bed morphology and controlling sediment transport processes in turbine-driven environments. At the same time, within the range of particle sizes studied in this study, increase in the bed material size mildly enhances the sediment deposition leading to higher sand bars.

\begin{figure} [H]
  \includegraphics[width=0.9\textwidth]{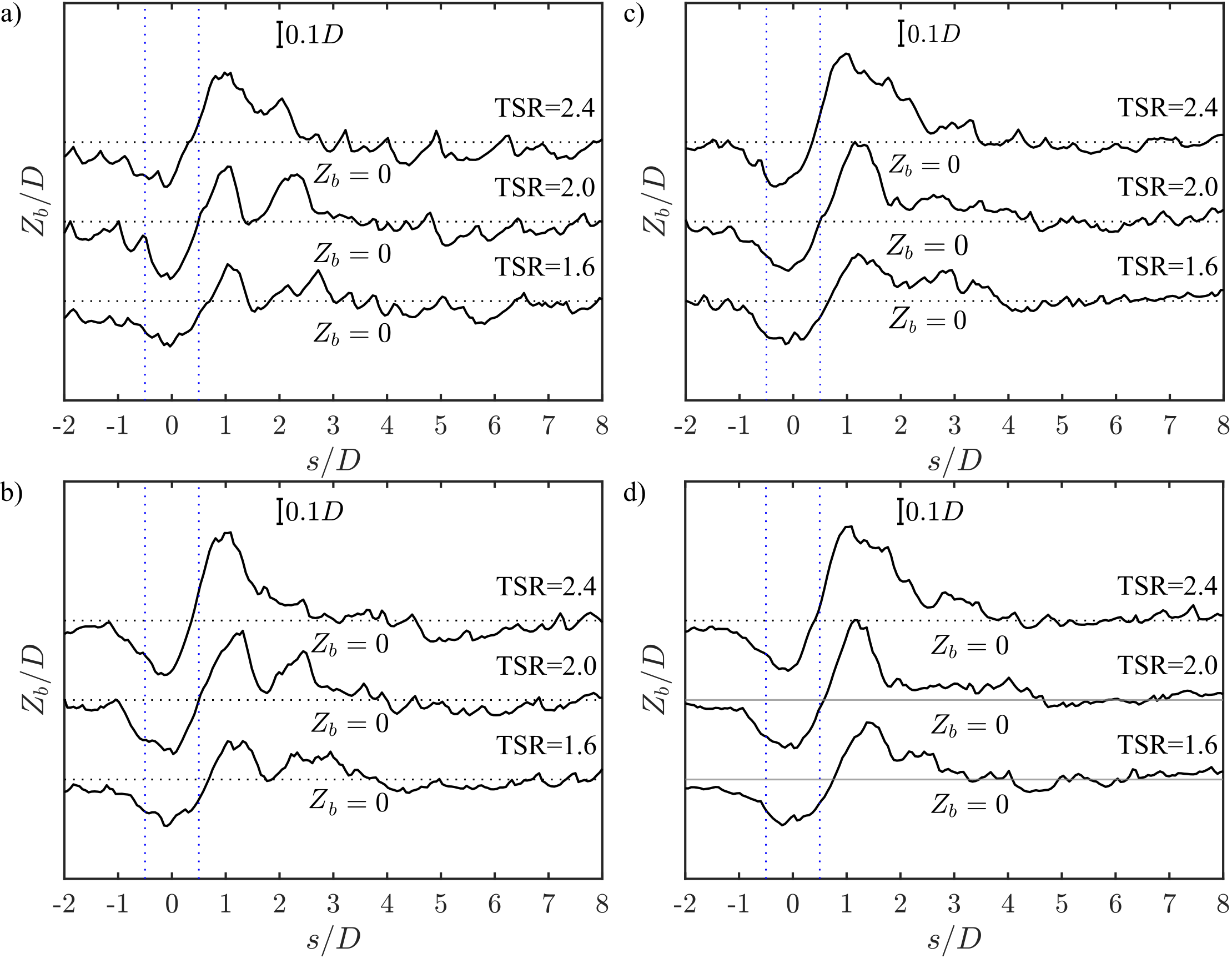}  
  \caption{Computed profiles of bed elevation along 's' vector at the dynamic equilibrium state for cases $4$ to $15$. The location of the vector is shown in \prettyref{Fig:4}. The center of the turbine is located at $s/D=0$, and the blue vertical dashed lines indicate the rotor region. The horizontal dashed lines represent the elevation of the initial flatbed, at z=0. (a) to (d) correspond to $d_{50}=0.35mm$, $0.7mm$, $1.05mm$, and $1.4mm$. The scale shows the vertical 
 dimensions. Flow is from left to right.}
  \label{Fig:5}
\end{figure}

In \prettyref{Fig:5}, we plot the profiles of the bed elevation along 's' vector (the coordinate of 's' vector is marked in \prettyref{Fig:4}). The bed profiles illustrate the numerically captured sand waves, scour holes and sand bars. We selected the coordinate of 's' vector to overlap with the trajectory of the larger and more prominent sand wave in the channel. As a result, 's' vector originates $2D$ upstream of the turbine and extends $8D$ downstream, forming an approximate angle of $8^\circ$ relative to the streamwise direction.
 As seen, there exist a wide range of sand waves ranging from a few centimeters to over $17cm$ in amplitude and about $15cm$ to over $1.8m$ in wavelength. These numerically captured sand waves are indeed ripples and dunes, respectively \cite{[119], [69]}. The ripples are visible throughout the channel as they migrate over the surface of larger sand waves, i.e., dunes, which also migrate downstream. The observed migrating sand waves mark the sediment mobilization and transport owing to the nearbed turbulence. 
 
The center of the turbine is located at $s/D=0$, while the blue vertical dashed lines indicate the rotor region. In all cases, the scour region can be seen in the near-wake region, i.e. $-1 < s/D < 1$, with the maximum scour depth occurring at $s/D \approx 0$, i.e., approximately beneath the turbine. This erosion seems to be driven by the local blades-induced turbulence and, consequently, the elevated bed shear stress and sediment transport. The sediment material picked up from the scour hole are mainly deposited immediately downstream, particularly in the region $1< s/D <3$, creating a single sand bar. The maximum height of the sand bar occurs approximately $1D$ downstream of the turbine, while the sand bar extends farther downstream to about $4D$ from the turbine. The location and positioning of the scour holes and sand bars seem independent of the TSR and grain size. However, an increase in either TSR  or a decrease in $d_{50}$ slightly intensifies the erosion and deposition resulting in greater scour depth and sediment deposition height. 


\subsection{Wake recovery}
\label{subsec:Wake recovery over rigid- and live-bed conditions}
\begin{figure} [H]
  \includegraphics[width=1.0\textwidth, height=0.85\textheight, keepaspectratio]{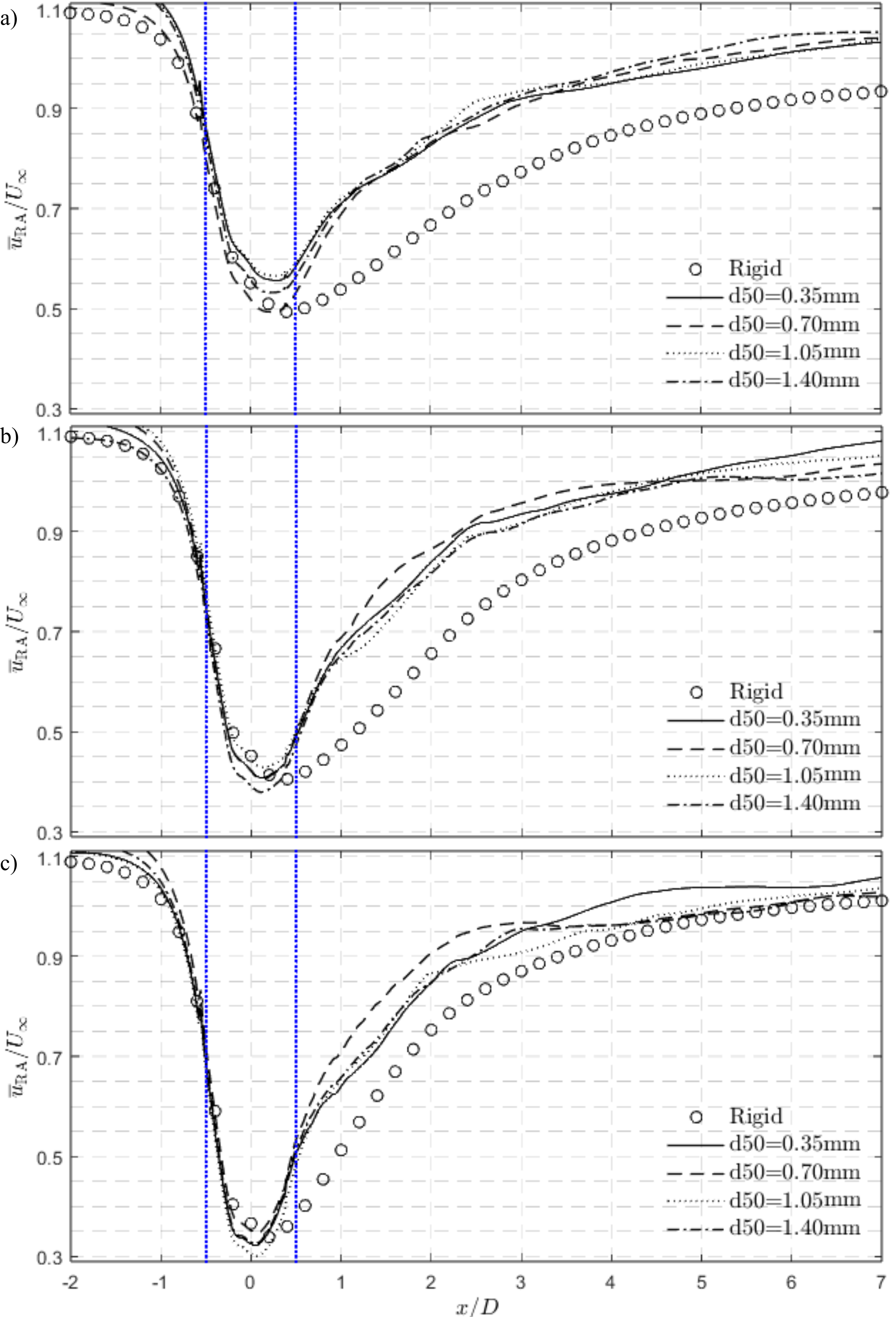}
  \caption{Variations of mean streamwise velocity averaged over the rotor swept area in the streamwise direction. Hollow circles represent the simulation results of the rigid bed case. The solid, dashed, dotted, and dash-dotted lines correspond to the cases under live bed conditions with different grain sizes. The blue vertical dashed lines indicate the turbine region. (a), (b), and (c) correspond to TSR of 1.6, 2.0, and 2.4, respectively.}
  \label{Fig:13}
\end{figure}

\noindent To examine the wake recovery of the turbine in different cases, we analyzed the longitudinal profile of the first-order turbulence statistic of the wake flow, i.e., the mean streamwise velocity, averaged over the swept area of the turbine as follows \cite{[69]}:
\begin{equation}
 \overline{u}_{RA} = \frac{4}{\pi D^2} \int_{A} \overline{U}_{stream} \, dA
 \label{eq:25}
\end{equation}

\noindent where $\overline{U}_{stream}$ is the mean streamwise velocity within the swept area of the turbine. 
Figure~\ref{Fig:13} depicts the longitudinal profile of mean streamwise velocity within the swept area. The circles mark the simulation results for the rigid bed cases, in which, the minimum mean streamwise velocity are $0.49$$U_{\infty}$, $0.39$$U_{\infty}$, and 0.34$U_{\infty}$ for the TSR of $1.6$, $2.0$, and $2.4$, respectively. This clearly suggests greater momentum deficits at higher TSR. In this case, the mean velocity starts reducing at about $1D$ upstream of the turbine. This trend continues to about $0.4D$ downstream of the turbine, where the maximum momentum deficit is observed. Following the maximum velocity deficit at $x/D=0.4$, the wake recovery process begins and progresses steadily until the velocity is nearly recovered at $x/D=7.0$, $x/D=5.5$, and $x/D=4.2$ downstream of the turbine for TSR of $1.6$, $2.0$, and $2.4$, respectively. These results demonstrate that, in the range of parameters studied in this work, increasing the TSR leads to both a greater momentum deficit and faster wake recovery, which is consistent with the findings of prior studies \cite{[117], [51], [39]}. This behavior could be also attributed to dynamic solidity, which is further discussed elsewhere \cite{[51], [39], [124]}. 

For a turbine with a constant chord blade, the dynamic solidity can be defined as \cite{[39]}:
\begin{equation}
\sigma_D = 1 - \frac{D}{2C} \frac{t_{blade}}{t_{\text{fluid}}} \,  = 1 - \frac{1}{2 \pi \sigma \Omega}
\label{eq:26}
\end{equation}
\noindent where $t_{blade} = \Gamma/N_bU_{\infty}\Omega$, is the time taken for the turbine blades to traverse the gaps between them, with $\Gamma$ representing the sum of the gaps around the perimeter of the turbine rotor, and the time required for a fluid particle with the bulk velocity to pass the same distance is $t_{fluid} = \Gamma/U_{\infty}$. In a sense, the ratio of $t_{blade}/t_{fluid}$ describes the concept of dynamic solidity \cite{[39]}. As TSR increases, this ratio -- i.e. dynamic solidity -- decreases, indicating that the flow particles have less time to traverse the length already covered by the blade. In other words, with increasing TSR, flow particles would have less time to respond to the blade's passage, making the blades appear more solid to the incident flow. Importantly, dynamic solidity has some limitations. For TSR $= 0$, the dynamic solidity has a singularity, and for TSR $<1/2 \pi \sigma$, it becomes negative, which has no physical meaning. Moreover, for TSR $>> 1$, the dynamic solidity approaches 1, corresponding to the geometrical solidity of a solid cylinder \cite{[39]}. The dynamic solidity of different cases with TSR of 1.6, 2.0, and 2.4 is ${\sigma}_D \ = 0.59$, 0.67, and 0.72, respectively. 
As seen in Fig. 11, as the dynamic solidity increases, the wake recovery speeds up, which is consistent with the findings of Araya et al. \cite{[39]}, who reported an increased rate of wake recovery for ${\sigma}_D > 0.65$.

Now we focus our attention on the wake recovery of the turbine under live bed conditions. In Fig. 11, different black lines show the longitudinal profiles of mean streamwise velocity over the swept area of the turbine. As seen, regardless of the bed material size, the maximum drop in momentum occurs at about $0.1D$ downstream of the turbine.  The mean streamwise velocity begins to recover until it fully recovers at approximately $x/D=3.0$ downstream of the turbine for all live bed conditions. Our observations indicate that, at any distance into the wake, the momentum deficit of the live-bed cases is significantly less than that of the rigid-bed case. In other words, sediment dynamics under live-bed conditions significantly reduces the velocity deficit in the wake flow. Further, the effect of TSR on the wake recovery deems similar to that of the rigid-bed case and the results of wake recovery of various sediment sizes seem convergent. Our finding concerning the wake recovery in presence of sediment transport is consistent with those of \citet{[69]} who conducted a similar study for the horizontal axis turbine. Although this work is limited to a single turbine, we argue that the smaller velocity deficits under live-bed conditions would reduce the wake-wake interaction of two longitudinally aligned turbines. In other words, in tidal farms with arrays of such turbines, sediment dynamics could be detrimental to the intensity of wake-wake interaction of longitudinally aligned turbines. Thus, the sediment transport could somewhat enhance the performance of tidal farms with arrays of vertical axis turbines.


\subsection{Turbine performance}

\noindent Now we focus on analyzing the efficiency of the turbine under the rigid- and live-bed conditions in terms of power generation. We note that the performance analysis is carried out using our simulation results for the single utility-scale turbine considering the impact of sediment dynamics on the turbine performance. To do so, we calculate the mean power coefficient of different cases, as follows \cite{[27], [40], [69]}:

\begin{equation}
C_p = \frac{\overline{P}}{P_{max}}
\label{eq:27}
\end{equation}

\noindent where $\overline{P}$ is the mean power extracted from the flow, and $P_{max}$ is the maximum power that can be obtained for a given bulk velocity of the flow passing through the turbine. These two parameters are defined as follows \cite{[27], [40], [69], [126]}:

\begin{equation}
\overline{P} = \overline{T} \times \Omega
\label{eq:28}
\end{equation}

\begin{equation}
P_{max} = 0.5 \rho {U^3}_{\infty} A
\label{eq:29}
\end{equation}

\noindent where $\overline{T}$ is the mean torque the flow applies to the shaft. The maximum power coefficient is limited to $16/27$, i.e., Betz’s limit \cite{[127]}.

\begin{figure} [H]
  \includegraphics[width=1.0\textwidth]{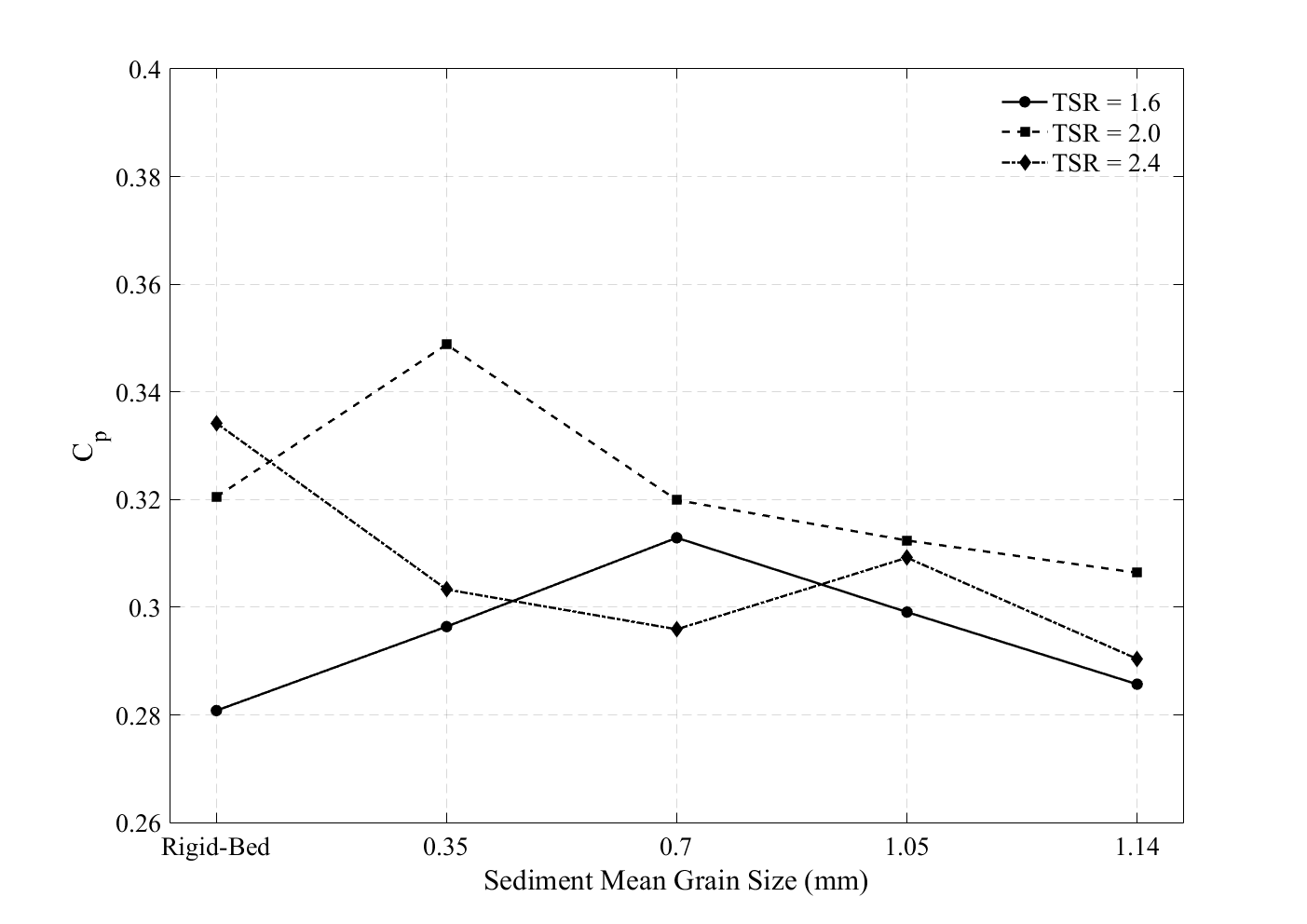}
  \caption{The mean power coefficient of the single utility-scale turbine under the rigid- and live-bed conditions.}
  \label{Fig:14}
\end{figure}

In \prettyref{Fig:14}, we plot the mean power coefficient across different bed conditions and TSRs. Starting with the power coefficient under rigid-bed conditions, i.e., the three data points on the left, it can be seen that the power coefficient decreases as TSR decreases. This is because the blades are experiencing greater dynamic stall at low TSR, which reduces the power coefficient. This observation is consistent with the findings of \citet{[27]}, who demonstrated that a decrease in TSR below the optimum value leads to reduced turbine efficiency. For example, as the TSR reduces from TSR$=2.4$ to TSR$=1.6$, the efficiency of the turbine decreases by about $5.34 \%$. This trend does not hold for live bed conditions, as the turbine's efficiency fluctuates with TSR and sediment particle size. These fluctuations in efficiency can be attributed to the changes in hydrodynamics around the turbine, induced by the migrating sand waves and the scour hole and sand bar development in the near wake region. These changes impact the torque applied to the rotor shaft, thereby affecting the power coefficient.

A comparison of turbine efficiency over the rigid and live beds highlights the influence of bed conditions and TSR on turbine performance.
More specifically, the average turbine efficiency over live beds increases from $29.85 \%$ to $32.04 \%$ as TSR rises from $1.6$ to $2.0$, but then declines to $29.97 \%$ when TSR increases to $2.4$. This trend suggests that while the additional simulations with a broader range of TSRs might be needed to precisely identify the optimal TSR for maximum turbine efficiency, it likely falls between $1.6$ and $2.4$ under live-bed conditions. At TSR $= 1.6$, turbine efficiency slightly improves by an average of $1.77 \%$ under the live-bed conditions. As TSR increases to $2.0$, the turbine reaches its highest efficiency over the finest sediment particle size ($d_{50} = 0.35mm$), showing a $2.6 \%$ increase. However, as the mean grain size increases, turbine efficiency declines, averaging a $0.89 \%$ efficiency reduction. At TSR $= 2.4$, the turbine experiences a more significant drop in mean power efficiency, with an average decrease of $3.44 \%$ across various sediment sizes.

Overall, for the range of the TSR studied in this work, the turbine's power coefficient under live-bed conditions seems to be about $1.7\%$ less than that under rigid-bed conditions. Despite this minor reduction in the efficiency of the single turbine, the live-bed condition was shown to enhance the wake recovery. This implies that in a tidal farm with arrays of turbines that is characterized with wake-wake interactions, live-bed conditions are expected to improve the overall performance of the tidal farms. Having said that, further studies with arrays of turbines is needed to further explore the impact of sediment dynamics on the efficiency of tidal farms.

\section{Conclusion}
\label{sec:5}
\noindent We employed a fully coupled LES and bed morphodynamics model, using the Virtual Flow Simulator (VFS-Geophysics) code, to investigate the performance of a utility-scale vertical-axis hydrokinetic turbine under live-bed conditions. Resolving the turbine and its components using the immersed-boundary method, the two-way interactions between turbine-induced flow structures and morphodynamics were numerically studied to gain insights into the mutual impact of turbine's wake flows and sediment transport, in terms of scour and sand bar development, bedform evolution, and turbine performance. To that end, a series of coupled flow and morphodynamics simulations with different tip speed ratios of the utility-scale turbine were carried out under live-bed conditions with considering a range of sediment particle sizes. As a baseline, the simulations were also conducted over the rigid-bed conditions. The simulation results were carefully analyzed to make sense of the wake behavior, power production, and sediment dynamics at various tip speed ratios and bed material sizes. 

Comparative analysis of turbine wake flow fields showed that the momentum deficit in the cases with live-bed conditions is somewhat less pronounced than that of the live-bed conditions. This was attributed to the formation of a jet-like flow beneath the turbine that injects high-momentum flow into the wake region, elevating the local velocity over the stoss side of the sand bar under the live-bed conditions. Moreover, our simulation results revealed that the bed deformation around the turbine leads to significant wake asymmetry, which is also intensified at higher tip speed ratios. In addition, contours of high turbulence kinetic energy in the wake were juxtaposed with the deep scour region around the vertical-axis turbine. 

Our morphodynamics results revealed the development of sand waves with a range of amplitude and wavelength throughout the bed. Throughout their development stages, and mainly when fully grown, these sand waves interact in a two-way manner with the wake flow. Also, locally elevated turbulence kinetic energy around the base of the turbine gives rise to a primary scour hole at the base of the turbine tower. The eroded bed material from the scour region deposits somewhat downstream to develop a primary sand bar with a unique geometry. As mentioned above, the flow over the stoss side of the sand bars accelerates to form the jet-like flow beneath the turbine. In cases with intense turbulence and high tip speed ratios, the crest of the sand bar is high enough to collide with the turbine's rotating blades.  

Our numerical observations concerning the wake recovery showed that
sediment transport substantially reduces the velocity deficit in the near wake region and thus enhances the wake recovery. Additionally, the turbine performance analysis revealed that the sediment transport under live-bed conditions is somewhat detrimental to the turbine performance, i.e., slightly decreasing the power coefficient by a minimal margin of about $1.7\%$. Past studies, however, have demonstrated that turbine-turbine wake interactions among arrays of turbines significantly reduce the overall efficiency of a tidal farm \cite{[148], [149]}. Hence, we argue that, by allowing for a faster wake recovery, sediment transport under live-bed conditions would potentially result in a greater power generation and a net reduction in the levelized cost of energy of the utility-scale tidal farms. In a future study, we will replicate this study by considering arrays of vertical-axis turbines to explore the impact of live bed conditions on the performance of turbine arrays.

\FloatBarrier
\section*{Acknowledgements}
\label{sec:acknowledge}
\noindent This work was supported by grants from the U.S. Department of Energy’s Office of Energy Efficiency and Renewable Energy (EERE) under the Water Power Technologies Office (WPTO) Award Numbers DE-EE0009450 and DE-EE00011379. Partial support was provided by NSF (grant number 2233986). The computational resources for the simulations of this study were partially provided by the Institute for Advanced Computational Science at Stony Brook University. The views expressed herein do not necessarily represent the view of the U.S. Department of Energy or the United States Government.

\section*{Author Contributions}

\noindent
\textbf{Mehrshad Gholami Anjiraki:} Conceptualization (equal); Data curation (equal); Formal analysis (equal); Investigation (equal); Methodology (equal); Visualization (equal); Writing – original draft (equal); Writing – review \& editing (equal). \textbf{Mustafa Meriç Aksen:} Investigation (equal); Visualization (equal); Writing – review \& editing (equal). \textbf{Jonathan Craig:} Investigation (equal); Methodology (equal); Writing – original draft (equal); Writing – review \& editing (equal). \textbf{Hossein Seyedzadeh:} Conceptualization (equal); Validation (equal); Writing – review \& editing (equal). \textbf{Ali Khosronejad:} Conceptualization (equal); Data curation (equal); Formal analysis (equal); Funding acquisition (lead); Investigation (equal); Methodology (equal); Project administration (lead); Resources (lead); Software (lead); Supervision (lead); Validation (equal); Visualization (equal); Writing – original draft (equal); Writing – review \& editing (equal).

\section*{Data Availability Statement}
\label{sec:7}
\noindent The software code (VFS-$3.1$ model) (\href{https://doi.org/10.5281/zenodo.15002824}{10.5281/zenodo.15002824}), 
along with the hydrodynamic results (\href{https://doi.org/10.5281/zenodo.15002375}{10.5281/zenodo.15002375}), 
power production data (\href{https://doi.org/10.5281/zenodo.15001388}{10.5281/zenodo.15001388}), 
wake recovery (\href{https://doi.org/10.5281/zenodo.15001454}{10.5281/zenodo.15001454}), 
the instantaneous morphodynamic results (\href{https://doi.org/10.5281/zenodo.15001934}{10.5281/zenodo.15001934}) for the test cases, and the channel and VAT surface files (\href{https://doi.org/10.5281/zenodo.15002280}{10.5281/zenodo.15002280}), 
are available in the Zenodo online repository.

\appendix
\section{Validation study}
\begin{figure} [H]
  \includegraphics[width=1\textwidth]{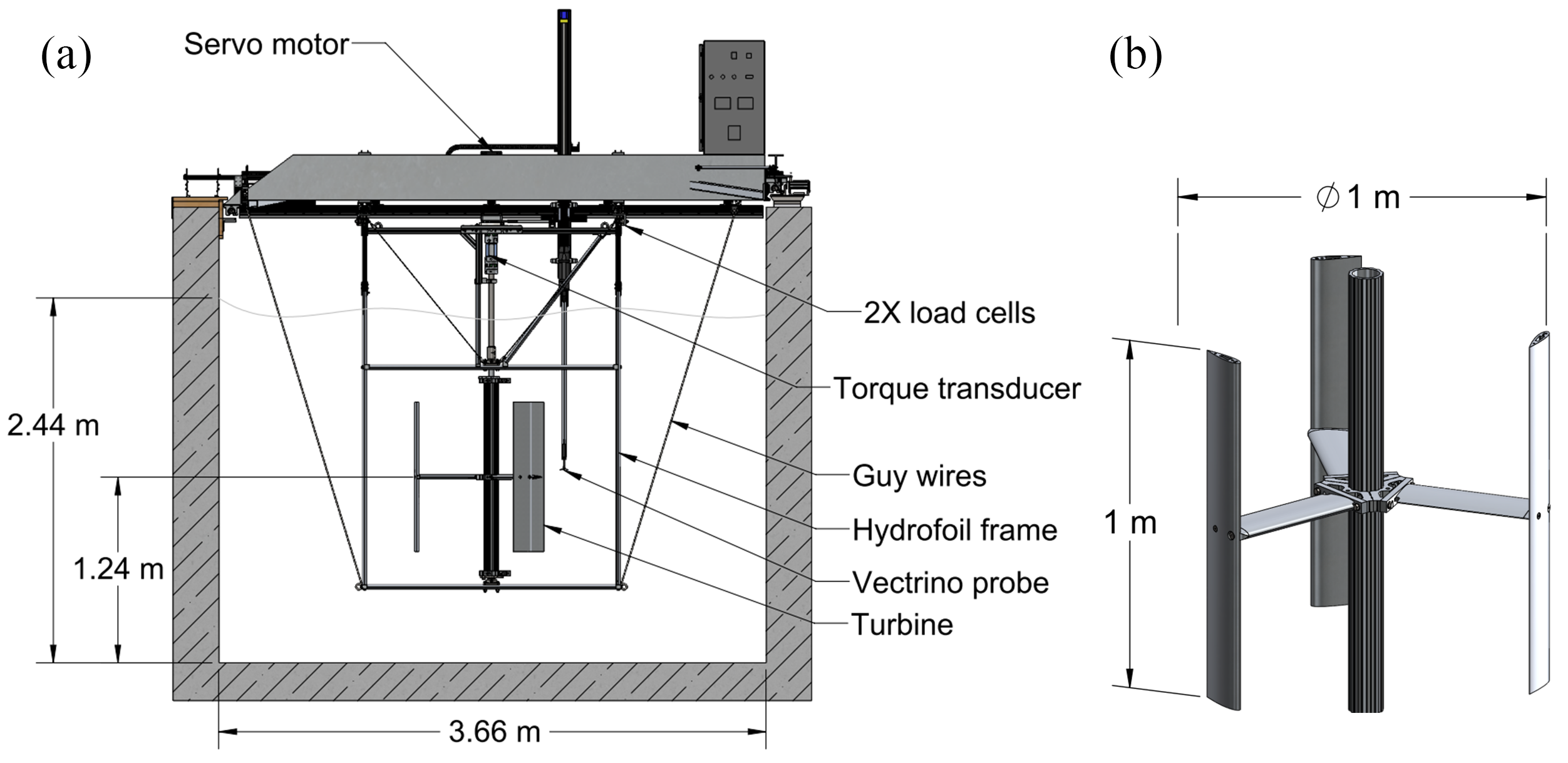}  
  \caption{Schematic of the channel and turbine setup in the tow tank. (a) Tow tank with the VAT installed, and (b) Turbine model with blades mounted along the midline of the turbine axis. Both the blades and struts have a NACA0020 hydrofoil profile. \cite{[130]}.}
  \label{Fig:A15}
\end{figure}

\noindent The coupled hydro- and morpho-dynamic model used in this study has undergone thorough validation against experimental data from both laboratory and field-scale studies \cite{[128], [64], [150], [151], [152], [153]}. Herein, the flow solver is validated using experimental data \cite{[130]} obtained from a laboratory test in a two tank involving flow around a high-solidity vertical axis turbine at the University of New Hampshire (UNH). The tow tank has a length of $36m$, a width of $3.66m$, and a depth of $2.44m$. Details of the experimental test can be found in \cite{[131],[132]}. The turbine model used in this study is the UNH Reference Vertical Axis Turbine (RVAT), which is similar to the Sandia National Labs/U.S. Department of Energy Reference Model $2$ (RM2) River Turbine \cite{[133]}. This turbine has three blades with a NACA0020 hydrofoil profile and a chord length of $14cm$. The turbine features a $1m$ span and a $1m$ diameter, with a blockage ratio of $11\%$, a solidity of $Nc/\pi D=0.13$, and a chord-to-radius ratio of $c/R=0.28$. A schematic of the experimental setup is provided in \prettyref{Fig:A15}.

\begin{figure} [H]
  \includegraphics[width=1\textwidth]{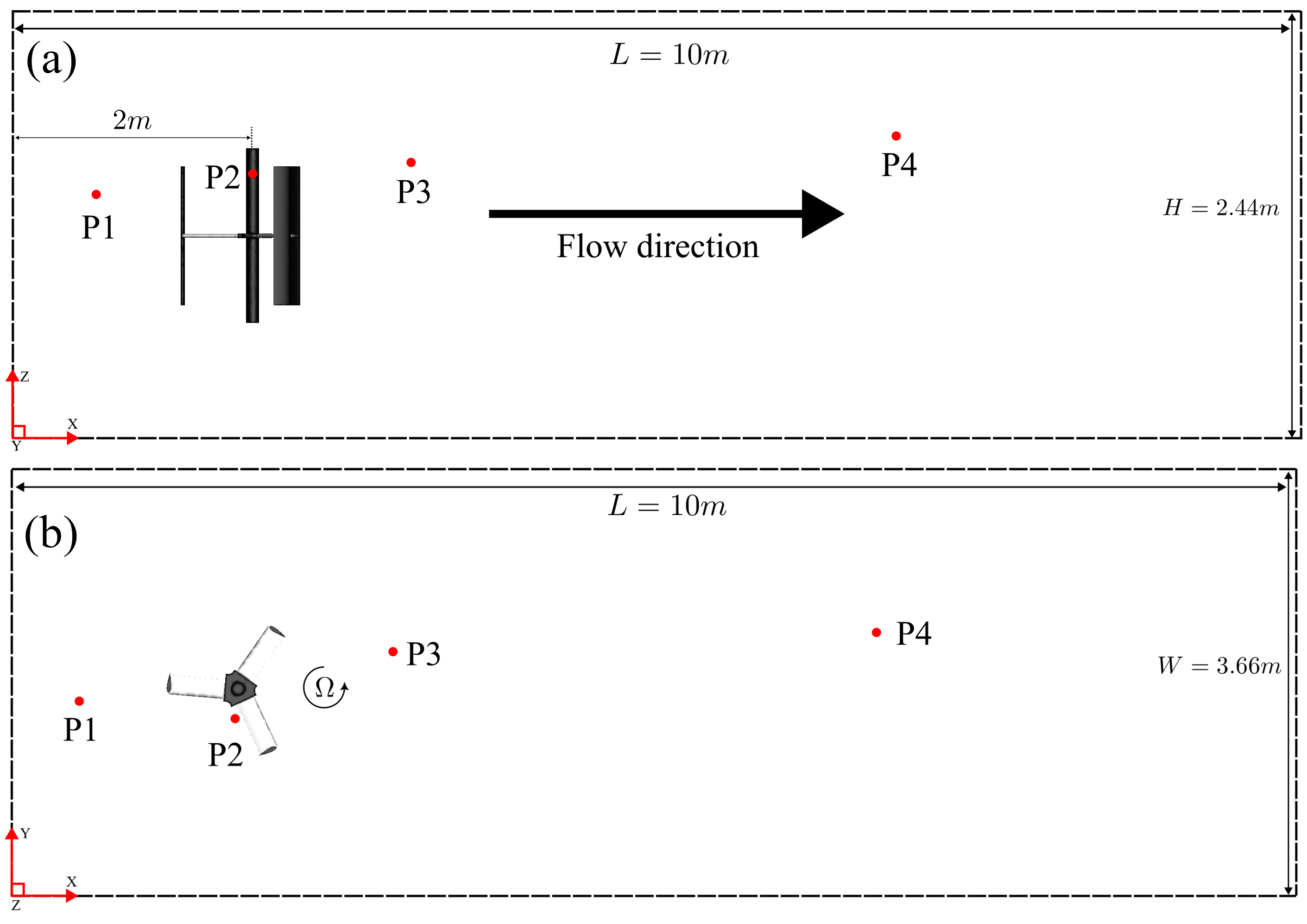}  
  \caption{Schematic of the simulation domain with the turbine from the side (a) and top (b) views. The rotor has a diameter of $1m$. The turbine is positioned $2m$ from the inlet. Points $P1$ to $P4$ show the probing locations to compare the numerical and measured results.}
  \label{Fig:A16}
\end{figure}

Different tow speeds were tested to obtain the optimum TSR ($=1.9$). At a Reynolds number $Re=0.8\times  10^{6}$ the power coefficient became independent of $Re$, leading to the selection of a bulk velocity of $U_{\infty}=1 m/s$, corresponding to $Re=1\times  10^{6}$ \cite{[130]}. The tow in the experimental setup corresponds to uniform inflow boundary conditions in the numerical simulations. The simulation domain matches the experimental setup in height and width, with a length of $10m$. The turbine is positioned $2m$ from the inlet. A schematic of the simulation domain with the turbine can be found in \prettyref{Fig:A16}.

The computational grid resolution is $N_{x}=1249$, $N_{y}=433$, and $N_{z}=309$, uniformly distributed in the x, y, and z directions, respectively, resulting in nearly $170$ million grid nodes. Near wall regions of the flow are modeled using the wall model. 

\begin{figure} [H]
  \includegraphics[width=1\textwidth]{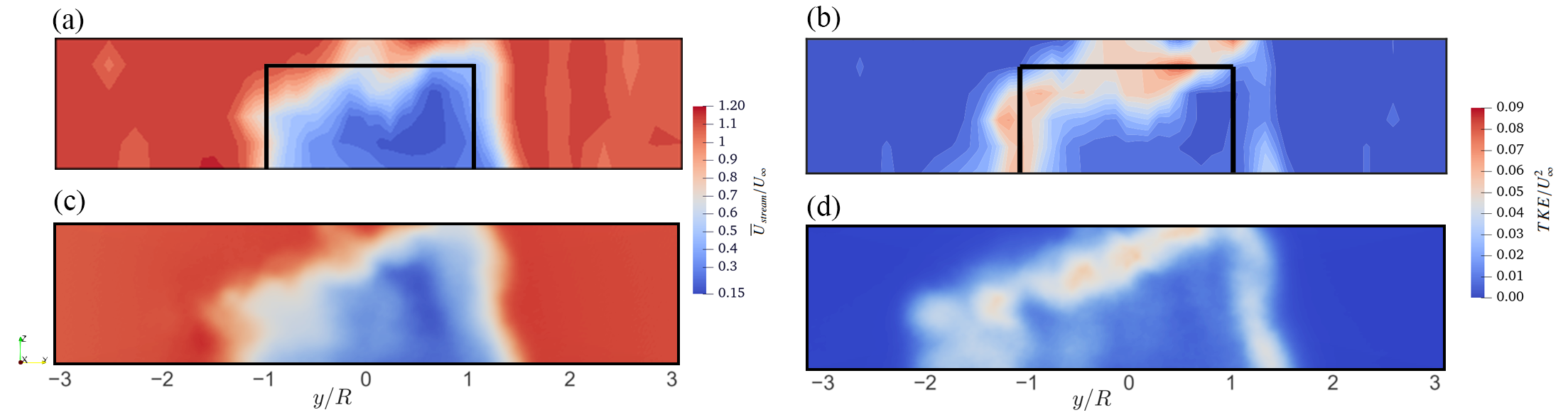}  
  \caption{Color maps of measured (a,c) and computed (c,d) normalized mean streamwise velocity (a,c) and turbulence kinetic energy (b,d) at $x/D = 1$ downstream of the vertical axis turbine.}
  \label{Fig:A19}
\end{figure}

In \prettyref{Fig:A19}, we present a qualitative comparison between the measured and computed normalized mean streamwise velocity and turbulence kinetic energy. The color maps of this figure are shown over the spanwise cross-planes at $x/D = 1$ downstream of the turbine.  As seen,  the comparison demonstrates that the experimental data and the LES-computed results are in reasonable qualitative agreement. 

\begin{figure} [H]
  \includegraphics[width=0.7\textwidth]{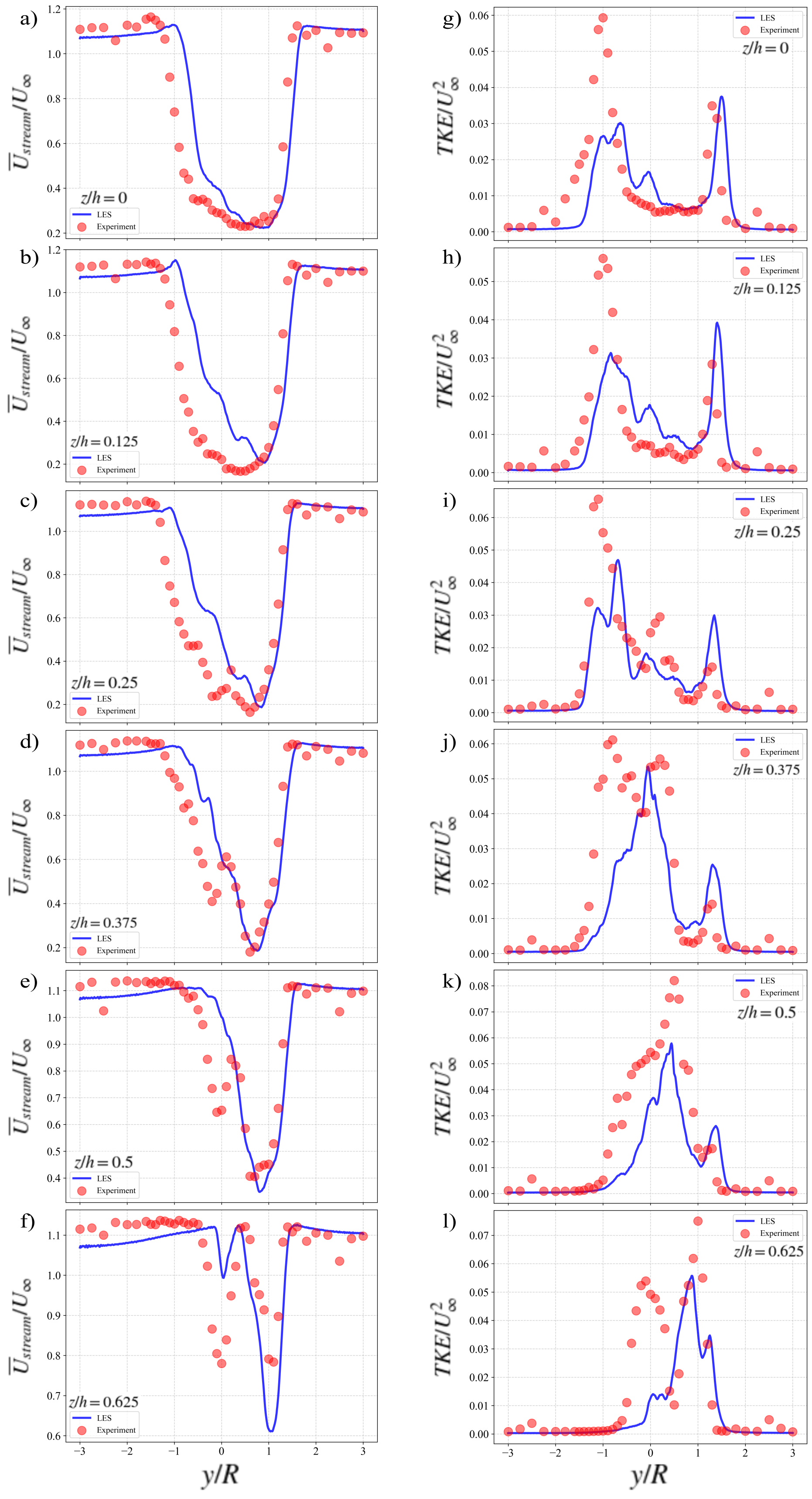}  
  \caption{Spanwise profiles of the normalized mean streamwise velocity (a-f) and turbulence kinetic energy (g-l) at different non-dimensional elevations of $z/H = 0, 0.125, 0.25, 0.375, 0.5, 0.625$. Elevations are normalized by the waterwater depth (\(h\)), and the spanwise length is normalized by the rotor diameter (\(R\)), following the experimental analysis in \cite{[130]}.}
  \label{Fig:A18}
\end{figure}

In \prettyref{Fig:A18}, we compare the wake flow field of the model against the measured mean streamwise velocity and turbulence kinetic energy. The spanwise profiles are taken at $x/D = 1$ downstream of the turbine at various water depth. The numerical results show reasonable agreement with the experimental data, accurately capturing the trends in wake deficit and asymmetry. However, discrepancies can be observed in the comparisons, which may be attributed to measurement uncertainties.

\section{Grid sensitivity analysis}
\noindent A grid independence study was performed to evaluate the sensitivity of the flow solver to the grid resolution and identify an optimal grid resolution that balances computational efficiency and accuracy. Three computational grid systems, denoted as grid A, B, and C, were tested. These grid systems are obtained by progressively refining the spatial resolution, as detailed in \prettyref{tab:B3}. The selected grid systems consist of 66 to 194 million computational nodes.

\begin{table}[h]
    \centering
    \caption{Details of the computational grid systems A, B, and C. The grid systems consist of $N_x, N_y$, and $N_z$ nodes in the streamwise, spanwise, and vertical directions, respectively. The spatial resolutions for the flow solver, normalized by the rotor diameter $D$, are denoted as $\Delta x$, $\Delta y$, and $\Delta z$. The minimum vertical grid spacing in wall units is expressed as $\Delta z^{+}$. The non-dimensional time step for the flow solver is $\Delta t_{*} = \Delta t U_{\infty}/{D}$, where $\Delta t$ is the dimensional time step.}
    \label{tab:grid_details}
    \begin{tabular}{lccc}
        \toprule
        Variable & Grid A & Grid B & Grid C \\
        \midrule
        Number of grid nodes & \(66 \times 10^6\) & \(107 \times 10^6\) & \(194 \times 10^6\) \\
        \(N_x, N_y, N_z\) & \(1561 \times 221 \times 193\) & \(1801 \times 261 \times 229\) & \(2253 \times 315 \times 277\) \\
        \(\Delta x, \Delta y, \Delta z\) & $0.012D$ & $0.01D$ & $0.008D$ \\
        \(\Delta t_{*}\) & $0.0005$ & $0.0005$ & $0.0005$ \\
        \(\Delta z^+\) & $730$ & $600$ & $475$ \\
        \bottomrule
    \end{tabular}
    \label{tab:B3}
\end{table}

\begin{figure} [H]
  \includegraphics[width=1\textwidth]{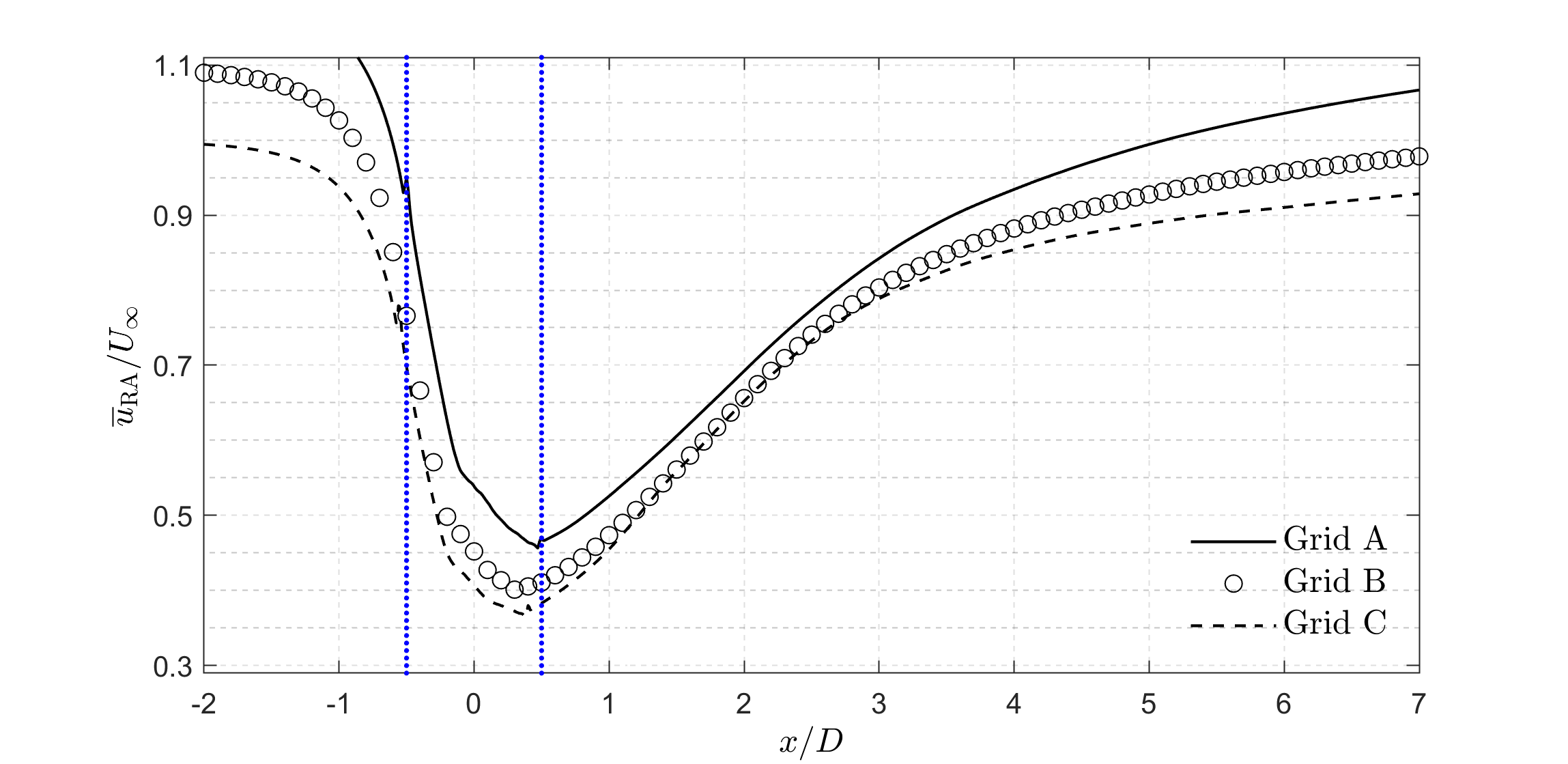}  
  \caption{Variations of normalized mean streamwise velocity averaged across the rotor swept area in the streamwise direction. Solid lines, hollow circles, and dashed lines correspond to grids A, B, and C, respectively. The blue vertical dashed lines indicate the rotor region.}
  \label{Fig:B20}
\end{figure}

\prettyref{Fig:B20} presents a comparison of the normalized mean streamwise velocity averaged over the rotor's swept area, \( u_{RA} \), for the rigid bed at TSR \(=2.0\) (i.e., case 2) using the three grid systems.  While relatively significant differences were observed between grids A and B upstream of the turbine, grids B and C exhibited better agreement, particularly in the far-field downstream of the turbine. Considering the balance between computational cost and accuracy, grid B was selected to perform the simulations of this study, as it significantly reduced the computational cost while maintaining accuracy compared to grid C.

\bibliographystyle{elsarticle-num-names} 
\bibliography{main.bib}

\begin{thebibliography}{127}
\expandafter\ifx\csname natexlab\endcsname\relax\def\natexlab#1{#1}\fi
\providecommand{\url}[1]{\texttt{#1}}
\providecommand{\href}[2]{#2}
\providecommand{\path}[1]{#1}
\providecommand{\DOIprefix}{doi:}
\providecommand{\ArXivprefix}{arXiv:}
\providecommand{\URLprefix}{URL: }
\providecommand{\Pubmedprefix}{pmid:}
\providecommand{\doi}[1]{\href{http://dx.doi.org/#1}{\path{#1}}}
\providecommand{\Pubmed}[1]{\href{pmid:#1}{\path{#1}}}
\providecommand{\bibinfo}[2]{#2}
\ifx\xfnm\relax \def\xfnm[#1]{\unskip,\space#1}\fi
\bibitem[{[1](2024)}]{[1]}
\bibinfo{title}{Renewables 2023}, \bibinfo{type}{Technical Report}, International Energy Agency, \bibinfo{year}{2024}. \URLprefix \url{https://www.iea.org/reports/renewables-2023}.
\bibitem[{[2](2024)}]{[2]}
\bibinfo{title}{{Renewables 2024 Global Status Report}}, \bibinfo{type}{Technical Report}, Renewable Energy Policy Network for the 21st Century, \bibinfo{year}{2024}. \URLprefix \url{https://www.ren21.net/reports/global-status-report/}.
\bibitem[{Wood(2024)}]{[3]}
\bibinfo{author}{J.~Wood}, \bibinfo{title}{{Massive expansion of renewable power opens door to achieving global tripling goal set at COP28}}, \bibinfo{howpublished}{https://www.iea.org/news/massive-expansion-of-renewable-power-opens-door-to-achieving-global-tripling-goal-set-at-cop28}, \bibinfo{year}{2024}.
\bibitem[{Khan et~al.(2009)Khan, Bhuyan, Iqbal, and Quaicoe}]{[4]}
\bibinfo{author}{M.~Khan}, \bibinfo{author}{G.~Bhuyan}, \bibinfo{author}{M.~Iqbal}, \bibinfo{author}{J.~Quaicoe},
\newblock \bibinfo{title}{Hydrokinetic energy conversion systems and assessment of horizontal and vertical axis turbines for river and tidal applications: A technology status review},
\newblock \bibinfo{journal}{Applied Energy} \bibinfo{volume}{86} (\bibinfo{year}{2009}) \bibinfo{pages}{1823–1835}. \DOIprefix\doi{10.1016/j.apenergy.2009.02.017}.
\bibitem[{Behrouzi et~al.(2016)Behrouzi, Nakisa, Maimun, and Ahmed}]{[5]}
\bibinfo{author}{F.~Behrouzi}, \bibinfo{author}{M.~Nakisa}, \bibinfo{author}{A.~Maimun}, \bibinfo{author}{Y.~M. Ahmed},
\newblock \bibinfo{title}{Global renewable energy and its potential in {Malaysia}: A review of hydrokinetic turbine technology},
\newblock \bibinfo{journal}{Renewable and Sustainable Energy Reviews} \bibinfo{volume}{62} (\bibinfo{year}{2016}) \bibinfo{pages}{1270--1281}. \DOIprefix\doi{https://doi.org/10.1016/j.rser.2016.05.020}.
\bibitem[{Kamal and Saini(2022)}]{[6]}
\bibinfo{author}{M.~M. Kamal}, \bibinfo{author}{R.~Saini},
\newblock \bibinfo{title}{A numerical investigation on the influence of savonius blade helicity on the performance characteristics of hybrid cross-flow hydrokinetic turbine},
\newblock \bibinfo{journal}{Renewable Energy} \bibinfo{volume}{190} (\bibinfo{year}{2022}) \bibinfo{pages}{788--804}. \DOIprefix\doi{https://doi.org/10.1016/j.renene.2022.03.155}.
\bibitem[{Neill et~al.(2021)Neill, Haas, Thiébot, and Yang}]{[7]}
\bibinfo{author}{S.~P. Neill}, \bibinfo{author}{K.~A. Haas}, \bibinfo{author}{J.~Thiébot}, \bibinfo{author}{Z.~Yang},
\newblock \bibinfo{title}{A review of tidal energy—resource, feedbacks, and environmental interactions},
\newblock \bibinfo{journal}{Journal of Renewable and Sustainable Energy} \bibinfo{volume}{13} (\bibinfo{year}{2021}). \DOIprefix\doi{10.1063/5.0069452}.
\bibitem[{Kilcher et~al.(2021)Kilcher, Fogarty, and Lawson}]{[8]}
\bibinfo{author}{L.~Kilcher}, \bibinfo{author}{M.~Fogarty}, \bibinfo{author}{M.~Lawson}, \bibinfo{title}{{Marine Energy in the United States: An Overview of Opportunities}}, \bibinfo{type}{Technical Report}, National Renewable Energy Laboratory, \bibinfo{year}{2021}. \URLprefix \url{https://www.nrel.gov/docs/fy21osti/78773.pdf}.
\bibitem[{Hussaina et~al.(2017)Hussaina, Arifb, and Aslamc}]{[9]}
\bibinfo{author}{A.~Hussaina}, \bibinfo{author}{S.~M. Arifb}, \bibinfo{author}{M.~Aslamc},
\newblock \bibinfo{title}{Emerging renewable and sustainable energy technologies: State of the art},
\newblock \bibinfo{journal}{Renewable and Sustainable Energy Reviews} \bibinfo{volume}{71} (\bibinfo{year}{2017}) \bibinfo{pages}{12--28}. \DOIprefix\doi{http://dx.doi.org/10.1016/j.rser.2016.12.033}.
\bibitem[{Jump et~al.(2020)Jump, Macleod, and Wills}]{[10]}
\bibinfo{author}{E.~Jump}, \bibinfo{author}{A.~Macleod}, \bibinfo{author}{T.~Wills},
\newblock \bibinfo{title}{Review of tidal turbine wake modelling methods—state of the art},
\newblock \bibinfo{journal}{International Marine Energy Journal} \bibinfo{volume}{3} (\bibinfo{year}{2020}). \DOIprefix\doi{https://doi.org/10.36688/imej.3.91‐100}.
\bibitem[{[11(2024)}]{[11]}
\bibinfo{title}{Hydropower explained}, \bibinfo{howpublished}{https://www.eia.gov/energyexplained/hydropower/tidal-power.php}, \bibinfo{year}{2024}.
\bibitem[{[12(2021)}]{[12]}
\bibinfo{title}{{Tidal Energy}}, \bibinfo{howpublished}{https://www.pnnl.gov/explainer-articles/tidal-energy}, \bibinfo{year}{2021}.
\bibitem[{Stansby and Ouro(2022)}]{[13]}
\bibinfo{author}{P.~Stansby}, \bibinfo{author}{P.~Ouro},
\newblock \bibinfo{title}{Modelling marine turbine arrays in tidal flows},
\newblock \bibinfo{journal}{Journal of Hydraulic Research} \bibinfo{volume}{60} (\bibinfo{year}{2022}) \bibinfo{pages}{1--18}. \DOIprefix\doi{10.1080/00221686.2021.2022032}.
\bibitem[{Fairley et~al.(2018)Fairley, Karunarathna, and Masters}]{[14]}
\bibinfo{author}{I.~Fairley}, \bibinfo{author}{H.~Karunarathna}, \bibinfo{author}{I.~Masters},
\newblock \bibinfo{title}{The influence of waves on morphodynamic impacts of energy extraction at a tidal stream turbine site in the {Pentland Firth}},
\newblock \bibinfo{journal}{Renewable Energy} \bibinfo{volume}{125} (\bibinfo{year}{2018}) \bibinfo{pages}{630--647}. \DOIprefix\doi{https://doi.org/10.1016/j.renene.2018.02.035}.
\bibitem[{Guillou et~al.(2019)Guillou, Thiébot, and Chapalain}]{[15]}
\bibinfo{author}{N.~Guillou}, \bibinfo{author}{J.~Thiébot}, \bibinfo{author}{G.~Chapalain},
\newblock \bibinfo{title}{Turbines’ effects on water renewal within a marine tidal stream energy site},
\newblock \bibinfo{journal}{Energy} \bibinfo{volume}{189} (\bibinfo{year}{2019}). \DOIprefix\doi{https://doi.org/10.1016/j.energy.2019.116113}.
\bibitem[{[16(2019)}]{[16]}
\bibinfo{title}{{Roosevelt Island Tidal Energy (RITE) Project Pilot}}, \bibinfo{howpublished}{https://tethys.pnnl.gov/project-sites/roosevelt-island-tidal-energy-rite-project-pilot}, \bibinfo{year}{2019}.
\bibitem[{[17(2024)}]{[17]}
\bibinfo{title}{{Cobscook Bay Tidal Energy Test Site}}, \bibinfo{howpublished}{https://tethys.pnnl.gov/project-sites/cobscook-bay-tidal-energy-project}, \bibinfo{year}{2024}.
\bibitem[{[18(2020)}]{[18]}
\bibinfo{title}{{A Tidal Hydrodynamic Model for Cook Inlet, Alaska, to Support Tidal Energy Resource Characterization}},
\newblock \bibinfo{journal}{Journal of Marine Science and Engineering} \bibinfo{volume}{8} (\bibinfo{year}{2020}). \DOIprefix\doi{https://doi.org/10.3390/jmse8040254}.
\bibitem[{Yang et~al.(2021)Yang, Wang, Branch, Xiao, and Deb}]{[19]}
\bibinfo{author}{Z.~Yang}, \bibinfo{author}{T.~Wang}, \bibinfo{author}{R.~Branch}, \bibinfo{author}{Z.~Xiao}, \bibinfo{author}{M.~Deb},
\newblock \bibinfo{title}{Tidal stream energy resource characterization in the {Salish Sea}},
\newblock \bibinfo{journal}{Renewable Energy} \bibinfo{volume}{172} (\bibinfo{year}{2021}) \bibinfo{pages}{188--208}. \DOIprefix\doi{https://doi.org/10.1016/j.renene.2021.03.028}.
\bibitem[{Chawdhary et~al.(2018)Chawdhary, Angelidis, Colby, Corren, Shen, and Sotiropoulos}]{[20]}
\bibinfo{author}{S.~Chawdhary}, \bibinfo{author}{D.~Angelidis}, \bibinfo{author}{J.~Colby}, \bibinfo{author}{D.~Corren}, \bibinfo{author}{L.~Shen}, \bibinfo{author}{F.~Sotiropoulos},
\newblock \bibinfo{title}{{Multiresolution Large-Eddy Simulation of an Array of Hydrokinetic Turbines in a Field-Scale River: The Roosevelt Island Tidal Energy Project in New York City}},
\newblock \bibinfo{journal}{Water Resources Research} \bibinfo{volume}{54} (\bibinfo{year}{2018}) \bibinfo{pages}{10188--10204}. \DOIprefix\doi{https://doi.org/10.1029/2018WR023345}.
\bibitem[{Saini and Saini(2019)}]{[21]}
\bibinfo{author}{G.~Saini}, \bibinfo{author}{R.~P. Saini},
\newblock \bibinfo{title}{A review on technology, configurations, and performance of cross‐flow hydrokinetic turbines},
\newblock \bibinfo{journal}{International Journal of Energy Research} \bibinfo{volume}{43} (\bibinfo{year}{2019}) \bibinfo{pages}{6639–6679}. \DOIprefix\doi{https://doi.org/10.1002/er.4625}.
\bibitem[{Ali et~al.(2015)Ali, Khan, Khalid, and Mehmood}]{[80]}
\bibinfo{author}{J.~Ali}, \bibinfo{author}{J.~Khan}, \bibinfo{author}{M.~S. Khalid}, \bibinfo{author}{N.~Mehmood},
\newblock \bibinfo{title}{Harnessing marine energy by horizontal axis marine turbines},
\newblock in: \bibinfo{booktitle}{{2015 12th International Bhurban Conference on Applied Sciences and Technology (IBCAST)}}, \bibinfo{publisher}{IEEE}, \bibinfo{address}{Islamabad, Pakistan}, \bibinfo{year}{2015}. \DOIprefix\doi{https://doi.org/10.1109/IBCAST.2015.7058548}.
\bibitem[{Uihlein and Magagna(2015)}]{[22]}
\bibinfo{author}{A.~Uihlein}, \bibinfo{author}{D.~Magagna}, \bibinfo{title}{2014 JRC Ocean Energy Status Report}, \bibinfo{type}{Technical Report}, European Commission, \bibinfo{year}{2015}. \URLprefix \url{https://op.europa.eu/en/publication-detail/-/publication/359b9147-ab4e-4639-b9db-17a6011a255f/language-en}. \DOIprefix\doi{10.2790/866387}.
\bibitem[{Massie et~al.(2019)Massie, Ouro, Stoesser, and Luo}]{[23]}
\bibinfo{author}{L.~Massie}, \bibinfo{author}{P.~Ouro}, \bibinfo{author}{T.~Stoesser}, \bibinfo{author}{Q.~Luo},
\newblock \bibinfo{title}{An actuator surface model to simulate vertical axis turbines},
\newblock \bibinfo{journal}{Energies} \bibinfo{volume}{12} (\bibinfo{year}{2019}). \URLprefix \url{https://www.mdpi.com/1996-1073/12/24/4741}. \DOIprefix\doi{10.3390/en12244741}.
\bibitem[{Le et~al.(2014)Le, Lee, Park, and Ko}]{[24]}
\bibinfo{author}{T.~Q. Le}, \bibinfo{author}{K.-S. Lee}, \bibinfo{author}{J.-S. Park}, \bibinfo{author}{J.~H. Ko},
\newblock \bibinfo{title}{Flow-driven rotor simulation of vertical axis tidal turbines: A comparison of helical and straight blades},
\newblock \bibinfo{journal}{International Journal of Naval Architecture and Ocean Engineering} \bibinfo{volume}{6} (\bibinfo{year}{2014}) \bibinfo{pages}{257--268}. \DOIprefix\doi{http://dx.doi.org/10.2478/IJNAOE-2013-0177}.
\bibitem[{Strom et~al.(2017)Strom, Brunton, and Polagye}]{[25]}
\bibinfo{author}{B.~Strom}, \bibinfo{author}{S.~L. Brunton}, \bibinfo{author}{B.~Polagye},
\newblock \bibinfo{title}{Intracycle angular velocity control of cross-flow turbines},
\newblock \bibinfo{journal}{Nature Energy} \bibinfo{volume}{2} (\bibinfo{year}{2017}). \DOIprefix\doi{https://doi.org/10.1038/nenergy.2017.103}.
\bibitem[{Nachtane et~al.(2020)Nachtane, M.Tarfaoui, Goda, and Rouway}]{[26]}
\bibinfo{author}{M.~Nachtane}, \bibinfo{author}{M.Tarfaoui}, \bibinfo{author}{I.~Goda}, \bibinfo{author}{M.~Rouway},
\newblock \bibinfo{title}{A review on the technologies, design considerations and numerical models of tidal current turbines},
\newblock \bibinfo{journal}{Renewable Energy} \bibinfo{volume}{157} (\bibinfo{year}{2020}) \bibinfo{pages}{1274--1288}. \DOIprefix\doi{https://doi.org/10.1016/j.renene.2020.04.155}.
\bibitem[{Ouro and Stoesser(2017)}]{[27]}
\bibinfo{author}{P.~Ouro}, \bibinfo{author}{T.~Stoesser},
\newblock \bibinfo{title}{An immersed boundary-based large-eddy simulation approach to predict the performance of vertical axis tidal turbines},
\newblock \bibinfo{journal}{Computers and Fluids} \bibinfo{volume}{152} (\bibinfo{year}{2017}) \bibinfo{pages}{74--87}. \DOIprefix\doi{http://dx.doi.org/10.1016/j.compfluid.2017.04.003}.
\bibitem[{Roberts et~al.(2016)Roberts, Thomas, Sewell, Khan, Balmain, and Gillman}]{[28]}
\bibinfo{author}{A.~Roberts}, \bibinfo{author}{B.~Thomas}, \bibinfo{author}{P.~Sewell}, \bibinfo{author}{Z.~Khan}, \bibinfo{author}{S.~Balmain}, \bibinfo{author}{J.~Gillman},
\newblock \bibinfo{title}{Current tidal power technologies and their suitability for applications in coastal and marine areas},
\newblock \bibinfo{journal}{Journal of Ocean Engineering and Marine Energy} \bibinfo{volume}{2} (\bibinfo{year}{2016}) \bibinfo{pages}{227--245}. \DOIprefix\doi{10.1007/s40722-016-0044-8}.
\bibitem[{Müller et~al.(2021)Müller, Muhawenimana, Wilson, and Ouro}]{[29]}
\bibinfo{author}{S.~Müller}, \bibinfo{author}{V.~Muhawenimana}, \bibinfo{author}{C.~A. Wilson}, \bibinfo{author}{P.~Ouro},
\newblock \bibinfo{title}{Experimental investigation of the wake characteristics behind twin vertical axis turbines},
\newblock \bibinfo{journal}{Energy Conversion and Ma} \bibinfo{volume}{247} (\bibinfo{year}{2021}). \DOIprefix\doi{https://doi.org/10.1016/j.enconman.2021.114768}.
\bibitem[{Müller et~al.(2023)Müller, Muhawenimana, Sonnino‐Sorisio, Wilson, Cable, and Ouro}]{[30]}
\bibinfo{author}{S.~Müller}, \bibinfo{author}{V.~Muhawenimana}, \bibinfo{author}{G.~Sonnino‐Sorisio}, \bibinfo{author}{C.~A. M.~E. Wilson}, \bibinfo{author}{J.~Cable}, \bibinfo{author}{P.~Ouro},
\newblock \bibinfo{title}{Fish response to the presence of hydrokinetic turbines as a sustainable energy solution},
\newblock \bibinfo{journal}{Scientific Reports} \bibinfo{volume}{13} (\bibinfo{year}{2023}). \DOIprefix\doi{https://doi.org/10.1038/s41598-023-33000-w}.
\bibitem[{[4b(2015)}]{[4b]}
\bibinfo{title}{Quadrennial Technology Review 2015}, \bibinfo{type}{Technical Report}, United States Department of Energy, \bibinfo{year}{2015}. \URLprefix \url{https://www.energy.gov/quadrennial-technology-review-2015}.
\bibitem[{[5b(2016)}]{[5b]}
\bibinfo{title}{Renewables 2016: Global Status Report}, \bibinfo{type}{Technical Report}, Renewable Energy Policy Network for the 21st Century, \bibinfo{year}{2016}. \URLprefix \url{https://www.ren21.net/gsr-2016/}.
\bibitem[{Bachant and Wosnik(2015)}]{[32]}
\bibinfo{author}{P.~Bachant}, \bibinfo{author}{M.~Wosnik},
\newblock \bibinfo{title}{Characterising the near-wake of a cross-flow turbine},
\newblock \bibinfo{journal}{Journal of Turbulence} \bibinfo{volume}{16} (\bibinfo{year}{2015}) \bibinfo{pages}{392--410}. \DOIprefix\doi{https://doi.org/10.1080/14685248.2014.1001852}.
\bibitem[{Saini and Saini(2019)}]{[34]}
\bibinfo{author}{G.~Saini}, \bibinfo{author}{R.~P. Saini},
\newblock \bibinfo{title}{A review on technology, configurations, and performance of cross‐flow hydrokinetic turbines},
\newblock \bibinfo{journal}{International Journal of Energy Research} \bibinfo{volume}{43} (\bibinfo{year}{2019}) \bibinfo{pages}{6639--6679}. \DOIprefix\doi{https://doi.org/10.1002/er.4625}.
\bibitem[{Brochier et~al.(2012)Brochier, Fraunie, Beguier, and Paraschivoiu}]{[35]}
\bibinfo{author}{G.~Brochier}, \bibinfo{author}{P.~Fraunie}, \bibinfo{author}{C.~Beguier}, \bibinfo{author}{I.~Paraschivoiu},
\newblock \bibinfo{title}{Water channel experiments of dynamic stall on darrieus wind turbine blades},
\newblock \bibinfo{journal}{Journal of Propulsion} \bibinfo{volume}{2} (\bibinfo{year}{2012}). \DOIprefix\doi{https://doi.org/10.2514/3.22927}.
\bibitem[{Kist(2012)}]{[36]}
\bibinfo{author}{S.~Kist}, \bibinfo{title}{{America’s First Ocean Energy Delivered to the Grid: ORPC Sells Tidal Power in Maine}}, \bibinfo{howpublished}{https://orpc.co/wp-content/uploads/2021/10/americas-first-ocean-energy-delivered-to-the-grid-orpc-sells-tidal-power-in-maine-sept.-13-2012-2012913155.pdf}, \bibinfo{year}{2012}.
\bibitem[{Bachant and Wosnik(2015)}]{[37]}
\bibinfo{author}{P.~Bachant}, \bibinfo{author}{M.~Wosnik},
\newblock \bibinfo{title}{Performance measurements of cylindrical- and spherical-helical cross-flow marine hydrokinetic turbines, with estimates of exergy efficiency},
\newblock \bibinfo{journal}{Renewable Energy} \bibinfo{volume}{74} (\bibinfo{year}{2015}) \bibinfo{pages}{318--325}. \DOIprefix\doi{http://dx.doi.org/10.1016/j.renene.2014.07.049}.
\bibitem[{Tescione et~al.(2014)Tescione, Ragni, He, {Simão Ferreira}, and {van Bussel}}]{[38]}
\bibinfo{author}{G.~Tescione}, \bibinfo{author}{D.~Ragni}, \bibinfo{author}{C.~He}, \bibinfo{author}{C.~{Simão Ferreira}}, \bibinfo{author}{G.~{van Bussel}},
\newblock \bibinfo{title}{Near wake flow analysis of a vertical axis wind turbine by stereoscopic particle image velocimetry},
\newblock \bibinfo{journal}{Renewable Energy} \bibinfo{volume}{70} (\bibinfo{year}{2014}) \bibinfo{pages}{47--61}. \DOIprefix\doi{https://doi.org/10.1016/j.renene.2014.02.042}, \bibinfo{note}{special issue on aerodynamics of offshore wind energy systems and wakes}.
\bibitem[{Araya et~al.(2017)Araya, Colonius, and Dabiri}]{[39]}
\bibinfo{author}{D.~B. Araya}, \bibinfo{author}{T.~Colonius}, \bibinfo{author}{J.~O. Dabiri},
\newblock \bibinfo{title}{Transition to bluff-body dynamics in the wake of vertical-axis wind turbines},
\newblock \bibinfo{journal}{Journal of Fluid Mechanics} \bibinfo{volume}{813} (\bibinfo{year}{2017}) \bibinfo{pages}{346--381}. \DOIprefix\doi{10.1017/jfm.2016.862}.
\bibitem[{Ouro et~al.(2019)Ouro, Runge, Luo, and Stoesser}]{[40]}
\bibinfo{author}{P.~Ouro}, \bibinfo{author}{S.~Runge}, \bibinfo{author}{Q.~Luo}, \bibinfo{author}{T.~Stoesser},
\newblock \bibinfo{title}{Three-dimensionality of the wake recovery behind a vertical axis turbine},
\newblock \bibinfo{journal}{Renewable Energy} \bibinfo{volume}{133} (\bibinfo{year}{2019}) \bibinfo{pages}{1066--1077}. \DOIprefix\doi{https://doi.org/10.1016/j.renene.2018.10.111}.
\bibitem[{Nguyen et~al.(2021)Nguyen, Balduzzi, and Goude}]{[41]}
\bibinfo{author}{M.~T. Nguyen}, \bibinfo{author}{F.~Balduzzi}, \bibinfo{author}{A.~Goude},
\newblock \bibinfo{title}{Effect of pitch angle on power and hydrodynamics of a vertical axis turbine},
\newblock \bibinfo{journal}{Ocean Engineering} \bibinfo{volume}{238} (\bibinfo{year}{2021}). \DOIprefix\doi{https://doi.org/10.1016/j.oceaneng.2021.109335}.
\bibitem[{Yang et~al.(2023)Yang, Gu, and Yeh}]{[42]}
\bibinfo{author}{M.-H. Yang}, \bibinfo{author}{Z.-T. Gu}, \bibinfo{author}{R.-H. Yeh},
\newblock \bibinfo{title}{Numerical and experimental analyses of the performance of a vertical axis turbine with controllable-blades for ocean current energy},
\newblock \bibinfo{journal}{Energy Conversion and Ma} \bibinfo{volume}{285} (\bibinfo{year}{2023}). \DOIprefix\doi{https://doi.org/10.1016/j.enconman.2023.117009}.
\bibitem[{Iida et~al.(2007)Iida, Kato, and Mizuno}]{[44]}
\bibinfo{author}{A.~Iida}, \bibinfo{author}{K.~Kato}, \bibinfo{author}{A.~Mizuno},
\newblock \bibinfo{title}{{Numerical Simulation of Unsteady Flow and Aerodynamic Performance of Vertical Axis Wind Turbines with LES}},
\newblock in: \bibinfo{booktitle}{16th Australasian Fluid Mechanics Conference}, \bibinfo{publisher}{Research Gate}, \bibinfo{address}{Crown Plaza, Gold Coast, Australia}, \bibinfo{year}{2007}.
\bibitem[{Li et~al.(2013)Li, Zhu, lin Xu, and Xiao}]{[45]}
\bibinfo{author}{C.~Li}, \bibinfo{author}{S.~Zhu}, \bibinfo{author}{Y.~lin Xu}, \bibinfo{author}{Y.~Xiao},
\newblock \bibinfo{title}{{2.5D} large eddy simulation of vertical axis wind turbine in consideration of high angle of attack flow},
\newblock \bibinfo{journal}{Renewable Energy} \bibinfo{volume}{51} (\bibinfo{year}{2013}) \bibinfo{pages}{317--330}. \DOIprefix\doi{http://dx.doi.org/10.1016/j.renene.2012.09.011}.
\bibitem[{Posa et~al.(2016)Posa, Parker, Leftwich, and Balaras}]{[46]}
\bibinfo{author}{A.~Posa}, \bibinfo{author}{C.~M. Parker}, \bibinfo{author}{M.~C. Leftwich}, \bibinfo{author}{E.~Balaras},
\newblock \bibinfo{title}{Wake structure of a single vertical axis wind turbine},
\newblock \bibinfo{journal}{International Journal of Heat and Fluid Flow} \bibinfo{volume}{61} (\bibinfo{year}{2016}) \bibinfo{pages}{75--84}. \DOIprefix\doi{https://doi.org/10.1016/j.ijheatfluidflow.2016.02.002}, \bibinfo{note}{sI\:TSFP9 special issue}.
\bibitem[{Shamsoddin and Porté-Agel(2016)}]{[47]}
\bibinfo{author}{S.~Shamsoddin}, \bibinfo{author}{F.~Porté-Agel},
\newblock \bibinfo{title}{{A Large-Eddy Simulation Study of Vertical Axis Wind Turbine Wakes in the Atmospheric Boundary Layer}},
\newblock \bibinfo{journal}{Energies} \bibinfo{volume}{9} (\bibinfo{year}{2016}). \DOIprefix\doi{10.3390/en9050366}.
\bibitem[{Abkar and Dabiri(2016)}]{[48]}
\bibinfo{author}{M.~Abkar}, \bibinfo{author}{J.~O. Dabiri},
\newblock \bibinfo{title}{Self-similarity and flow characteristics of vertical-axis wind turbine wakes: an {LES} study},
\newblock \bibinfo{journal}{Journal of Turbulence} \bibinfo{volume}{18} (\bibinfo{year}{2016}) \bibinfo{pages}{373--389}. \DOIprefix\doi{https://doi.org/10.1080/14685248.2017.1284327}.
\bibitem[{Elkhoury et~al.(2015)Elkhoury, Kiwata, and Aoun}]{[49]}
\bibinfo{author}{M.~Elkhoury}, \bibinfo{author}{T.~Kiwata}, \bibinfo{author}{E.~Aoun},
\newblock \bibinfo{title}{Experimental and numerical investigation of a three-dimensional vertical-axis wind turbine with variable-pitch},
\newblock \bibinfo{journal}{Journal of Wind Engineering and Industrial Aerodynamics} \bibinfo{volume}{139} (\bibinfo{year}{2015}) \bibinfo{pages}{111--123}. \DOIprefix\doi{https://doi.org/10.1016/j.jweia.2015.01.004}.
\bibitem[{Posa(2019)}]{[50]}
\bibinfo{author}{A.~Posa},
\newblock \bibinfo{title}{Wake characterization of coupled configurations of vertical axis wind turbines using {Large Eddy Simulation}},
\newblock \bibinfo{journal}{International Journal of Heat and Fluid Flow} \bibinfo{volume}{75} (\bibinfo{year}{2019}) \bibinfo{pages}{27--43}. \DOIprefix\doi{https://doi.org/10.1016/j.ijheatfluidflow.2018.11.008}.
\bibitem[{Posa(2020)}]{[51]}
\bibinfo{author}{A.~Posa},
\newblock \bibinfo{title}{Dependence of the wake recovery downstream of a {Vertical Axis Wind Turbine} on its dynamic solidity},
\newblock \bibinfo{journal}{Journal of Wind Engineering and Industrial Aerodynamics} \bibinfo{volume}{202} (\bibinfo{year}{2020}). \DOIprefix\doi{https://doi.org/10.1016/j.jweia.2020.104212}.
\bibitem[{Reddy et~al.(2022)Reddy, Bhosale, and Saini}]{[52]}
\bibinfo{author}{K.~B. Reddy}, \bibinfo{author}{A.~C. Bhosale}, \bibinfo{author}{R.~Saini},
\newblock \bibinfo{title}{Performance parameters of lift-based vertical axis hydrokinetic turbines - {A} review},
\newblock \bibinfo{journal}{Ocean Engineering} \bibinfo{volume}{266} (\bibinfo{year}{2022}). \DOIprefix\doi{https://doi.org/10.1016/j.oceaneng.2022.113089}.
\bibitem[{Harries et~al.(2016)Harries, Kwan, Brammer, and Falconer}]{[53]}
\bibinfo{author}{T.~Harries}, \bibinfo{author}{A.~Kwan}, \bibinfo{author}{J.~Brammer}, \bibinfo{author}{R.~Falconer},
\newblock \bibinfo{title}{Physical testing of performance characteristics of a novel drag-driven vertical axis tidal stream turbine; with comparisons to a conventional {Savonius}},
\newblock \bibinfo{journal}{International Journal of Marine Energy} \bibinfo{volume}{14} (\bibinfo{year}{2016}). \DOIprefix\doi{10.1016/j.ijome.2016.01.008}.
\bibitem[{Bhuyan and Biswas(2014)}]{[54]}
\bibinfo{author}{S.~Bhuyan}, \bibinfo{author}{A.~Biswas},
\newblock \bibinfo{title}{Investigations on self-starting and performance characteristics of simple h and hybrid h-savonius vertical axis wind rotors},
\newblock \bibinfo{journal}{Energy Conversion and Ma} \bibinfo{volume}{87} (\bibinfo{year}{2014}) \bibinfo{pages}{859--867}. \DOIprefix\doi{http://dx.doi.org/10.1016/j.enconman.2014.07.056}.
\bibitem[{Karakaya et~al.(2024)Karakaya, Bor, and Elçi}]{[55]}
\bibinfo{author}{D.~Karakaya}, \bibinfo{author}{A.~Bor}, \bibinfo{author}{S.~Elçi},
\newblock \bibinfo{title}{{Numerical Analysis of Three Vertical Axis Turbine Designs for Improved Water Energy Efficiency}},
\newblock \bibinfo{journal}{Energies} \bibinfo{volume}{17} (\bibinfo{year}{2024}). \DOIprefix\doi{https://doi.org/10.3390/en17061398}.
\bibitem[{Hill and Mirko~Musa(2016)}]{[56]}
\bibinfo{author}{C.~Hill}, \bibinfo{author}{M.~G. Mirko~Musa},
\newblock \bibinfo{title}{Interaction between instream axial flow hydrokinetic turbines and uni-directional flow bedforms},
\newblock \bibinfo{journal}{Renewable Energy} \bibinfo{volume}{86} (\bibinfo{year}{2016}) \bibinfo{pages}{409--421}. \DOIprefix\doi{https://doi.org/10.1016/j.renene.2015.08.019}.
\bibitem[{Shields et~al.(2011)Shields, Woolf, Grist, Kerr, Jackson, Harris, Bell, Beharie, Want, Osalusi, Gibb, and Side}]{[57]}
\bibinfo{author}{M.~A. Shields}, \bibinfo{author}{D.~K. Woolf}, \bibinfo{author}{E.~P. Grist}, \bibinfo{author}{S.~A. Kerr}, \bibinfo{author}{A.~Jackson}, \bibinfo{author}{R.~E. Harris}, \bibinfo{author}{M.~C. Bell}, \bibinfo{author}{R.~Beharie}, \bibinfo{author}{A.~Want}, \bibinfo{author}{E.~Osalusi}, \bibinfo{author}{S.~W. Gibb}, \bibinfo{author}{J.~Side},
\newblock \bibinfo{title}{Marine renewable energy: The ecological implications of altering the hydrodynamics of the marine environment},
\newblock \bibinfo{journal}{Ocean and Coastal Management} \bibinfo{volume}{54} (\bibinfo{year}{2011}) \bibinfo{pages}{2--9}. \DOIprefix\doi{https://doi.org/10.1016/j.ocecoaman.2010.10.036}.
\bibitem[{Jacobson et~al.(2012)Jacobson, Amaral, Castro-Santos, Giza, Haro, G.~E.~Hecker, and Perkins}]{[58]}
\bibinfo{author}{P.~T. Jacobson}, \bibinfo{author}{S.~V. Amaral}, \bibinfo{author}{T.~Castro-Santos}, \bibinfo{author}{D.~J. Giza}, \bibinfo{author}{A.~Haro}, \bibinfo{author}{B.~J.~M. G.~E.~Hecker}, \bibinfo{author}{N.~Perkins}, \bibinfo{title}{Environmental effects of hydrokinetic turbines on fish: Desktop and laboratory flume studies}, \bibinfo{type}{Technical Report}, Electric Power Research Institute, \bibinfo{year}{2012}.
\bibitem[{Lee and Oh(2018)}]{[59]}
\bibinfo{author}{J.~Lee}, \bibinfo{author}{J.~Oh},
\newblock \bibinfo{title}{A study on the characteristics of organic matter and nutrients released from sediments into agricultural reservoirs},
\newblock \bibinfo{journal}{Water} \bibinfo{volume}{10} (\bibinfo{year}{2018}).
\bibitem[{Hauer et~al.(2018)Hauer, Kail, Schmütz, and Sendzimir}]{[60]}
\bibinfo{author}{C.~Hauer}, \bibinfo{author}{J.~Kail}, \bibinfo{author}{C.~Schmütz}, \bibinfo{author}{J.~Sendzimir},
\newblock \bibinfo{title}{The role of sediment and sediment dynamics in the aquatic environment},
\newblock \bibinfo{journal}{Riverine Ecosystem Management}  (\bibinfo{year}{2018}) \bibinfo{pages}{123--145}.
\bibitem[{Ross et~al.(2021)Ross, Sottolichio, Huybrechts, and Brunet}]{[61]}
\bibinfo{author}{L.~Ross}, \bibinfo{author}{A.~Sottolichio}, \bibinfo{author}{N.~Huybrechts}, \bibinfo{author}{P.~Brunet},
\newblock \bibinfo{title}{Tidal turbines in the estuarine environment: From identifying optimal location to environmental impact.},
\newblock \bibinfo{journal}{Renewable Energy} \bibinfo{volume}{169} (\bibinfo{year}{2021}) \bibinfo{pages}{700--713}.
\bibitem[{Musa et~al.(2018)Musa, Heisel, and Guala}]{[62]}
\bibinfo{author}{M.~Musa}, \bibinfo{author}{M.~Heisel}, \bibinfo{author}{M.~Guala},
\newblock \bibinfo{title}{Predictive model for local scour downstream of hydrokinetic turbines in erodible channels},
\newblock \bibinfo{journal}{Phys. Rev. Fluids} \bibinfo{volume}{3} (\bibinfo{year}{2018}) \bibinfo{pages}{024606}. \DOIprefix\doi{10.1103/PhysRevFluids.3.024606}.
\bibitem[{Copping et~al.(2016)Copping, Sather, Hanna, Whiting, Zydlewski, Staines, Gill, Hutchison, O'Hagan, Simas, Bald, Sparling, Wood, and Madsen}]{[63]}
\bibinfo{author}{A.~Copping}, \bibinfo{author}{N.~Sather}, \bibinfo{author}{L.~Hanna}, \bibinfo{author}{J.~Whiting}, \bibinfo{author}{G.~Zydlewski}, \bibinfo{author}{G.~Staines}, \bibinfo{author}{A.~Gill}, \bibinfo{author}{I.~Hutchison}, \bibinfo{author}{A.~M. O'Hagan}, \bibinfo{author}{T.~Simas}, \bibinfo{author}{J.~Bald}, \bibinfo{author}{C.~Sparling}, \bibinfo{author}{J.~Wood}, \bibinfo{author}{E.~Madsen}, \bibinfo{title}{Annex IV 2016 State of the Science Report: Environmental Effects of Marine Renewable Energy Development Around the World}, \bibinfo{type}{Technical Report}, \bibinfo{year}{2016}. \URLprefix \url{http://tethys.pnnl.gov/publications/state-of-the-science-2016}.
\bibitem[{Yang et~al.(2017)Yang, Khosronejad, and Sotiropoulos}]{[64]}
\bibinfo{author}{X.~Yang}, \bibinfo{author}{A.~Khosronejad}, \bibinfo{author}{F.~Sotiropoulos},
\newblock \bibinfo{title}{Large-eddy simulation of a hydrokinetic turbine mounted on an erodible bed},
\newblock \bibinfo{journal}{Renewable Energy} \bibinfo{volume}{113} (\bibinfo{year}{2017}) \bibinfo{pages}{1419--1433}. \DOIprefix\doi{http://dx.doi.org/10.1016/j.renene.2017.07.007}.
\bibitem[{Deng et~al.(2024)Deng, Zhang, and Lin}]{[65]}
\bibinfo{author}{X.~Deng}, \bibinfo{author}{J.~Zhang}, \bibinfo{author}{X.~Lin},
\newblock \bibinfo{title}{Proposal of actuator line-immersed boundary coupling model for tidal stream turbine modeling with hydrodynamics upon scouring morphology},
\newblock \bibinfo{journal}{Energy} \bibinfo{volume}{292} (\bibinfo{year}{2024}) \bibinfo{pages}{130451}. \DOIprefix\doi{https://doi.org/10.1016/j.energy.2024.130451}.
\bibitem[{Azrulhishama et~al.(2018)Azrulhishama, Jamaluddinb, Azric, and Yusoff}]{[66]}
\bibinfo{author}{E.~A. Azrulhishama}, \bibinfo{author}{Z.~Z. Jamaluddinb}, \bibinfo{author}{M.~A. Azric}, \bibinfo{author}{S.~B.~M. Yusoff},
\newblock \bibinfo{title}{Potential evaluation of vertical axis hydrokinetic turbine implementation in equatorial river},
\newblock in: \bibinfo{booktitle}{International Conference on Energy, Electrical and Power Engineering}, \bibinfo{publisher}{Journal of Physics: Conference Series}, \bibinfo{year}{2018}. \DOIprefix\doi{10.1088/1742-6596/1072/1/012002}.
\bibitem[{Jeona et~al.(2024)Jeona, Kima, Kim, and Kang}]{[67]}
\bibinfo{author}{J.~Jeona}, \bibinfo{author}{Y.~Kima}, \bibinfo{author}{D.~Kim}, \bibinfo{author}{S.~Kang},
\newblock \bibinfo{title}{Flume experiments for flow around debris accumulation at a bridge},
\newblock \bibinfo{journal}{Journal of Civil Engineering} \bibinfo{volume}{28} (\bibinfo{year}{2024}) \bibinfo{pages}{1049--1061}. \DOIprefix\doi{10.1007/s12205-024-1442-4}.
\bibitem[{Aksen et~al.(2024)Aksen, Flora, Seyedzadeh, Anjiraki, and Khosronejad}]{[68]}
\bibinfo{author}{M.~M. Aksen}, \bibinfo{author}{K.~Flora}, \bibinfo{author}{H.~Seyedzadeh}, \bibinfo{author}{M.~G. Anjiraki}, \bibinfo{author}{A.~Khosronejad},
\newblock \bibinfo{title}{On the impact of debris accumulation on power production of marine hydrokinetic turbines: Insights gained via {LES}},
\newblock \bibinfo{journal}{Theoretical and Applied Mechanics Letters} \bibinfo{volume}{14} (\bibinfo{year}{2024}). \DOIprefix\doi{https://doi.org/10.1016/j.taml.2024.100524}.
\bibitem[{Aksen et~al.(2025)Aksen, Seyedzadeh, Anjiraki, Craig, Flora, Santoni, Sotiropoulos, and Khosronejad}]{[69]}
\bibinfo{author}{M.~M. Aksen}, \bibinfo{author}{H.~Seyedzadeh}, \bibinfo{author}{M.~G. Anjiraki}, \bibinfo{author}{J.~Craig}, \bibinfo{author}{K.~Flora}, \bibinfo{author}{C.~Santoni}, \bibinfo{author}{F.~Sotiropoulos}, \bibinfo{author}{A.~Khosronejad},
\newblock \bibinfo{title}{Large eddy simulation of a utility-scale horizontal axis turbine with woody debris accumulation under live bed conditions},
\newblock \bibinfo{journal}{Renewable Energy} \bibinfo{volume}{239} (\bibinfo{year}{2025}). \DOIprefix\doi{https://doi.org/10.1016/j.renene.2024.122110}.
\bibitem[{Cada et~al.(2007)Cada, Ahlgrimm, Bahleda, Bigford, Stavrakas, Hall, Moursund, and Sale}]{[71]}
\bibinfo{author}{G.~Cada}, \bibinfo{author}{J.~Ahlgrimm}, \bibinfo{author}{M.~Bahleda}, \bibinfo{author}{T.~Bigford}, \bibinfo{author}{S.~D. Stavrakas}, \bibinfo{author}{D.~Hall}, \bibinfo{author}{R.~Moursund}, \bibinfo{author}{M.~Sale},
\newblock \bibinfo{title}{{Potential Impacts of Hydrokinetic and Wave energy Conversion Technologies on Aquatic environments}},
\newblock \bibinfo{journal}{Fisheries} \bibinfo{volume}{32} (\bibinfo{year}{2007}) \bibinfo{pages}{174--181}. \DOIprefix\doi{https://doi.org/10.1577/1548-8446(2007)32[174:PIOHAW]2.0.CO;2}.
\bibitem[{Wang et~al.(2022)Wang, Tan, Chen, Fan, and Liu}]{[70]}
\bibinfo{author}{C.~Wang}, \bibinfo{author}{L.~Tan}, \bibinfo{author}{M.~Chen}, \bibinfo{author}{H.~Fan}, \bibinfo{author}{D.~Liu},
\newblock \bibinfo{title}{A review on synergy of cavitation and sediment erosion in hydraulic machinery},
\newblock \bibinfo{journal}{Frontiers in Energy Research}  (\bibinfo{year}{2022}). \DOIprefix\doi{https://doi.org/10.3389/fenrg.2022.1047984}.
\bibitem[{Lin et~al.(2019)Lin, Zhang, Wang, Zhang, Liu, and Zhang}]{[72]}
\bibinfo{author}{X.~Lin}, \bibinfo{author}{J.~Zhang}, \bibinfo{author}{R.~Wang}, \bibinfo{author}{J.~Zhang}, \bibinfo{author}{W.~Liu}, \bibinfo{author}{Y.~Zhang},
\newblock \bibinfo{title}{Scour around a mono-pile foundation of a horizontal axis tidal stream turbine under steady current},
\newblock \bibinfo{journal}{oe} \bibinfo{volume}{192} (\bibinfo{year}{2019}). \DOIprefix\doi{10.1016/j.oceaneng.2019.106571}.
\bibitem[{Hill et~al.(2016)Hill, Kozarek, Sotiropoulos, and Guala}]{[73]}
\bibinfo{author}{C.~Hill}, \bibinfo{author}{J.~Kozarek}, \bibinfo{author}{F.~Sotiropoulos}, \bibinfo{author}{M.~Guala},
\newblock \bibinfo{title}{Hydrodynamics and sediment transport in a meandering channel with a model axial-flow hydrokinetic turbine},
\newblock \bibinfo{journal}{Water Resources Research}  (\bibinfo{year}{2016}) \bibinfo{pages}{860--879}. \DOIprefix\doi{https://doi.org/10.1002/2015WR017949}.
\bibitem[{Chen et~al.(2017)Chen, Hashim, Othman, and Motamedi}]{[74]}
\bibinfo{author}{L.~Chen}, \bibinfo{author}{R.~Hashim}, \bibinfo{author}{F.~Othman}, \bibinfo{author}{S.~Motamedi},
\newblock \bibinfo{title}{Experimental study on scour profile of pile-supported horizontal axis tidal current turbine},
\newblock \bibinfo{journal}{Renewable Energy} \bibinfo{volume}{114} (\bibinfo{year}{2017}) \bibinfo{pages}{744--754}. \DOIprefix\doi{https://doi.org/10.1016/j.renene.2017.07.026}.
\bibitem[{Ramírez-Mendoza et~al.(2018)Ramírez-Mendoza, Amoudry, Thorne, Cooke, McLelland, Jordan, Simmons, Parsons, and Murdoch}]{[75]}
\bibinfo{author}{R.~Ramírez-Mendoza}, \bibinfo{author}{L.~Amoudry}, \bibinfo{author}{P.~Thorne}, \bibinfo{author}{R.~Cooke}, \bibinfo{author}{S.~McLelland}, \bibinfo{author}{L.~Jordan}, \bibinfo{author}{S.~Simmons}, \bibinfo{author}{D.~Parsons}, \bibinfo{author}{L.~Murdoch},
\newblock \bibinfo{title}{Laboratory study on the effects of hydro kinetic turbines on hydrodynamics and sediment dynamics},
\newblock \bibinfo{journal}{Renewable Energy} \bibinfo{volume}{129} (\bibinfo{year}{2018}) \bibinfo{pages}{271--284}. \DOIprefix\doi{https://doi.org/10.1016/j.renene.2018.05.094}.
\bibitem[{Musa et~al.(2019)Musa, Hill, and Guala}]{[76]}
\bibinfo{author}{M.~Musa}, \bibinfo{author}{C.~Hill}, \bibinfo{author}{M.~Guala},
\newblock \bibinfo{title}{Interaction between hydrokinetic turbine wakes and sediment dynamics: array performance and geomorphic effects under different siting strategies and sediment transport conditions},
\newblock \bibinfo{journal}{Renewable Energy} \bibinfo{volume}{138} (\bibinfo{year}{2019}) \bibinfo{pages}{738--753}. \DOIprefix\doi{https://doi.org/10.1016/j.renene.2019.02.009}.
\bibitem[{Vybulkova(2013)}]{[77]}
\bibinfo{author}{L.~Vybulkova}, \bibinfo{title}{A study of the wake of an isolated tidal turbine with application to its effects on local sediment transport}, Ph.D. thesis, University of Glasgow, \bibinfo{year}{2013}.
\bibitem[{Lee et~al.(2019)Lee, Musa, Feist, Gao, Shen, and Guala}]{[78]}
\bibinfo{author}{J.~Lee}, \bibinfo{author}{M.~Musa}, \bibinfo{author}{C.~Feist}, \bibinfo{author}{J.~Gao}, \bibinfo{author}{L.~Shen}, \bibinfo{author}{M.~Guala},
\newblock \bibinfo{title}{{Wake Characteristics and Power Performance of a Drag-Driven in-Bank Vertical Axis Hydrokinetic Turbine}},
\newblock \bibinfo{journal}{12}  (\bibinfo{year}{2019}). \DOIprefix\doi{https://doi.org/10.3390/en12193611}.
\bibitem[{Gao et~al.(2022)Gao, Liu, Lee, Zheng, Guala, and Shen}]{[79]}
\bibinfo{author}{J.~Gao}, \bibinfo{author}{H.~Liu}, \bibinfo{author}{J.~Lee}, \bibinfo{author}{Y.~Zheng}, \bibinfo{author}{M.~Guala}, \bibinfo{author}{L.~Shen},
\newblock \bibinfo{title}{{Large-eddy simulation and Co-Design strategy for a drag-type vertical axis hydrokinetic turbine in open channel flows}},
\newblock \bibinfo{journal}{Renewable Energy} \bibinfo{volume}{181} (\bibinfo{year}{2022}) \bibinfo{pages}{1305--1316}. \DOIprefix\doi{https://doi.org/10.1016/j.renene.2021.09.119}.
\bibitem[{Khosronejad et~al.(2023)Khosronejad, Limaye, Zhang, Kang, Yang, and Sotiropoulos}]{[107]}
\bibinfo{author}{A.~Khosronejad}, \bibinfo{author}{A.~B. Limaye}, \bibinfo{author}{Z.~Zhang}, \bibinfo{author}{S.~Kang}, \bibinfo{author}{X.~Yang}, \bibinfo{author}{F.~Sotiropoulos},
\newblock \bibinfo{title}{{On the Morphodynamics of a Wide Class of Large-Scale Meandering Rivers: Insights Gained by Coupling LES With Sediment-Dynamics}},
\newblock \bibinfo{journal}{Journal of Advances in Modeling Earth Systems} \bibinfo{volume}{15} (\bibinfo{year}{2023}). \DOIprefix\doi{https://doi.org/10.1029/2022MS003257}.
\bibitem[{Seyedzadeh et~al.(2023)Seyedzadeh, Oaks, Craig, Aksen, Sanz, and Khosronejad}]{[147]}
\bibinfo{author}{H.~Seyedzadeh}, \bibinfo{author}{W.~Oaks}, \bibinfo{author}{J.~Craig}, \bibinfo{author}{M.~Aksen}, \bibinfo{author}{M.~S. Sanz}, \bibinfo{author}{A.~Khosronejad},
\newblock \bibinfo{title}{Lagrangian dynamics of particle transport in oral and nasal breathing},
\newblock \bibinfo{journal}{Physics of Fluids} \bibinfo{volume}{35} (\bibinfo{year}{2023}).
\bibitem[{Kang et~al.(2012)Kang, Borazjani, Colby, and Sotiropoulos}]{[143]}
\bibinfo{author}{S.~Kang}, \bibinfo{author}{I.~Borazjani}, \bibinfo{author}{J.~A. Colby}, \bibinfo{author}{F.~Sotiropoulos},
\newblock \bibinfo{title}{Numerical simulation of 3d flow past a real-life marine hydrokinetic turbine},
\newblock \bibinfo{journal}{Advances in Water Resources} \bibinfo{volume}{39} (\bibinfo{year}{2012}) \bibinfo{pages}{33--43}. \URLprefix \url{https://www.sciencedirect.com/science/article/pii/S0309170811002533}. \DOIprefix\doi{https://doi.org/10.1016/j.advwatres.2011.12.012}.
\bibitem[{Khosronejad and Sotiropoulos(2014)}]{[101]}
\bibinfo{author}{A.~Khosronejad}, \bibinfo{author}{F.~Sotiropoulos},
\newblock \bibinfo{title}{Numerical simulation of sand waves in a turbulent open channel flow},
\newblock \bibinfo{journal}{Journal of Fluid Mechanics} \bibinfo{volume}{753} (\bibinfo{year}{2014}) \bibinfo{pages}{150–216}. \DOIprefix\doi{10.1017/jfm.2014.335}.
\bibitem[{Khosronejad and Sotiropoulos(2017)}]{[113]}
\bibinfo{author}{A.~Khosronejad}, \bibinfo{author}{F.~Sotiropoulos},
\newblock \bibinfo{title}{On the genesis and evolution of barchan dunes: morphodynamics},
\newblock \bibinfo{journal}{Journal of Fluid Mechanics} \bibinfo{volume}{815} (\bibinfo{year}{2017}) \bibinfo{pages}{117–148}. \DOIprefix\doi{10.1017/jfm.2016.880}.
\bibitem[{Khosronejad et~al.(2014)Khosronejad, Kozarek, and Sotiropoulos}]{[128]}
\bibinfo{author}{A.~Khosronejad}, \bibinfo{author}{J.~L. Kozarek}, \bibinfo{author}{F.~Sotiropoulos},
\newblock \bibinfo{title}{{Simulation-Based Approach for Stream Restoration Structure Design: Model Development and Validation}},
\newblock \bibinfo{journal}{Journal of Hydraulic Engineering} \bibinfo{volume}{140} (\bibinfo{year}{2014}) \bibinfo{pages}{04014042}. \DOIprefix\doi{10.1061/(ASCE)HY.1943-7900.0000904}.
\bibitem[{Zhang et~al.(2024)Zhang, Anjiraki, Seyedzadeh, Sotiropoulos, and Khosronejad}]{[135]}
\bibinfo{author}{Z.~Zhang}, \bibinfo{author}{M.~G. Anjiraki}, \bibinfo{author}{H.~Seyedzadeh}, \bibinfo{author}{F.~Sotiropoulos}, \bibinfo{author}{A.~Khosronejad}, \bibinfo{title}{Toward ultra-efficient high-fidelity prediction of bed morphodynamics of large-scale meandering rivers using a novel les-trained machine learning approach}, \bibinfo{year}{2024}. \URLprefix \url{https://arxiv.org/abs/2407.18359}. \href{http://arxiv.org/abs/2407.18359}{{\tt arXiv:2407.18359}}.
\bibitem[{Flora and Khosronejad(2024)}]{[100]}
\bibinfo{author}{K.~Flora}, \bibinfo{author}{A.~Khosronejad},
\newblock \bibinfo{title}{{Uncertainty quantification of bank vegetation impacts on the flood flow field in the American River, California, using large-eddy simulations}},
\newblock \bibinfo{journal}{Earth Surface Processes and Landforms} \bibinfo{volume}{49} (\bibinfo{year}{2024}) \bibinfo{pages}{967--979}. \DOIprefix\doi{https://doi.org/10.1002/esp.5745}.
\bibitem[{Germano et~al.(1991)Germano, Piomelli, Moin, and Cabot}]{[103]}
\bibinfo{author}{M.~Germano}, \bibinfo{author}{U.~Piomelli}, \bibinfo{author}{P.~Moin}, \bibinfo{author}{W.~H. Cabot},
\newblock \bibinfo{title}{{A dynamic subgrid‐scale eddy viscosity model}},
\newblock \bibinfo{journal}{Physics of Fluids A: Fluid Dynamics} \bibinfo{volume}{3} (\bibinfo{year}{1991}) \bibinfo{pages}{1760--1765}. \DOIprefix\doi{10.1063/1.857955}.
\bibitem[{Smagorinsky(1963)}]{[104]}
\bibinfo{author}{J.~Smagorinsky},
\newblock \bibinfo{title}{General circulation experiments with the primitive equations: I. the basic experiment},
\newblock \bibinfo{journal}{Monthly Weather Review} \bibinfo{volume}{91} (\bibinfo{year}{1963}) \bibinfo{pages}{99 -- 164}. \DOIprefix\doi{10.1175/1520-0493(1963)091<0099:GCEWTP>2.3.CO;2}.
\bibitem[{Yang and Sotiropoulos(2018)}]{[136]}
\bibinfo{author}{X.~Yang}, \bibinfo{author}{F.~Sotiropoulos},
\newblock \bibinfo{title}{A new class of actuator surface models for wind turbines},
\newblock \bibinfo{journal}{Wind Energy} \bibinfo{volume}{21} (\bibinfo{year}{2018}) \bibinfo{pages}{285--302}. \URLprefix \url{https://onlinelibrary.wiley.com/doi/abs/10.1002/we.2162}. \DOIprefix\doi{https://doi.org/10.1002/we.2162}. \href{http://arxiv.org/abs/https://onlinelibrary.wiley.com/doi/pdf/10.1002/we.2162}{{\tt arXiv:https://onlinelibrary.wiley.com/doi/pdf/10.1002/we.2162}}.
\bibitem[{Yang et~al.(2015)Yang, Sotiropoulos, Conzemius, Wachtler, and Strong}]{[137]}
\bibinfo{author}{X.~Yang}, \bibinfo{author}{F.~Sotiropoulos}, \bibinfo{author}{R.~J. Conzemius}, \bibinfo{author}{J.~N. Wachtler}, \bibinfo{author}{M.~B. Strong},
\newblock \bibinfo{title}{Large-eddy simulation of turbulent flow past wind turbines/farms: the virtual wind simulator (vwis)},
\newblock \bibinfo{journal}{Wind Energy} \bibinfo{volume}{18} (\bibinfo{year}{2015}) \bibinfo{pages}{2025--2045}. \URLprefix \url{https://onlinelibrary.wiley.com/doi/abs/10.1002/we.1802}. \DOIprefix\doi{https://doi.org/10.1002/we.1802}. \href{http://arxiv.org/abs/https://onlinelibrary.wiley.com/doi/pdf/10.1002/we.1802}{{\tt arXiv:https://onlinelibrary.wiley.com/doi/pdf/10.1002/we.1802}}.
\bibitem[{Yang et~al.(2012)Yang, Kang, and Sotiropoulos}]{[138]}
\bibinfo{author}{X.~Yang}, \bibinfo{author}{S.~Kang}, \bibinfo{author}{F.~Sotiropoulos},
\newblock \bibinfo{title}{Computational study and modeling of turbine spacing effects in infinite aligned wind farms},
\newblock \bibinfo{journal}{Physics of Fluids} \bibinfo{volume}{24} (\bibinfo{year}{2012}) \bibinfo{pages}{115107}. \URLprefix \url{https://doi.org/10.1063/1.4767727}. \DOIprefix\doi{10.1063/1.4767727}. \href{http://arxiv.org/abs/https://pubs.aip.org/aip/pof/article-pdf/doi/10.1063/1.4767727/14131050/115107\_1\_online.pdf}{{\tt arXiv:https://pubs.aip.org/aip/pof/article-pdf/doi/10.1063/1.4767727/14131050/115107\_1\_online.pdf}}.
\bibitem[{Sørensen et~al.(2002)Sørensen, Michelsen, and Schreck}]{[139]}
\bibinfo{author}{N.~N. Sørensen}, \bibinfo{author}{J.~A. Michelsen}, \bibinfo{author}{S.~Schreck},
\newblock \bibinfo{title}{Navier–stokes predictions of the nrel phase vi rotor in the nasa ames 80 ft × 120 ft wind tunnel},
\newblock \bibinfo{journal}{Wind Energy} \bibinfo{volume}{5} (\bibinfo{year}{2002}) \bibinfo{pages}{151--169}. \URLprefix \url{https://onlinelibrary.wiley.com/doi/abs/10.1002/we.64}. \DOIprefix\doi{https://doi.org/10.1002/we.64}. \href{http://arxiv.org/abs/https://onlinelibrary.wiley.com/doi/pdf/10.1002/we.64}{{\tt arXiv:https://onlinelibrary.wiley.com/doi/pdf/10.1002/we.64}}.
\bibitem[{Johansen et~al.(2002)Johansen, Sørensen, Michelsen, and Schreck}]{[140]}
\bibinfo{author}{J.~Johansen}, \bibinfo{author}{N.~N. Sørensen}, \bibinfo{author}{J.~A. Michelsen}, \bibinfo{author}{S.~Schreck},
\newblock \bibinfo{title}{Detached-eddy simulation of flow around the nrel phase vi blade},
\newblock \bibinfo{journal}{Wind Energy} \bibinfo{volume}{5} (\bibinfo{year}{2002}) \bibinfo{pages}{185--197}. \URLprefix \url{https://onlinelibrary.wiley.com/doi/abs/10.1002/we.63}. \DOIprefix\doi{https://doi.org/10.1002/we.63}. \href{http://arxiv.org/abs/https://onlinelibrary.wiley.com/doi/pdf/10.1002/we.63}{{\tt arXiv:https://onlinelibrary.wiley.com/doi/pdf/10.1002/we.63}}.
\bibitem[{Sezer~Uzol and Long(2006)}]{[141]}
\bibinfo{author}{N.~Sezer~Uzol}, \bibinfo{author}{L.~Long},
\newblock \bibinfo{title}{3-d time-accurate cfd simulations of wind turbine rotor flow fields},
\newblock \bibinfo{journal}{American Institute of Aeronautics and Astronautics} \bibinfo{volume}{394} (\bibinfo{year}{2006}). \DOIprefix\doi{10.2514/6.2006-394}.
\bibitem[{Zahle et~al.(2009)Zahle, Sørensen, and Johansen}]{[142]}
\bibinfo{author}{F.~Zahle}, \bibinfo{author}{N.~N. Sørensen}, \bibinfo{author}{J.~Johansen},
\newblock \bibinfo{title}{Wind turbine rotor-tower interaction using an incompressible overset grid method},
\newblock \bibinfo{journal}{Wind Energy} \bibinfo{volume}{12} (\bibinfo{year}{2009}) \bibinfo{pages}{594--619}. \URLprefix \url{https://onlinelibrary.wiley.com/doi/abs/10.1002/we.327}. \DOIprefix\doi{https://doi.org/10.1002/we.327}. \href{http://arxiv.org/abs/https://onlinelibrary.wiley.com/doi/pdf/10.1002/we.327}{{\tt arXiv:https://onlinelibrary.wiley.com/doi/pdf/10.1002/we.327}}.
\bibitem[{Gilmanov and Sotiropoulos(2005)}]{[144]}
\bibinfo{author}{A.~Gilmanov}, \bibinfo{author}{F.~Sotiropoulos},
\newblock \bibinfo{title}{A hybrid cartesian/immersed boundary method for simulating flows with 3d, geometrically complex, moving bodies},
\newblock \bibinfo{journal}{Journal of Computational Physics} \bibinfo{volume}{207} (\bibinfo{year}{2005}) \bibinfo{pages}{457--492}. \URLprefix \url{https://www.sciencedirect.com/science/article/pii/S0021999105000379}. \DOIprefix\doi{https://doi.org/10.1016/j.jcp.2005.01.020}.
\bibitem[{Ge and Sotiropoulos(2007)}]{[145]}
\bibinfo{author}{L.~Ge}, \bibinfo{author}{F.~Sotiropoulos},
\newblock \bibinfo{title}{A numerical method for solving the 3d unsteady incompressible navier–stokes equations in curvilinear domains with complex immersed boundaries},
\newblock \bibinfo{journal}{Journal of Computational Physics} \bibinfo{volume}{225} (\bibinfo{year}{2007}) \bibinfo{pages}{1782--1809}. \URLprefix \url{https://www.sciencedirect.com/science/article/pii/S0021999107000873}. \DOIprefix\doi{https://doi.org/10.1016/j.jcp.2007.02.017}.
\bibitem[{Borazjani et~al.(2008)Borazjani, Ge, and Sotiropoulos}]{[146]}
\bibinfo{author}{I.~Borazjani}, \bibinfo{author}{L.~Ge}, \bibinfo{author}{F.~Sotiropoulos},
\newblock \bibinfo{title}{Curvilinear immersed boundary method for simulating fluid structure interaction with complex 3d rigid bodies},
\newblock \bibinfo{journal}{Journal of Computational Physics} \bibinfo{volume}{227} (\bibinfo{year}{2008}) \bibinfo{pages}{7587--7620}. \URLprefix \url{https://www.sciencedirect.com/science/article/pii/S0021999108002490}. \DOIprefix\doi{https://doi.org/10.1016/j.jcp.2008.04.028}.
\bibitem[{Kang et~al.(2011)Kang, Lightbody, Hill, and Sotiropoulos}]{[102]}
\bibinfo{author}{S.~Kang}, \bibinfo{author}{A.~Lightbody}, \bibinfo{author}{C.~Hill}, \bibinfo{author}{F.~Sotiropoulos},
\newblock \bibinfo{title}{High-resolution numerical simulation of turbulence in natural waterways},
\newblock \bibinfo{journal}{Advances in Water Resources} \bibinfo{volume}{34} (\bibinfo{year}{2011}) \bibinfo{pages}{98--113}. \DOIprefix\doi{https://doi.org/10.1016/j.advwatres.2010.09.018}.
\bibitem[{Khosronejad et~al.(2011)Khosronejad, Kang, Borazjani, and Sotiropoulos}]{[112]}
\bibinfo{author}{A.~Khosronejad}, \bibinfo{author}{S.~Kang}, \bibinfo{author}{I.~Borazjani}, \bibinfo{author}{F.~Sotiropoulos},
\newblock \bibinfo{title}{Curvilinear immersed boundary method for simulating coupled flow and bed morphodynamic interactions due to sediment transport phenomena},
\newblock \bibinfo{journal}{Advances in Water Resources} \bibinfo{volume}{34} (\bibinfo{year}{2011}) \bibinfo{pages}{829--843}. \DOIprefix\doi{https://doi.org/10.1016/j.advwatres.2011.02.017}.
\bibitem[{Van~Rijn(1993)}]{[105]}
\bibinfo{author}{L.~C. Van~Rijn}, \bibinfo{title}{Principles of sediment transport in rivers, estuaries, and coastal seas}, \bibinfo{publisher}{Aqua Publications}, \bibinfo{year}{1993}.
\bibitem[{Paola and Voller(2005)}]{[106]}
\bibinfo{author}{C.~Paola}, \bibinfo{author}{V.~R. Voller},
\newblock \bibinfo{title}{A generalized exner equation for sediment mass balance},
\newblock \bibinfo{journal}{Journal of Geophysical Research: Earth Surface} \bibinfo{volume}{110} (\bibinfo{year}{2005}). \DOIprefix\doi{https://doi.org/10.1029/2004JF000274}.
\bibitem[{Borazjani et~al.(2008)Borazjani, Ge, and Sotiropoulos}]{[108]}
\bibinfo{author}{I.~Borazjani}, \bibinfo{author}{L.~Ge}, \bibinfo{author}{F.~Sotiropoulos},
\newblock \bibinfo{title}{Curvilinear immersed boundary method for simulating fluid structure interaction with complex {3D} rigid bodies},
\newblock \bibinfo{journal}{Journal of Computational Physics} \bibinfo{volume}{227} (\bibinfo{year}{2008}) \bibinfo{pages}{7587--7620}. \DOIprefix\doi{https://doi.org/10.1016/j.jcp.2008.04.028}.
\bibitem[{Soulsby and Whitehouse(1997)}]{[109]}
\bibinfo{author}{R.~L. Soulsby}, \bibinfo{author}{R.~J. Whitehouse}, \bibinfo{title}{{Threshold of Sediment Motion in Coastal Environments}}, \bibinfo{publisher}{Centre for Advanced Engineering, University of Canterbury}, \bibinfo{address}{Christchurch, N.Z.}, \bibinfo{year}{1997}. \URLprefix \url{https://search.informit.org/doi/10.3316/informit.929741720399033}.
\bibitem[{Chou and Fringer(2010)}]{[110]}
\bibinfo{author}{Y.-J. Chou}, \bibinfo{author}{O.~B. Fringer},
\newblock \bibinfo{title}{A model for the simulation of coupled flow-bed form evolution in turbulent flows},
\newblock \bibinfo{journal}{Journal of Geophysical Research: Oceans} \bibinfo{volume}{115} (\bibinfo{year}{2010}). \URLprefix \url{https://agupubs.onlinelibrary.wiley.com/doi/abs/10.1029/2010JC006103}. \DOIprefix\doi{https://doi.org/10.1029/2010JC006103}. \href{http://arxiv.org/abs/https://agupubs.onlinelibrary.wiley.com/doi/pdf/10.1029/2010JC006103}{{\tt arXiv:https://agupubs.onlinelibrary.wiley.com/doi/pdf/10.1029/2010JC006103}}.
\bibitem[{Khosronejad et~al.(2007)Khosronejad, Rennie, Neyshabouri, and Townsend}]{[111]}
\bibinfo{author}{A.~Khosronejad}, \bibinfo{author}{C.~D. Rennie}, \bibinfo{author}{S.~A. A.~S. Neyshabouri}, \bibinfo{author}{R.~D. Townsend},
\newblock \bibinfo{title}{{3D Numerical Modeling of Flow and Sediment Transport in Laboratory Channel Bends}},
\newblock \bibinfo{journal}{Journal of Hydraulic Engineering} \bibinfo{volume}{133} (\bibinfo{year}{2007}) \bibinfo{pages}{1123--1134}. \DOIprefix\doi{10.1061/(ASCE)0733-9429(2007)133:10(1123)}.
\bibitem[{Valentine(2019)}]{[134]}
\bibinfo{author}{P.~C. Valentine}, \bibinfo{title}{Sediment classification and the characterization, identification, and mapping of geologic substrates for the glaciated Gulf of Maine seabed and other terrains, providing a physical framework for ecological research and seabed management}, \bibinfo{type}{Scientific Investigations Report} \bibinfo{number}{2019-5073}, U.S. Geological Survey, \bibinfo{address}{Reston, VA}, \bibinfo{year}{2019}. \URLprefix \url{https://pubs.usgs.gov/publication/sir20195073}. \DOIprefix\doi{10.3133/sir20195073}.
\bibitem[{Fujisawa and Shibuya(2001)}]{[120]}
\bibinfo{author}{N.~Fujisawa}, \bibinfo{author}{S.~Shibuya},
\newblock \bibinfo{title}{Observations of dynamic stall on darrieus wind turbine blades},
\newblock \bibinfo{journal}{Journal of Wind Engineering and Industrial Aerodynamics} \bibinfo{volume}{89} (\bibinfo{year}{2001}) \bibinfo{pages}{201--214}. \DOIprefix\doi{https://doi.org/10.1016/S0167-6105(00)00062-3}.
\bibitem[{Posa and Balaras(2018)}]{[116]}
\bibinfo{author}{A.~Posa}, \bibinfo{author}{E.~Balaras},
\newblock \bibinfo{title}{{Large Eddy Simulation of an isolated vertical axis wind turbine}},
\newblock \bibinfo{journal}{Journal of Wind Engineering and Industrial Aerodynamics} \bibinfo{volume}{172} (\bibinfo{year}{2018}) \bibinfo{pages}{139--151}. \URLprefix \url{https://www.sciencedirect.com/science/article/pii/S0167610517305123}. \DOIprefix\doi{https://doi.org/10.1016/j.jweia.2017.11.004}.
\bibitem[{{Simao Ferreira} et~al.(2008){Simao Ferreira}, {van Kuik}, {van Bussel}, and Scarano}]{[121]}
\bibinfo{author}{C.~{Simao Ferreira}}, \bibinfo{author}{G.~{van Kuik}}, \bibinfo{author}{G.~{van Bussel}}, \bibinfo{author}{F.~Scarano},
\newblock \bibinfo{title}{Visualization by {PIV} of dynamic stall on vertical axis wind turbine},
\newblock \bibinfo{journal}{Experiments in Fluids: experimental methods and their applications to fluid flow} \bibinfo{volume}{46} (\bibinfo{year}{2008}) \bibinfo{pages}{97--108}.
\bibitem[{Posa(2020)}]{[117]}
\bibinfo{author}{A.~Posa},
\newblock \bibinfo{title}{{Influence of Tip Speed Ratio on wake features of a Vertical Axis Wind Turbine}},
\newblock \bibinfo{journal}{Journal of Wind Engineering and Industrial Aerodynamics} \bibinfo{volume}{197} (\bibinfo{year}{2020}) \bibinfo{pages}{104076}. \URLprefix \url{https://www.sciencedirect.com/science/article/pii/S0167610519306774}. \DOIprefix\doi{https://doi.org/10.1016/j.jweia.2019.104076}.
\bibitem[{Khosronejad et~al.(2017)Khosronejad, Diplas, and Sotiropoulos}]{[122]}
\bibinfo{author}{A.~Khosronejad}, \bibinfo{author}{P.~Diplas}, \bibinfo{author}{F.~Sotiropoulos},
\newblock \bibinfo{title}{Simulation-based optimization of in–stream structures design: bendway weirs},
\newblock \bibinfo{journal}{Environmental Fluid Mechanics} \bibinfo{volume}{17} (\bibinfo{year}{2017}) \bibinfo{pages}{79--109}. \URLprefix \url{https://doi.org/10.1007/s10652-016-9452-5}. \DOIprefix\doi{10.1007/s10652-016-9452-5}.
\bibitem[{van Rijn(1984)}]{[119]}
\bibinfo{author}{L.~C. van Rijn},
\newblock \bibinfo{title}{{Sediment Transport, Part III: Bed forms and Alluvial Roughness}},
\newblock \bibinfo{journal}{Journal of Hydraulic Engineering} \bibinfo{volume}{110} (\bibinfo{year}{1984}) \bibinfo{pages}{1733--1754}. \DOIprefix\doi{10.1061/(ASCE)0733-9429(1984)110:12(1733)}.
\bibitem[{Ouro and Stoesser(2017)}]{[124]}
\bibinfo{author}{P.~Ouro}, \bibinfo{author}{T.~Stoesser},
\newblock \bibinfo{title}{Wake generated downstream of a vertical axis tidal turbine},
\newblock \bibinfo{year}{2017}.
\bibitem[{Liu et~al.(2021)Liu, Yu, and Zhu}]{[126]}
\bibinfo{author}{K.~Liu}, \bibinfo{author}{M.~Yu}, \bibinfo{author}{W.~Zhu},
\newblock \bibinfo{title}{Performance analysis of vertical axis water turbines under single-phase water and two-phase open channel flow conditions},
\newblock \bibinfo{journal}{Ocean Engineering} \bibinfo{volume}{238} (\bibinfo{year}{2021}) \bibinfo{pages}{109769}. \URLprefix \url{https://www.sciencedirect.com/science/article/pii/S0029801821011355}. \DOIprefix\doi{https://doi.org/10.1016/j.oceaneng.2021.109769}.
\bibitem[{Betz(1966)}]{[127]}
\bibinfo{author}{A.~Betz}, \bibinfo{title}{{Introduction to the Theory of Flow Machines}}, \bibinfo{publisher}{Pergamon Press}, \bibinfo{year}{1966}. \URLprefix \url{https://books.google.com/books?id=3-JSAAAAMAAJ}.
\bibitem[{Lee et~al.(2020)Lee, Kim, Khosronejad, and Kang}]{[148]}
\bibinfo{author}{J.~Lee}, \bibinfo{author}{Y.~Kim}, \bibinfo{author}{A.~Khosronejad}, \bibinfo{author}{S.~Kang},
\newblock \bibinfo{title}{Experimental study of the wake characteristics of an axial flow hydrokinetic turbine at different tip speed ratios},
\newblock \bibinfo{journal}{Ocean Engineering} \bibinfo{volume}{196} (\bibinfo{year}{2020}) \bibinfo{pages}{106777}. \URLprefix \url{https://www.sciencedirect.com/science/article/pii/S0029801819308777}. \DOIprefix\doi{https://doi.org/10.1016/j.oceaneng.2019.106777}.
\bibitem[{Liu et~al.(2022)Liu, Li, Yang, Xu, Kang, and Khosronejad}]{[149]}
\bibinfo{author}{X.~Liu}, \bibinfo{author}{Z.~Li}, \bibinfo{author}{X.~Yang}, \bibinfo{author}{D.~Xu}, \bibinfo{author}{S.~Kang}, \bibinfo{author}{A.~Khosronejad},
\newblock \bibinfo{title}{Large-eddy simulation of wakes of waked wind turbines},
\newblock \bibinfo{journal}{Energies} \bibinfo{volume}{15} (\bibinfo{year}{2022}). \URLprefix \url{https://www.mdpi.com/1996-1073/15/8/2899}. \DOIprefix\doi{10.3390/en15082899}.
\bibitem[{Khosronejad et~al.(2020{\natexlab{a}})Khosronejad, Flora, and Kang}]{[150]}
\bibinfo{author}{A.~Khosronejad}, \bibinfo{author}{K.~Flora}, \bibinfo{author}{S.~Kang},
\newblock \bibinfo{title}{Effect of inlet turbulent boundary conditions on scour predictions of coupled les and morphodynamics in a field-scale river: Bankfull flow conditions},
\newblock \bibinfo{journal}{Journal of Hydraulic Engineering} \bibinfo{volume}{146} (\bibinfo{year}{2020}{\natexlab{a}}) \bibinfo{pages}{04020020}. \URLprefix \url{https://ascelibrary.com/doi/abs/10.1061/%28ASCE%29HY.1943-7900.0001719}. \DOIprefix\doi{10.1061/(ASCE)HY.1943-7900.0001719}.
\bibitem[{Khosronejad et~al.(2020{\natexlab{b}})Khosronejad, Diplas, Angelidis, Zhang, Heydari, and Sotiropoulos}]{[151]}
\bibinfo{author}{A.~Khosronejad}, \bibinfo{author}{P.~Diplas}, \bibinfo{author}{D.~Angelidis}, \bibinfo{author}{Z.~Zhang}, \bibinfo{author}{N.~Heydari}, \bibinfo{author}{F.~Sotiropoulos},
\newblock \bibinfo{title}{Scour depth prediction at the base of longitudinal walls: a combined experimental, numerical, and field study},
\newblock \bibinfo{journal}{Environmental Fluid Mechanics} \bibinfo{volume}{20} (\bibinfo{year}{2020}{\natexlab{b}}) \bibinfo{pages}{459--478}. \URLprefix \url{https://doi.org/10.1007/s10652-019-09704-x}. \DOIprefix\doi{10.1007/s10652-019-09704-x}.
\bibitem[{Khosronejad et~al.(2020{\natexlab{c}})Khosronejad, Flora, Zhang, and Kang}]{[152]}
\bibinfo{author}{A.~Khosronejad}, \bibinfo{author}{K.~Flora}, \bibinfo{author}{Z.~Zhang}, \bibinfo{author}{S.~Kang},
\newblock \bibinfo{title}{Large-eddy simulation of flash flood propagation and sediment transport in a dry-bed desert stream},
\newblock \bibinfo{journal}{International Journal of Sediment Research} \bibinfo{volume}{35} (\bibinfo{year}{2020}{\natexlab{c}}) \bibinfo{pages}{576--586}. \URLprefix \url{https://www.sciencedirect.com/science/article/pii/S1001627920300068}. \DOIprefix\doi{https://doi.org/10.1016/j.ijsrc.2020.02.002}.
\bibitem[{Khosronejad et~al.(2018)Khosronejad, Kozarek, Diplas, Hill, Jha, Chatanantavet, Heydari, and Sotiropoulos}]{[153]}
\bibinfo{author}{A.~Khosronejad}, \bibinfo{author}{J.~L. Kozarek}, \bibinfo{author}{P.~Diplas}, \bibinfo{author}{C.~Hill}, \bibinfo{author}{R.~Jha}, \bibinfo{author}{P.~Chatanantavet}, \bibinfo{author}{N.~Heydari}, \bibinfo{author}{F.~Sotiropoulos},
\newblock \bibinfo{title}{Simulation-based optimization of in-stream structures design: rock vanes},
\newblock \bibinfo{journal}{Environmental Fluid Mechanics} \bibinfo{volume}{18} (\bibinfo{year}{2018}) \bibinfo{pages}{695--738}. \URLprefix \url{https://doi.org/10.1007/s10652-018-9579-7}. \DOIprefix\doi{10.1007/s10652-018-9579-7}.
\bibitem[{Bachant and Wosnik(2016)}]{[130]}
\bibinfo{author}{P.~Bachant}, \bibinfo{author}{M.~Wosnik},
\newblock \bibinfo{title}{Effects of reynolds number on the energy conversion and near-wake dynamics of a high solidity vertical-axis cross-flow turbine},
\newblock \bibinfo{journal}{Energies} \bibinfo{volume}{9} (\bibinfo{year}{2016}) \bibinfo{pages}{73}. \DOIprefix\doi{10.3390/en9020073}.
\bibitem[{Bachant and Wosnik(2014)}]{[131]}
\bibinfo{author}{P.~Bachant}, \bibinfo{author}{M.~Wosnik}, \bibinfo{title}{{UNH-RVAT baseline performance and near-wake measurements: Reduced dataset and processing code}}, \bibinfo{howpublished}{fig\textbf{share}. http://dx.doi.org/10.6084/m9.figshare.1080781}, \bibinfo{year}{2014}. \URLprefix \url{http://dx.doi.org/10.6084/m9.figshare.1080781}. \DOIprefix\doi{10.6084/m9.figshare.1080781}.
\bibitem[{[13(2015)}]{[132]}
\bibinfo{title}{{UNH-RVAT Reynolds number dependence experiment: Reduced dataset and processing code}, author = {Peter Bachant and Martin Wosnik}}, \bibinfo{howpublished}{fig\textbf{share}. http://dx.doi.org/10.6084/m9.figshare.1286960}, \bibinfo{year}{2015}. \URLprefix \url{http://dx.doi.org/10.6084/m9.figshare.1286960}. \DOIprefix\doi{10.6084/m9.figshare.1286960}.
\bibitem[{Neary et~al.(2014)Neary, Lawson, Previsic, Copping, Hallett, Labonte, Rieks, and Murray}]{[133]}
\bibinfo{author}{V.~S. Neary}, \bibinfo{author}{M.~Lawson}, \bibinfo{author}{M.~Previsic}, \bibinfo{author}{A.~Copping}, \bibinfo{author}{K.~C. Hallett}, \bibinfo{author}{A.~Labonte}, \bibinfo{author}{J.~Rieks}, \bibinfo{author}{D.~Murray},
\newblock \bibinfo{title}{Methodology for design and economic analysis of marine energy conversion ({MEC}) technologies}  (\bibinfo{year}{2014}).

\end{thebibliography}

\end{document}